\documentclass[usenatbib]{mn2e}
\usepackage{natbib}
\usepackage{epsfig}
\usepackage{amsmath}

\title[Water towards methanol: 6$^{\circ}$ to 20$^{\circ}$]{A search for water masers associated with class II methanol masers - I. Longitude range 6$^{\circ}$ to 20$^{\circ}$}
\author[A. M. Titmarsh et al.]{A. M. Titmarsh$^{1,2}$\thanks{E-mail:
Anita.Titmarsh@utas.edu.au}, S. P. Ellingsen$^{1}$, S. L. Breen$^{2}$, J. L. Caswell$^{2}$, M. A. Voronkov$^{2}$ \\
$^{1}$School of Mathematics and Physics, University of Tasmania, Private Bag 37, Hobart, Tasmania 7001, Australia \\
$^{2}$CSIRO Astronomy and Space Science, Australia Telescope National Facility, PO Box 76, Epping, NSW 1710} 
\begin{document}

\date{Received 2014 April 1; accepted 2014 July 3.}

\pagerange{\pageref{firstpage}--\pageref{lastpage}} \pubyear{2014}

\maketitle

\label{firstpage}

\begin{abstract}

The Australia Telescope Compact Array has been used to search for 22-GHz water masers towards the 119 6.7-GHz methanol masers detected in the Methanol Multi-Beam survey between Galactic longitudes $6^{\circ}$ and $20^{\circ}$; we find water masers associated with 55 ($\sim$46 per cent). Methanol masers with associated water masers have a higher mean integrated luminosity than those without and there is a general trend for sources with more luminous 6.7-GHz methanol masers to be associated with more luminous water maser emission. We have inspected the GLIMPSE three colour images of the regions surrounding the masers and cross-matched the maser positions with existing catalogues of Extended Green Objects and Infrared Dark Clouds. We find more Extended Green Objects at sites where both methanol and water masers are present than at sites with only methanol masers, but no significant difference in the fraction embedded within Infrared Dark Clouds. Analysis of the 1.1-mm dust emission shows dust clumps associated with masers that have greater flux densities and higher column densities than those without. Dust clumps associated with both water and 6.7-GHz methanol masers are generally the most compact clumps followed by those associated with only methanol then the clumps without associated maser emission. We conclude that protostars with both methanol and water masers are often older than those with only methanol, however, we suggest that the evolutionary phase traced by water masers is not as well defined as for 6.7-GHz methanol masers.

\end{abstract}


\section{Introduction}
\label{sec:intro}
Some of the best signposts of high-mass star formation are interstellar masers. They are common, intense and, being observable at radio frequencies, they do not suffer the high extinction that affects other frequency bands. The most common maser species are water, methanol and hydroxyl, and along with maser pumping models, they provide us with valuable information about the physical conditions at sites of star formation. Water masers are collisionally pumped and are the most widespread of the maser species known \citep[e.g.][]{walsh11}, being present not only in association with young high mass stars, but also in other environments such as dense circumstellar shells around evolved stars, often tracing shocked gas and outflows.

However, in regions of massive star formation, methanol masers are especially useful. The different methanol transitions have been empirically divided into two categories \citep[class I and class II; ][]{batrla87,menten91}. The pumping mechanism of class I methanol masers is predominantly from collisions with molecular hydrogen \citep{cragg92} and they are typically observed to be distributed on scales of a few to 10s of arcseconds within a star formation region \citep{voronkov06,voronkov14}.  Class II methanol masers (such as the $5_1 - 6_0$ A$^+$ transition at 6.7-GHz) are pumped by far-infrared radiation \citep{sobolev94,cragg05} and are observed exclusively at sites of high-mass star formation \citep[]{minier03,xu08,bartkiewicz11,breen13}. This is because methanol is produced in high abundance in relatively restricted circumstances. It forms on the mantles of dust grains at temperatures $<$~10~K during the cold-core phase of high-mass star formation \citep{taquet13}, is released into the gas phase as the temperature rises, but then rapidly depleted through gas phase reactions. Water molecules, on the other hand, are found in a much wider variety of conditions and the 22-GHz maser is excited in high velocity outflows as well as in the vicinity of the YSO, and also around late-type stars. According to the model by \cite{elitzur89} they need dissociative shocks in dense gas with temperatures up to $\sim$400~K and preshock densities $\sim 10^7$~cm$^{-3}$ to form. Methanol masers can form in similar densities ($10^7$~cm$^{-3}$ - 10$^9$~cm$^{-3}$) but they are favoured by lower temperatures \citep[100 K - 150 K;][]{cragg02,cragg05}.

Masers can be associated with shock tracers such as extended emission in the 4.5~$\mu$m band, often referred to as either Extended Green Objects \citep[EGOs;][]{cyganowski08}, or ``green fuzzies'' \citep{chambers09} as the 4.5~$\mu$m emission is usually coloured green in \textit{Spitzer} Galactic Legacy Infrared Midplane Survey Extraordinaire \citep[GLIMPSE]{benjamin03} three colour images. The 4.5~$\mu$m band covers the wavelength range of a number of H$_{2}$ and CO spectral lines which are excited by shocks. The presence of strong, extended emission in this band has been shown to be a good tracer of outflows from protostellar objects.

Residing within the dusty molecular cores where high-mass stars are formed, the presence and/or absence of various maser transitions have been proposed to trace different evolutionary stages in their formation \citep{ellingsen13,breen10ev,ellingsen07}.  The masers are sometimes  associated with ultra-compact H{\sc ii} regions \citep{phillips98,walsh98} and more often at a younger stage, embedded within infra-red dark clouds \citep{ellingsen06}. \cite{breen10ev} show, through statistical analysis of maser presence/absence and multi-wavelength continuum observations, that the presence of 12.2-GHz methanol and OH masers signpost star forming regions that are more evolved than those showing just 6.7-GHz methanol masers. The largest uncertainties remaining in the evolutionary sequence relate to water masers and class I methanol masers. Despite a number of previous studies encompassing both methanol and water masers \citep[e.g.][]{beuther02,szymczak05,xu08}, the characteristics of the objects associated with these masers are still not well understood, mainly due to poor spatial resolution and biases in the target sources.

The Methanol Multibeam (MMB) survey is an unbiased survey of the Galactic Plane covering $\pm 2^{\circ}$ latitude for the 6.7-GHz methanol maser transition \citep{green09} and the catalogues covering the longitude range $l = 186^{\circ} - 20^{\circ}$ have been published \citep[][]{caswell10,caswell11,green10,green12}. The Parkes 64 m dish performed the initial search of the southern Galactic Plane and accurate positions for the sites of maser emission were obtained with interferometric observations carried out with the Australia Telescope Compact Array (ATCA). For sources further north, some MMB positions have been obtained with the Multi-Element Radio Linked Interferometer Network (MERLIN).

Here we report results of follow-up 22-GHz water maser observations made with the ATCA towards the MMB detections in the $l=6^{\circ} - 20^{\circ}$ range. The MMB sources in the range $l = 310^{\circ}$ through the Galactic Centre to $6^{\circ}$ have also been observed for water maser emission and will be the subject of future publications. Although the water observations are targeted (rather than an unbiased search), we will be able to combine the results of these sensitive, high resolution observations, with the less sensitive, but statistically complete HOPS survey \citep{walsh11}, to properly quantify the water maser transition within the maser-based evolutionary scheme.

\section[]{Observations and data reduction}

The water maser observations were carried out using the ATCA antennas in the H214 and H168 array configurations. These hybrid array configurations have both East - West and North - South baselines and so provide better \textit{uv} coverage for equatorial sources. 

Although the observations were made with a relatively compact array, configurations having longer baselines would not necessarily equate to greater astrometric accuracy. The absolute astrometric accuracy for the ATCA is about 0.4 arcseconds \citep[set by the astrometric accuracy of the phase calibrators and the degree to which they are point sources for the array configuration and the frequency of the observation;][]{caswell97}. The astrometric accuracy for these observations is estimated to range from 0.4 - 2.0 arcseconds (as some of the observations were made in poor weather). Weather is a major consideration when observing water masers because of absorption and emission in the 6$_{16}$ - 5$_{23}$ transition by the water vapour in the Earth's atmosphere. 

The primary beam for the ATCA at 22~GHz is 2.1~arcminutes and the synthesised beams of the H214 and H168 configurations are $\sim$~9.6 and $\sim$~12.4 arcseconds respectively. Some of our sources were observed by \cite{breen10oh} with the ATCA and our positions agreed with the positions they obtained to within $\sim$ 2 arcseconds. Uncertainty is also introduced as water maser clusters are typically spread over linear scales of around 30~mpc \citep[1 arcsecond angular scale at 6~kpc; see][]{forster89}; in some regions the linear scales are as much as 5 time greater \citep[4 arcsecond angular scale at 8~kpc; see][]{reid88}.

The observations were conducted over two epochs, 2010 November 2-3 (taken with the H214 array configuration) and 2011 August 10 (with the H168 array). Our targets consisted of all 6.7-GHz methanol masers in the $l = 6^{\circ} - 20^{\circ}$ region, including those which had previously been observed. This is because water masers are known to have extremely variable flux densities \citep[e.g.][]{breen10oh} and we wish to have a statistically complete sample of masers at one epoch. 

We observed groups of six 6.7-GHz methanol maser targets which were close together on the sky for 1.5~minutes each before observing a phase calibrator for 1.5~minutes. Each methanol maser position was observed at least four times over an hour angle range of at least six hours to ensure sufficient \textit{uv} coverage for imaging giving a total on-source time of at least six minutes. Observations of PKS B1934-638 were taken for primary flux density calibration (expected to be accurate to $\sim$~20 per cent), PKS B1253-055 was used for bandpass calibration and PKS B1730-30 was used for phase calibration of the targets in the range $l=6^{\circ} - 15^{\circ}$ and PKS B1829-106 for $l=15^{\circ} - 20^{\circ}$. 

The Compact Array Broadband Backend \citep[CABB;][]{wilson11} was configured with two bands to observe the 22.235-GHz water maser transition in the first band and the ammonia (1,1) and (2,2) transitions in the second  band centred between the two transitions at 23.708 GHz. Each band had 64~MHz bandwidth with 2048 spectral channels corresponding to a velocity width of 0.42~km~s$^{-1}$ and 0.39~km~s$^{-1}$ for the water and ammonia transitions respectively, velocity resolutions for uniform weighting of the spectral channels of 0.50~km~s$^{-1}$ and 0.47~km~s$^{-1}$ and velocity coverages of $>$ 800 km~s$^{-1}$. The rms noise in an individual spectral channel for these observations ranged from $\sim$40 mJy to $\sim$80 mJy or $\sim$100 mJy - $\sim$160 mJy depending on weather conditions and total time on-source. In addition 2 x 2 GHz continuum bands with 32 x 64~MHz channels were available for the 2011 August observations. The results of the ammonia and continuum observations will be reported in future publications.

The water maser data were reduced in {\sc miriad} \cite[]{sault95} using standard techniques for ATCA spectral line data. We inspected the image cubes within an arcminute radius of each methanol maser position to find the associated water masers and the spectra were produced from these image cubes by integrating the emission at each maser site. 

\section{Results}

In this paper we report the 22-GHz water masers detected towards 6.7-GHz methanol masers in the $l=6^{\circ} - 20^{\circ}$ range. We found that 55 of the 119 ($\sim$ 46 per cent) 6.7-GHz methanol masers had an associated water maser. 

Previous high resolution, large surveys of masers with the ATCA \citep[e.g.][]{breen10oh,caswell10} have used a three arcsecond criteria to establish if different maser species are associated with the same object. In Fig 1, we show a histogram of the separations between methanol maser targets and our detected water masers.  This provides excellent empirical justification for adopting 3 arcseconds as an appropriate threshold for real association, with slightly larger separations requiring individual consideration. For the most distant sources in our sample 3~arcseconds corresponds to a linear scale of $\sim$200~mpc. Water masers 6.795-0.257 and 10.342-0.142 are separated from their nearest methanol targets by 3.4 and 3.2 arcseconds respectively. The \cite{breen10oh} positions for these two masers are a little closer to the methanol masers than the positions obtained in these observations; inspection of the \textit{Spitzer} Galactic Legacy Infrared Midplane Survey Extraordinaire (GLIMPSE) three colour images point to the emission coming from the same object and both water masers have emission at similar velocities to that of their target methanol maser.

\begin{figure}
\includegraphics[width=3in]{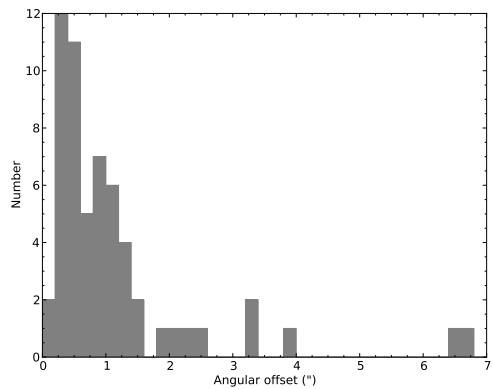}
\caption{Separations between methanol maser targets and our detected water masers. Some water masers further than 7 arcseconds from their targets were also detected. We have zoomed in on this axis for clarity.}
\label{fig:sep}
\end{figure}

Water masers associated with 6.7-GHz methanol maser targets are presented in Table \ref{table:assoc_sources}:  column 1 is the source name (Galactic Longitude and Latitude) of the target 6.7-GHz methanol maser; columns 2 and 3 give the position of the water maser in Right Ascension and Declination (J2000); column 4 the peak velocity; column 5 and 6 the minimum and maximum velocities of the emission; column 7 the peak flux densities; column 8 the integrated flux densities; column 9 the peak velocity of the associated 6.7-GHz methanol masers; column 10 the angular offsets; column 11 the epoch of the observations; column 12 distances are mostly from \cite{green11}, of which a few are astrometric distances but the majority are the kinematic distances that they found from H{\sc i} self-absorption; the remainder are determined from the 6.7-GHz methanol peak velocities using the \cite{reid09} rotation curve. Column 13 lists other associated maser species from surveys by \citet[]{breen12} and \citet[]{caswell98,caswell13} and associations with Infrared Dark Clouds and Extended Green Objects. The spectra taken with the ATCA are presented in Figure \ref{fig:assoc}. Individual sources of interest are discussed further in Section \ref{sec:indiv}.

The methanol maser sites observed for which we found no associated water emission are listed in Table \ref{table:notass_sources} along with the 5$\sigma$ detection limits for each source. Note that 6 of the sites in Table \ref{table:notass_sources} have shown water emission at other epochs \citep{breen10oh,pillai06b}, as identified in the 4th column (associations) and remarked on in Section \ref{sec:indiv}. Thus although our single epoch search resulted in detections toward 55 of the 119 sites searched, the extension to a multi-epoch search, which partially compensates for maser variability over several years, increases this to at least 61/119 (51 per cent).

\begin{table*}
\caption{Positions and parameters of 55 water masers associated with 6.7-GHz methanol masers. Epochs are coded 1, 2, 3 and 4 for 2010 November 2, 2010 November 3, 2011 August 9 and 2011 August 10 respectively. Associations with 12.2-GHz methanol masers (Breen et al., 2012) are indicated with an `m' and associations with the 1665~MHz transition of OH masers (Caswell, 1998; Caswell et al. 2013) are indicated with an `o'. From inspection of the GLIMPSE infrared images, masers associated with an Infrared Dark Cloud or Extended Green Object are indicated with an `i' or `e' respectively. Sources outside of the GLIMPSE survey range are indicated with an `*'. Distances estimates are from Green et al. (2011) where available, otherwise they are calculated using the near kinematic distance (these are in italics).}
\begin{tabular}{l l l l l l l l l l l l l l l}
\hline
MMB Target & \multicolumn{2}{l}{Equatorial Coordinates} & & & & & S$_{int}$ & \multicolumn{3}{l}{6.7-GHz methanol} & & & & \\
Name & RA & DEC & V$_{pk}$ & V$_{l}$ & V$_{h}$ & S$_{pk}$ & (Jy km & V$_{pk}$ & V$_{l}$ & V$_{h}$ & Offsets & Epoch & Dist. & Assoc. \\
(l, b) & (h m s) & ($^{\circ}$ ' ") & \multicolumn{3}{c}{(km s$^{-1}$)} & (Jy) & s$^{-1}$) & \multicolumn{3}{c}{(km s$^{-1}$)} & (") & & (kpc) & \\
\hline
06.189--0.358    & 18 01 02.16  &   --23 47 10.4  &  -29.4    & -48.3 & -24.7 &  13.1   & 36.5    & -30.2 & -37.5 & -27.1   & 0.4 & 3  & 5.1   &  ie  \\ 
06.588--0.192    & 18 01 16.07  &   --23 21 25.7  &  -1.4     &  -3.5 &  -0.1 &  1.6    & 3.0     & 5.0   & 3.5 & 7.0   & 1.5 & 3  & 1.4   &  --  \\
06.610--0.082    & 18 00 54.03  &   --23 17 02.8  &  2.9      &  -4.2 &   7.1 &  3.7    & 21.8    & 0.8   & -6.6 & 7.5 & 0.3 & 2  & \textit{0.3}   &  m   \\
06.795--0.257    & 18 01 57.53  &   --23 12 33.0  &  14.7     &  -3.0 &  19.3 &  19.8   & 122.5   & 16.3  & 12.1 & 31.4   & 3.4 & 2  & 3.1   & o \\ 
07.166+0.131     & 18 01 17.47  &   --22 41 43.7  &  76.3     &  71.2 &  81.7 &  3.6    & 9.1     & 85.7  & 74.5 & 91.0  & 0.2 & 3  & 11.6  &  --  \\
07.601--0.139    & 18 03 14.40  &   --22 27 00.7  &  148.5    & 145.1 & 154.0 &  1.4    & 9.7     & 154.7 & 151.0 & 156.5   & 0.3 & 3  & 7.4   &  --  \\
07.632--0.109    & 18 03 11.59  &   --22 24 32.0  &  155.3    & 143.4 & 156.5 &  1.18   & 3.8     & 157.0 & 146.5 & 158.9  & 0.6 & 3  & 7.4   &  --  \\
08.139+0.226     & 18 03 00.83  &   --21 48 10.7  &   27.2    &  10.3 &  31.8 &  0.4    & 4.1     & 19.9  & 18.8 & 21.8  & 1.5 & 2  & 3.2   &  m   \\
08.317--0.096    & 18 04 36.09  &   --21 48 20.2  &   47.3    &  31.6 &  48.5 &  0.5    & 2.3     & 47.1  & 44.0 & 49.2 & 1.2 & 3  & 11.7  &  i   \\
08.669--0.356    & 18 06 18.97  &   --21 37 32.2  &  34.2     &  22.7 &  49.3 &  14.4   & 114.9   & 39.0  & 35.8 & 39.7  & 0.2 & 2  & 4.4   &  om  \\
08.832--0.028    & 18 05 25.65  &   --21 19 24.8  &  -9.8     & -26.6 &   9.9 &  12.3   & 93.3    & -3.8  & -6.0 & 5.9  & 0.3 & 3  & 5.2   &  e   \\
09.215--0.202    & 18 06 52.83  &   --21 04 27.1  &  36.3     &  34.6 &  39.6 &  2.7    & 12.7    & 45.6  & 36.0 & 50.0 & 0.4 & 3  & \textit{4.6}   &  i   \\
09.621+0.196	 & 18 06 14.77  &   --20 31 33.8  &  -6.0     & -14.8 &  11.7 &  18.2   & 222.2   & 1.3   & -4.8 & 8.9  & 2.1 & 2  & 5.2   &  om  \\ 
09.986--0.028    & 18 07 50.14  &   --20 18 56.4  &  52.5     &  41.1 &  63.0 &  12.9   & 72.8    & 42.2  & 40.6 & 51.8  & 0.3 & 2  & 12.0  &  mi  \\
10.287--0.125    & 18 08 49.33  &   --20 05 58.9  &  13.8     &   2.8 &  15.0 &  1.0    & 7.8     & 4.5   & 1.5 & 6.0 & 0.4 & 2  & \textit{0.7}   &  mie \\
10.320--0.259    & 18 09 23.28  &   --20 08 06.6  &  43.1     &  40.5 &  46.4 &  5.4    & 13.2    & 39.0  & 35.0 & 39.6  & 0.3 & 3  & \textit{3.9}   &  --  \\
10.342--0.142    & 18 08 59.96  &   --20 03 39.1  &  10.0     &  -1.8 &  65.1 &  10.5   & 191.0   & 14.8  & 6.0 & 18.0  & 3.2 & 2  & \textit{1.8}   &  --  \\ 
10.444--0.018    & 18 08 44.88  &   --19 54 38.4  &  71.5     &  65.6 &  81.2 &  12.6   & 75.0    & 73.4  & 67.6 & 79.0 & 0.3 & 2  & 11.0   &  omi \\
10.472+0.027     & 18 08 38.54  &   --19 51 50.1  &  87.6     &  18.4 & 286.8 &  75.3   & 1337.1  & 75.1  & 57.5 & 77.6 & 0.9 & 2  & 8.5    &  om  \\
10.724--0.334    & 18 10 29.99  &   --19 49 06.1  &   -21.6   & -25.4 & -18.2 &  0.8    & 2.4     & -2.1  & -2.5 & -1.6  & 0.9 & 3  & 5.2    &  --  \\
10.822--0.103    & 18 09 50.45  &   --19 37 13.9  &   63.5    &  61.3 &  67.6 &  0.4    & 1.7     & 72.1  & 68.0 & 74.0  & 1.0 & 3  & \textit{5.3}   &  --  \\
10.886+0.123	 & 18 09 07.95  &   --19 27 24.0  &  13.8     &  11.2 &  26.8 &  21.1   & 68.9    & 17.2  & 14.0 & 22.5   & 2.2 & 3  & 2.5    &  e   \\ 
10.958+0.022     & 18 09 39.26  &   --19 26 27.7  &  24.4     &  18.5 &  27.3 &  5.9    & 19.9    & 24.5  & 23.0 & 25.5   & 0.8 & 2  & 13.5   &  --  \\
11.034+0.062     & 18 09 39.86  &   --19 21 20.2  &   17.7    &   5.4 &  22.7 &  0.5    & 4.7     & 20.6  & 15.2 & 21.0  & 0.4 & 2  & 2.4    &  o   \\
12.199--0.033    & 18 12 23.42  &   --18 22 51.2  &  47.4     &  44.0 &  55.3 &  0.9    & 3.9     & 49.3  & 48.2 & 57.1   & 0.6 & 4 & 12.0   &  --  \\
12.209--0.102    & 18 12 39.76  &   --18 24 18.0  &  22.0     & -13.4 &  41.7 &  41.3   & 1170.7  & 19.8  & 16.0 & 22.0  & 1.4 & 2  & \textit{2.3}   &  om   \\
12.265--0.051    & 18 12 35.37  &   --18 19 52.1  &  59.2     &  56.6 &  63.4 &  2.6    & 11.7    & 68.3  & 58.0 & 70.9   & 0.5 & 4 & 11.5   &  --  \\
12.681--0.182    & 18 13 54.75  &   --18 01 46.5  &   59.9    &  45.1 &  74.2 &  702.0  & 2491.9  & 57.5  & 50.0 & 62.0   & 0.0 & 2  & \textit{4.5}   &  om  \\
12.889+0.489     & 18 11 51.39  &   --17 31 28.8  &  29.9     &  23.9 &  32.8 &  12.3   & 44.8    & 39.2  & 28.0 & 43.0   & 0.8 & 2  & 2.3    &  om  \\
12.904--0.031    & 18 13 48.27  &   --17 45 39.7  &  66.2     &  60.7 &  69.1 &  13.5   & 70.4    & 59.1  & 55.8 & 61.0   & 0.9 & 4 & \textit{4.5}   &  e  \\
12.909--0.260    & 18 14 39.51  &   --17 52 01.2  &  37.5     &  33.3 &  42.1 &  4.7    & 19.2    & 39.9  & 34.7 & 47.0   & 1.3 & 2  & \textit{3.7}   &  ome \\
13.657--0.599    & 18 17 24.25  &   --17 22 12.6  &  47.7     &  34.5 &  55.6 &  10.1   & 95.0    & 51.3  & 45.0 & 52.7   & 0.2 & 4 & 12.3   &  o   \\
14.101+0.087     & 18 15 45.80  &   --16 39 09.3  &  8.5      &  -1.5 &  10.6 &  14.5   & 43.7    & 15.4  & 4.4 & 16.6  & 0.1 & 4 & 5.4    &  --  \\
14.604+0.017     & 18 17 01.14  &   --16 14 39.1  &  27.4     &  16.0 &  35.3 &  8.2    & 86.9    & 24.7  & 22.1 & 35.8  & 1.2 & 4 & 2.8    &  m   \\
15.094+0.192     & 18 17 20.84  &   --15 43 45.7  &  28.3     & -14.6 &  32.0 &  13.2   & 99.9    & 25.8  & 22.5 & 26.5 & 0.8 & 4 & 13.8   &  --  \\
15.665--0.499    & 18 20 59.79  &   --15 33 09.9  &  -6.8     &  -9.3 &  -1.3 &  11.3   & 58.9    & -2.9  & -5.0 & -2.0   & 0.7 & 1  & 16.7   &  e   \\
16.585--0.051    & 18 21 09.16  &   --14 31 49.0  &  64.6     &  54.0 &  72.1 &  26.7   & 196.2   & 62.1  & 52.0 & 69.5   & 0.7 & 2  & 4.3    &  om   \\
16.831+0.079     & 18 21 09.52  &   --14 15 08.8  &  41.1     &  38.1 &  84.5 &  1.2    & 13.2    & 58.7  & 57.2 & 69.4  & 0.3 & 1  & 11.8   &  --  \\
16.864--2.159    & 18 29 24.37  &   --15 16 04.5  &  6.1      & -12.4 &  25.4 &  6.8    & 40.3    & 15.0  & 14.0 & 20.0   & 0.6 & 1  & 1.70   &  om * \\
17.021--2.403    & 18 30 36.31  &   --15 14 28.1  &  21.1     &  15.2 &  27.4 &  5.3    & 25.3    & 23.6  & 17.0 & 25.0  & 0.4 & 1  & 2.0    &  --* \\
17.638+0.157	 & 18 22 26.47  &   --13 30 11.6  &  19.7     &  12.9 &  29.3 &  7.4    & 55.5    & 20.8  & 20.0 & 22.0  & 2.5 & 2  & 2.0    &  o   \\ 
18.341+1.768     & 18 17 58.19  &   --12 07 25.2  &   13.4    &   5.7 &  16.3 &  0.9    & 11.4    & 28.1  & 26.0 & 32.0  & 1.0 & 1  & \textit{2.2}   &  --* \\
18.661+0.034     & 18 24 51.10  &   --12 39 21.9  &  78.4     &  21.0 &  86.3 &  1.3    & 16.7    & 79.0  & 76.0 & 83.0 & 0.5 & 1  & 11.2   &  --  \\
18.735--0.227    & 18 25 56.46  &   --12 42 48.0  &  33.7     &  18.9 &  52.6 &  54.4   & 590.0   & 38.2  & 36.3 & 38.5   & 1.4 & 1  & 13.0   &  --  \\
18.999--0.239    & 18 26 29.32  &   --12 29 07.8  &  -11.9    & -27.0 &   4.9 &  5.6    & 76.4    & 69.4  & 65.0 & 69.8 & 1.4 & 1  & 4.3    &  i   \\
19.009--0.029    & 18 25 44.80  &   --12 22 45.4  &  67.4     &  65.3 &  68.2 &  0.5    & 2.4     & 55.4  & 53.0 & 63.0    & 0.8 & 1  & 12.0   &  e  \\
19.267+0.349     & 18 24 52.36  &   --11 58 28.0  &  27.2     &  13.6 &  31.3 &  7.5    & 37.9    & 16.3  & 12.5 & 17.5  & 0.3 & 1  & 14.5   &  --  \\
19.472+0.170n    & 18 25 54.69  &   --11 52 33.9  &  21.6     & -84.9 &  25.8 &  9.2    & 72.3    & 21.7  & 17.0 & 23.0 & 1.0 & 1  & 1.8    &  oi   \\
19.486+0.151     & 18 26 00.42  &   --11 52 22.0  &  26.0     &  22.2 &  29.3 &  2.7    & 10.2    & 20.9  & 19.0 & 27.5   & 0.7 & 1  & 2.0    &  o   \\
19.496+0.115     & 18 26 09.28  &   --11 52 51.7  &   125.0   & 107.3 & 128.8 &  0.9    & 6.9     & 121.3 & 120.0 & 122.0   & 1.8 & 1  & 9.8    &  ie  \\
19.609--0.234    & 18 27 38.06  &   --11 56 37.2  &  42.5     &  20.1 &  69.8 &  69.7   & 1321.7  & 40.2  & 36.0 & 42.0     & 1.2 & 2  & \textit{2.9}   &  om   \\
19.612--0.134    & 18 27 16.56  &   --11 53 37.8  &  56.8     &  53.8 &  58.5 &  1.7    & 6.0     & 56.5  & 49.0 & 61.0   & 0.7 & 1  & 12.1   &  om   \\
19.614+0.011     & 18 26 45.23  &   --11 49 31.9  &  35.1     &  32.1 &  36.7 &  2.7    & 8.6     & 32.9  & 30.8 & 34.8  & 0.5 & 1  & 13.2   &  --  \\
19.701--0.267    & 18 27 55.47  &   --11 52 40.3  &  42.1     &  36.2 &  58.9 &  1.6    & 14.1    & 43.8  & 41.5 & 46.5  & 0.7 & 1  & 12.6   &  i   \\
19.884--0.534    & 18 29 14.36  &   --11 50 22.7  &  44.1     &   0.7 &  49.6 &  21.7   & 178.4   & 46.8  & 46.0 & 48.0  & 0.2 & 1  & 3.3    &  e   \\
\end{tabular}
\label{table:assoc_sources}
\end{table*}

\begin{table*}
\caption{This table lists all the 6.7-GHz methanol masers where no associated water maser emission was detected. Column 1 is the name of the target methanol maser given in Galactic coordinates; column 2 is the 5 sigma detection limit; column 3 is the epoch of observation coded 1, 2, 3 and 4 for 2010 November 2, 2010 November 3, 2011 August 9 and 2011 August 10 respectively. Column 4 gives associations with Infrared Dark Clouds or Extended Green Objects indicated with an `i' or `e' respectively from inspection of the GLIMPSE infrared images. Sources outside of the GLIMPSE survey range are indicated with an `*'. Also in this column, water masers detected in Pillai et al. (2006b) or the Breen et al. (2010a) 2003 or 2004 observations but not in ours are indicated with a `w' and associations with the 1665~MHz transition of OH masers (Caswell, 1998; Caswell et al. 2013) are indicated with an `o'. Column 5 is the distance to the methanol maser. Distances estimates are from Green et al. (2011) where available, otherwise they are calculated using the near kinematic distance (these are in italics).}
\begin{tabular}{ccccc @{\hskip 2cm} ccccc}
\hline
MMB Target & Det. & Epoch & Assoc. & & MMB Target & Det. & Epoch & Assoc. &  \\
Source Name &  lim. & & & Distance & Source Name &  lim. & & & Distance \\
(l, b) & (mJy) & & & (kpc) & (l, b) & (mJy) & & & (kpc) \\
\hline
06.368--0.052	&  40  & 3  & -- & 7.4             &  14.390--0.020	&  100 & 4 & -- & 13.6 \\            
06.539--0.108	&  50  & 3  & -- & 13.9            &  14.457--0.143	&  110 & 4 & -- & 3.6  \\            
06.881+0.093	&  50  & 3  & -- & 17.3            &  14.490+0.014	&  110 & 4 & i  & 2.3 \\             
08.683--0.368	&  50  & 3  & oe  & 4.5             &  14.521+0.155	&  115 & 4 & --  & 5.5 \\             
08.872--0.493	&  45  & 3  & -- & 3.4             &  14.631--0.577	&  150 & 4 & ie & 13.7  \\           
09.619+0.193	&  50  & 3  & o  & 5.2             &  14.991--0.121	&  140 & 4 & --  & 12.3 \\            
10.205--0.345	&  40  & 3  & -- & 1.4             &  15.034--0.677	&  150 & 4 & o  & 2.3  \\            
10.299--0.146	&  50  & 3  & -- & \textit{2.7}    &  15.607--0.255	&  160 & 4 & -- & 11.7 \\            
10.323--0.160	&  75  & 2  & w & \textit{1.6}    &  16.112--0.303	&  60  & 1  & --  & 3.0 \\             
10.356--0.148	&  50  & 3  & i  & \textit{4.6}    &  16.302--0.196	&  45  & 1  & -- & 3.8 \\             
10.480+0.033	&  75  & 2  & woi  & 11.4            &  16.403--0.181	&  55  & 1  & --  & 12.9 \\            
10.627--0.384	&  50  & 3  & -- & \textit{0.2}    &  16.662--0.331	&  70  & 1  & -- & 12.7 \\            
10.629--0.333	&  40  & 3  & i  & 5.2             &  16.855+0.641	&  80  & 1  & -- & 13.8  \\           
11.109--0.114	&  35  & 3  & w  & 13.2            &  16.976--0.005	&  60  & 1  & -- & 15.3 \\            
11.497--1.485	&  75  & 2  & w* & 1.6             &  17.029--0.071	&  65  & 1  & --  & 10.8 \\            
11.903--0.102	&  50  & 3  & o  & 12.9            &  17.862+0.074	&  50  & 1  & -- & 10.2 \\            
11.904--0.141	&  75  & 2  & wo  & 4.0             &  18.073+0.077	&  70  & 1  & -- & \textit{3.6} \\    
11.936--0.150	&  150 & 4 & -- & 12.2            &  18.159+0.094	&  50  & 1  & -- & 12.0 \\            
11.936--0.616	&  150 & 4 & -- & 3.7             &  18.262--0.244	&  75  & 1  & -- & 4.7 \\             
11.992--0.272	&  150 & 4 & -- & 11.7            &  18.440+0.045	&  70  & 1  & -- & 11.8 \\            
12.025--0.031	&  125 & 4 & o  & 11.1            &  18.460--0.004	&  45  & 1  & oi  & 3.5 \\             
12.112--0.126	&  150 & 4 & -- & 4.1             &  18.667+0.025	&  45  & 1  & e  & 11.2 \\            
12.181--0.123   &  150 & 4 & -- & \textit{3.1}    &  18.733--0.224	&  80  & 1  & -- & 12.5 \\            
12.202--0.120	&  55  & 2  & i  & 3.2             &  18.834--0.300	&  60  & 1  & -- & 3.1 \\             
12.203--0.107	&  55  & 2  & w & \textit{2.9}    &  18.874+0.053	&  60  & 1  & --  & 12.9 \\            
12.526+0.016	&  150 & 4 & -- & 12.6            &  18.888--0.475	&  80  & 1  & ie & 3.8  \\            
12.625--0.017	&  100 & 4 & -- & 2.7             &  19.249+0.267	&  80  & 1  & i  & 14.3 \\            
12.776+0.128	&  150 & 4 & -- & 13.2            &  19.365--0.030	&  75  & 1  & e  & 2.3 \\             
13.179+0.061	&  150 & 4 & i  & 4.1             &  19.472+0.170	&  60  & 1  & i  & 1.8 \\             
13.696--0.156	&  140 & 4 & -- & 10.9            &  19.612--0.120	&  75  & 1  & -- & 12.2 \\            
13.713--0.083	&  100 & 4 & -- & 4.0             &  19.667+0.114	&  50  & 1  & -- & 14.4 \\            
14.230--0.509	&  120 & 4 & -- & \textit{2.6}    &  19.755--0.128	&  80  & 1  & -- & 9.9 \\             
\end{tabular}
\label{table:notass_sources}
\end{table*}

\begin{figure*}
\includegraphics[width=2.2in]{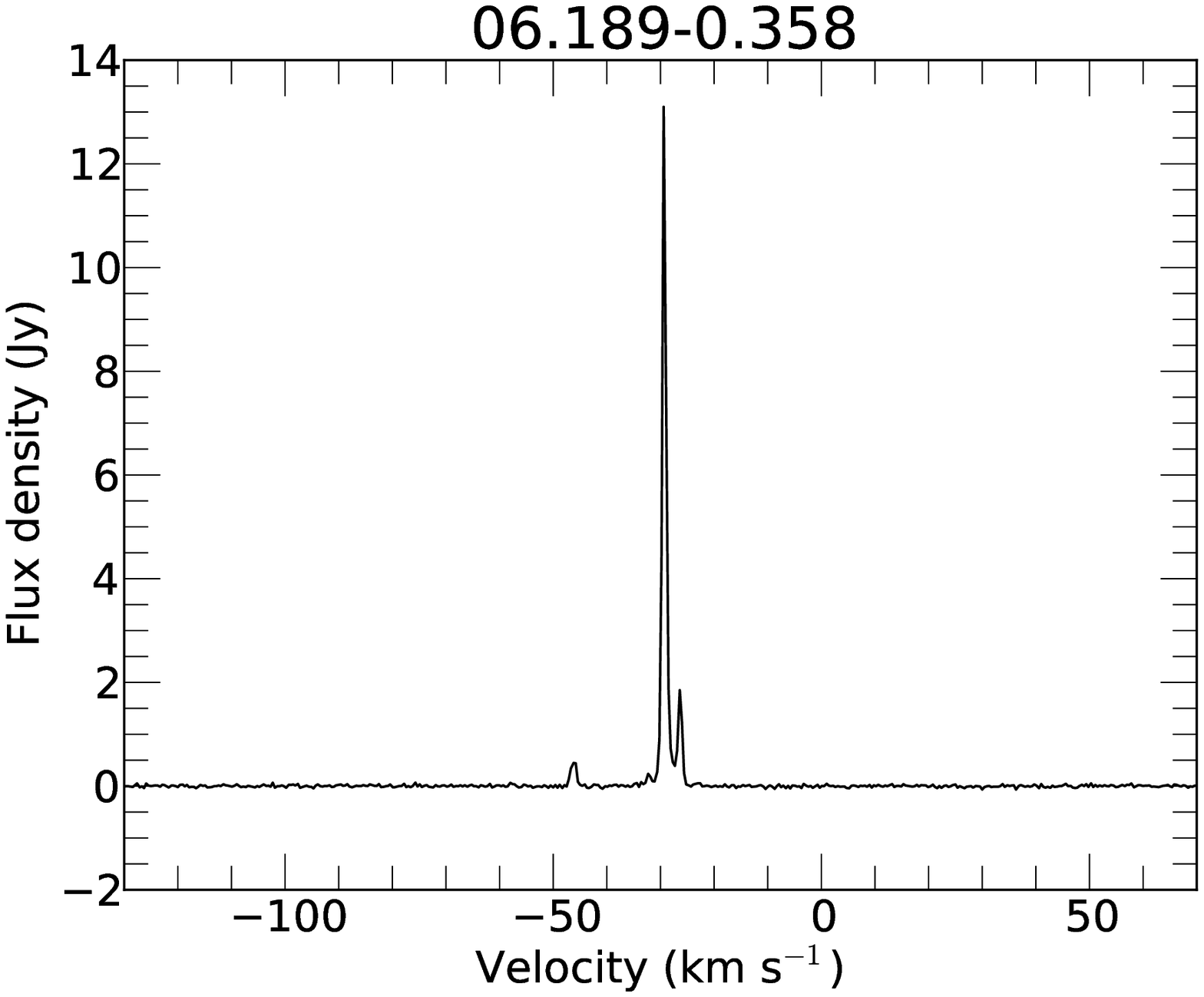}
\includegraphics[width=2.2in]{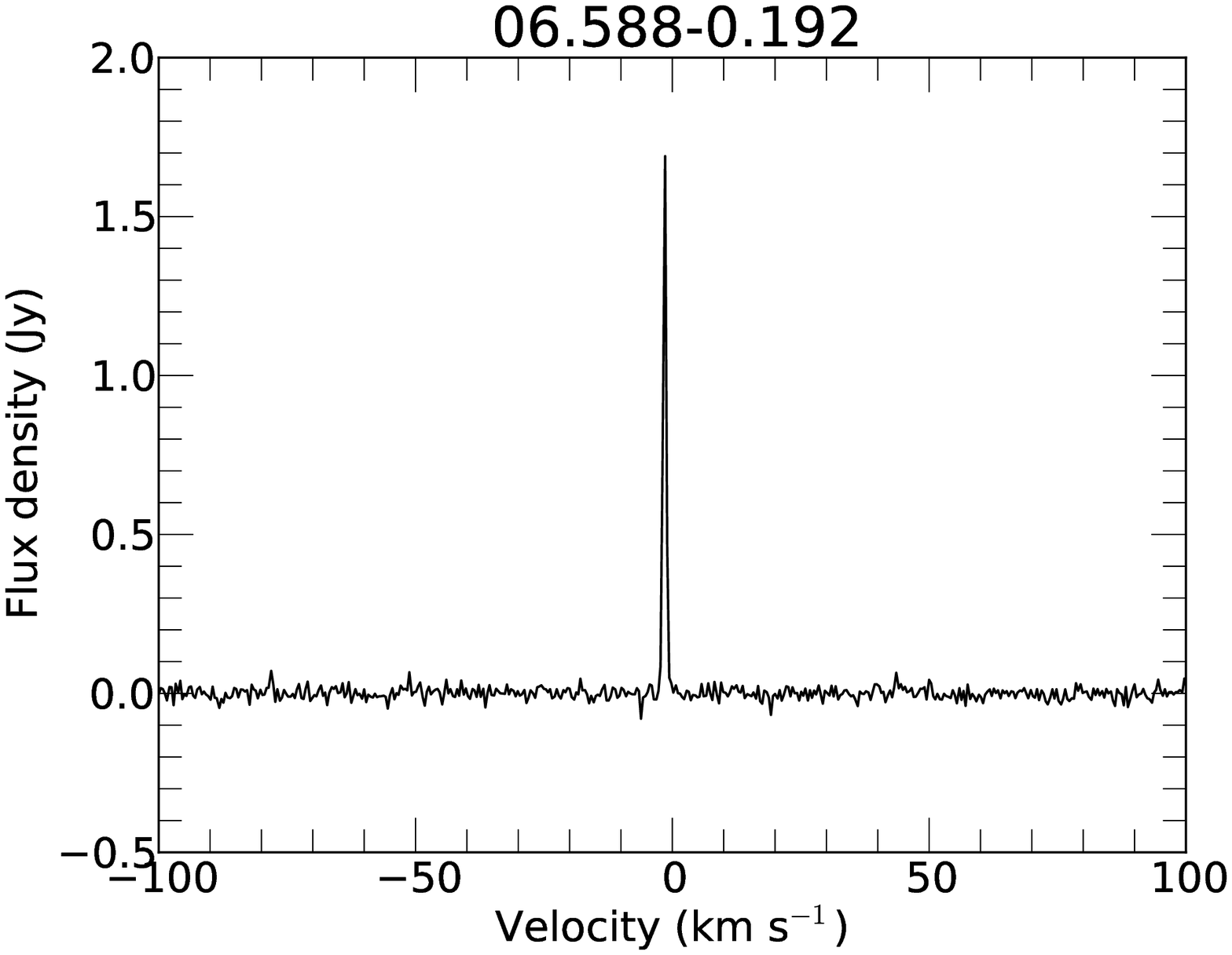}
\includegraphics[width=2.2in]{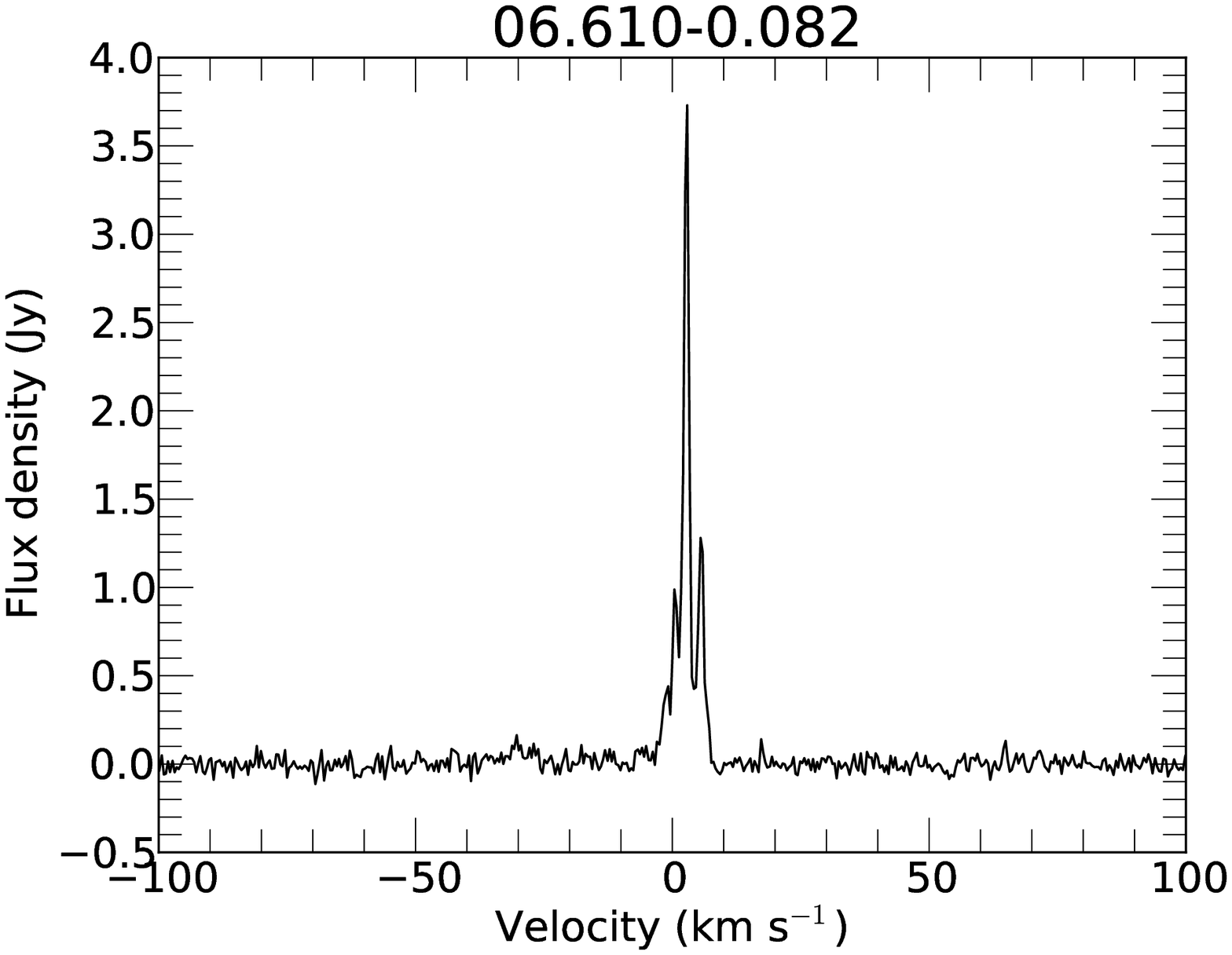}
\includegraphics[width=2.2in]{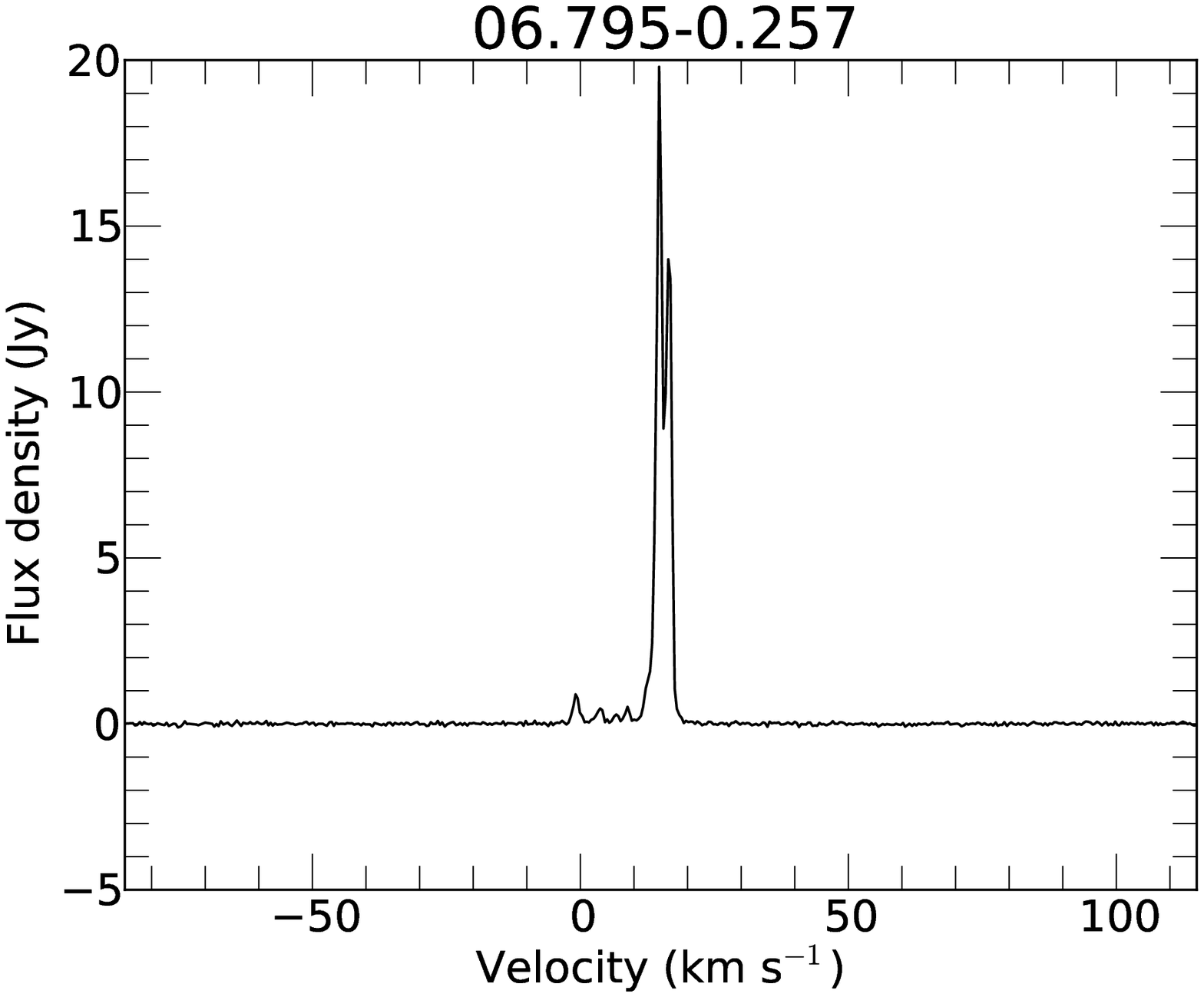}
\includegraphics[width=2.2in]{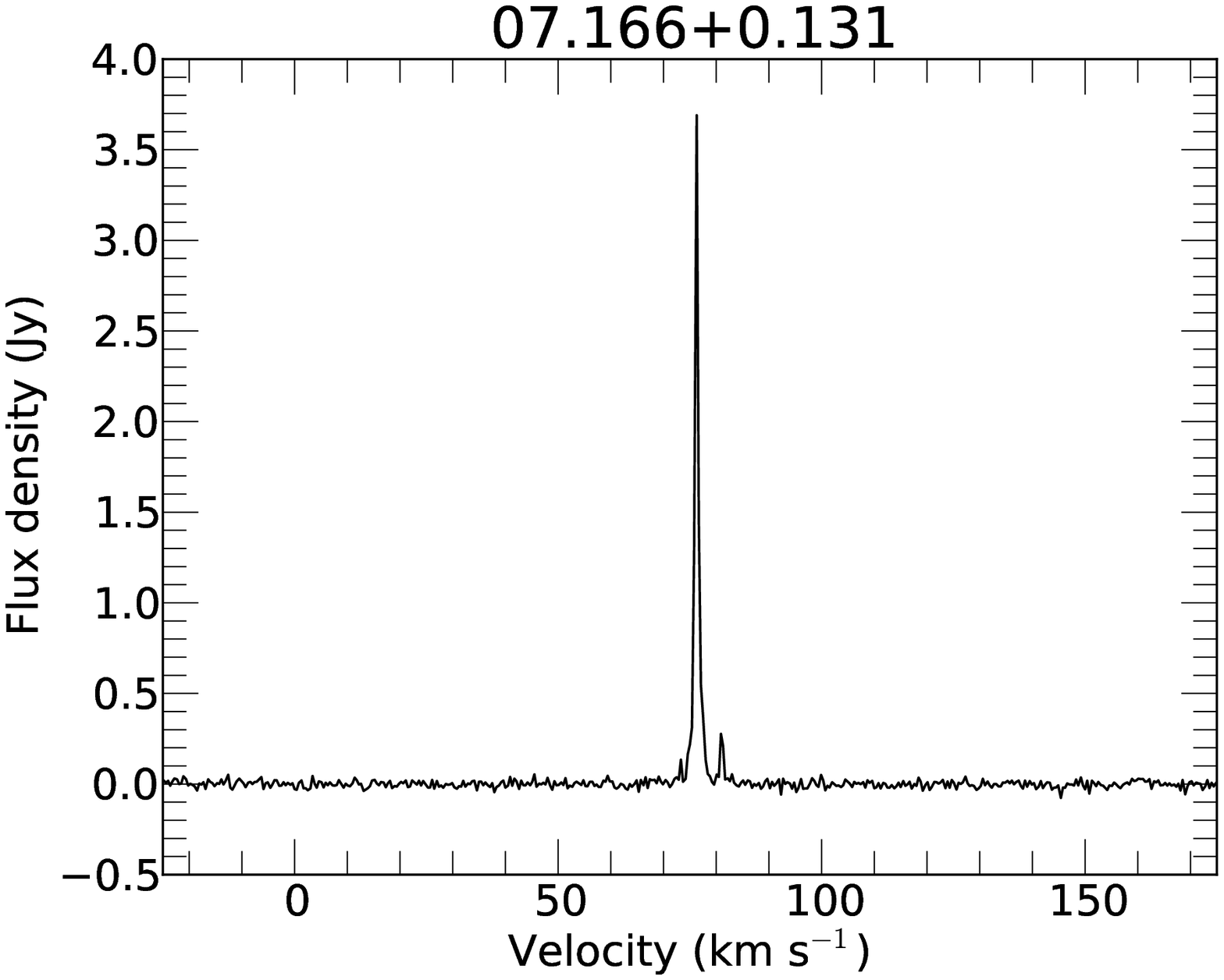}
\includegraphics[width=2.2in]{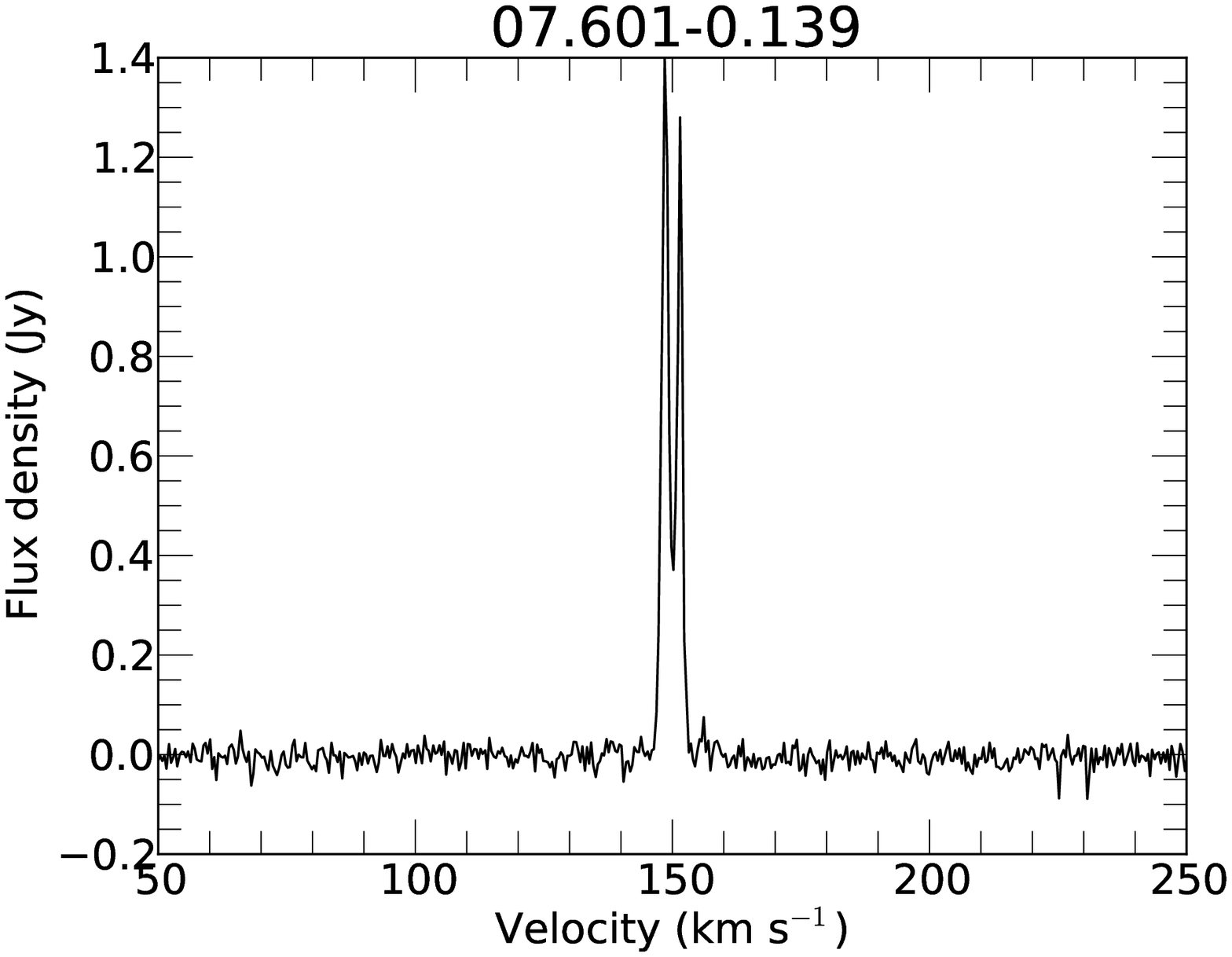}
\includegraphics[width=2.2in]{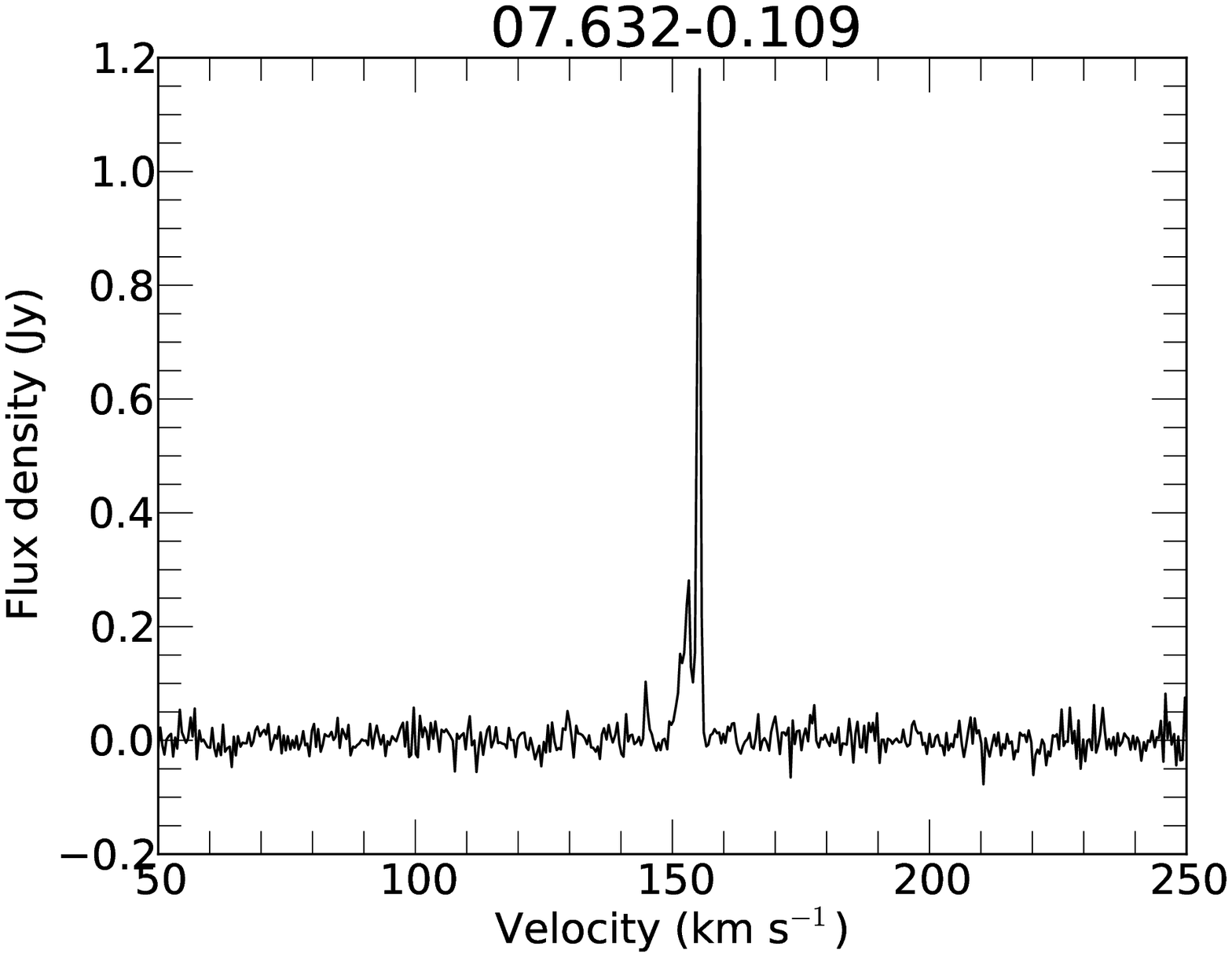}
\includegraphics[width=2.2in]{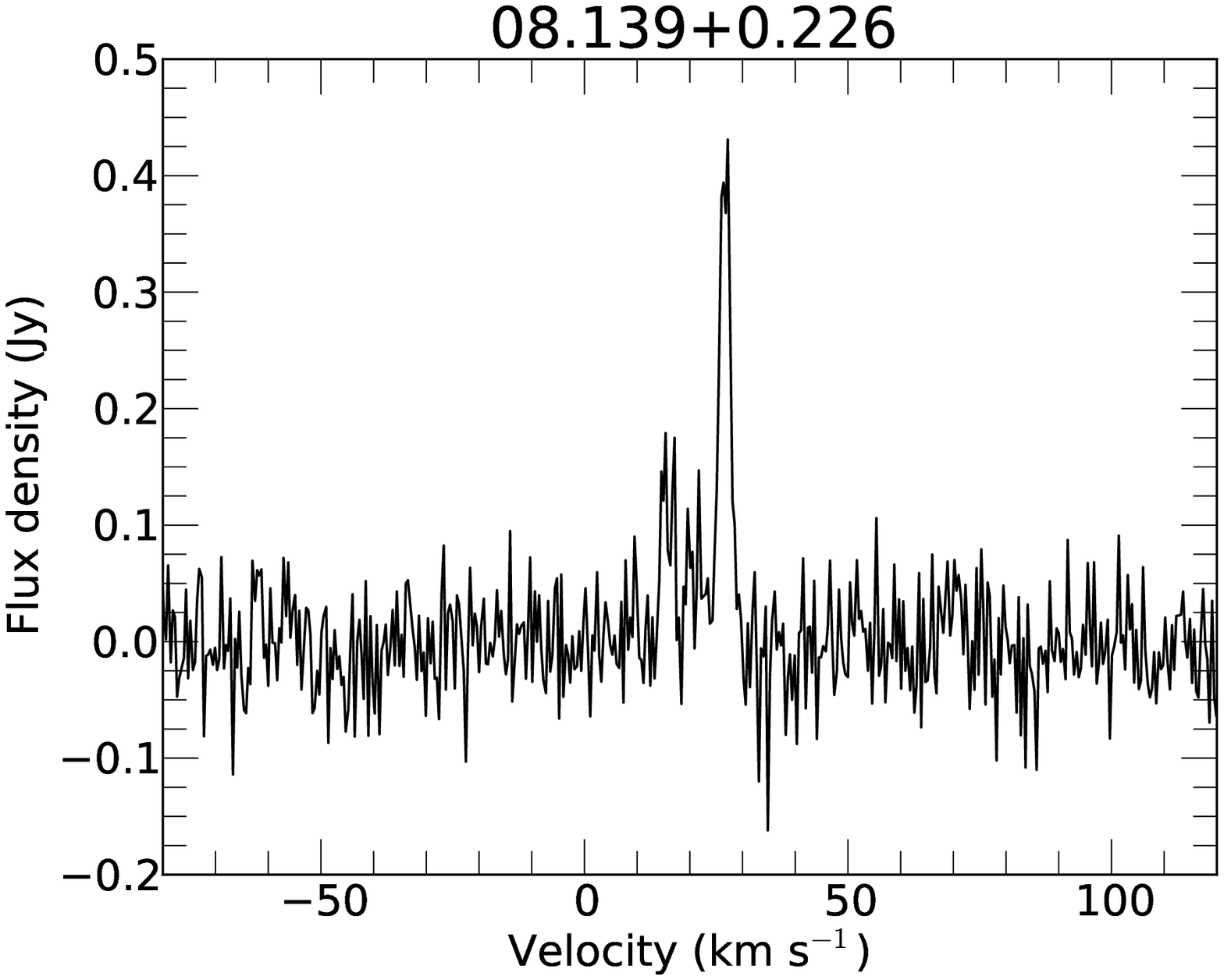}
\includegraphics[width=2.2in]{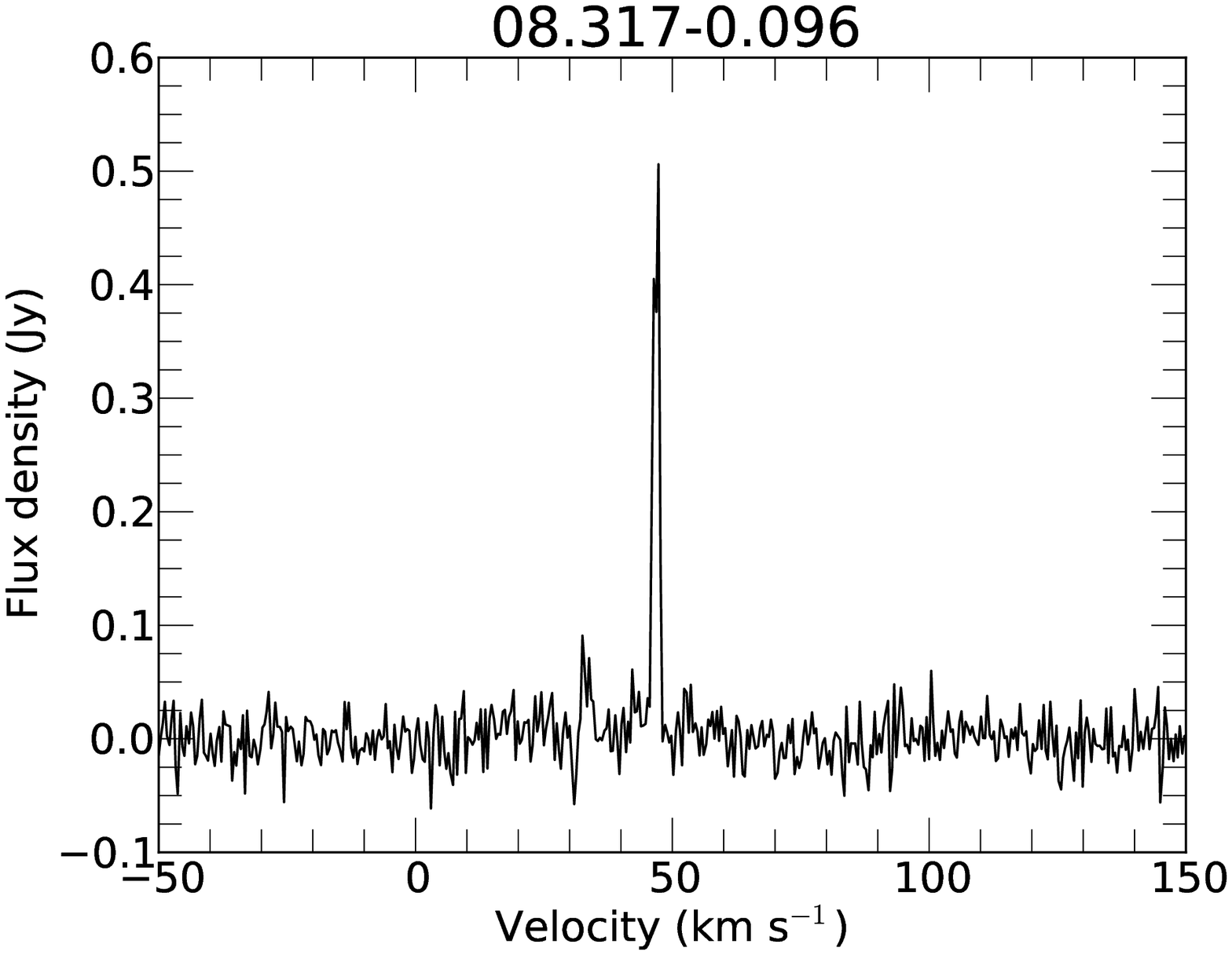}
\includegraphics[width=2.2in]{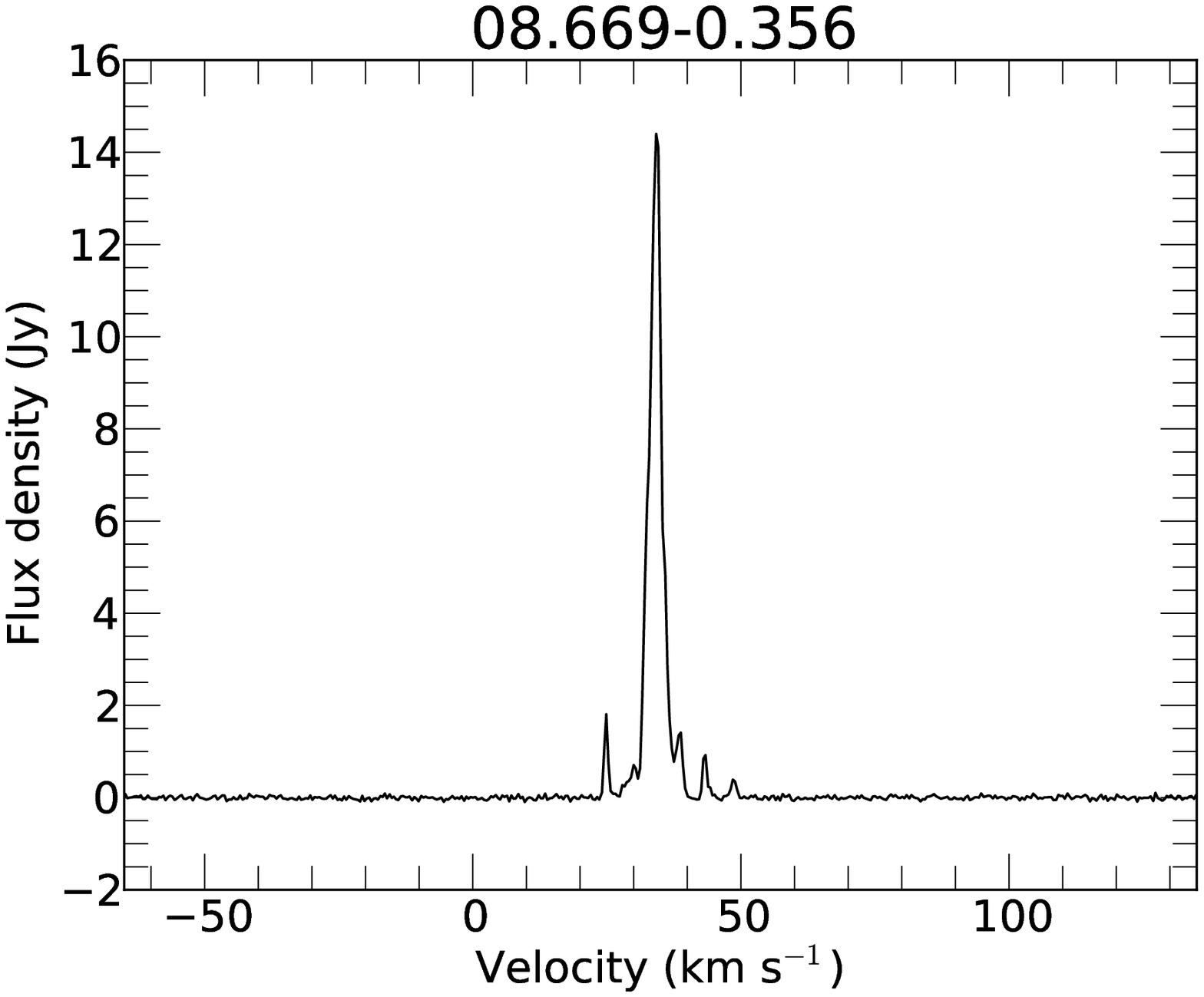}
\includegraphics[width=2.2in]{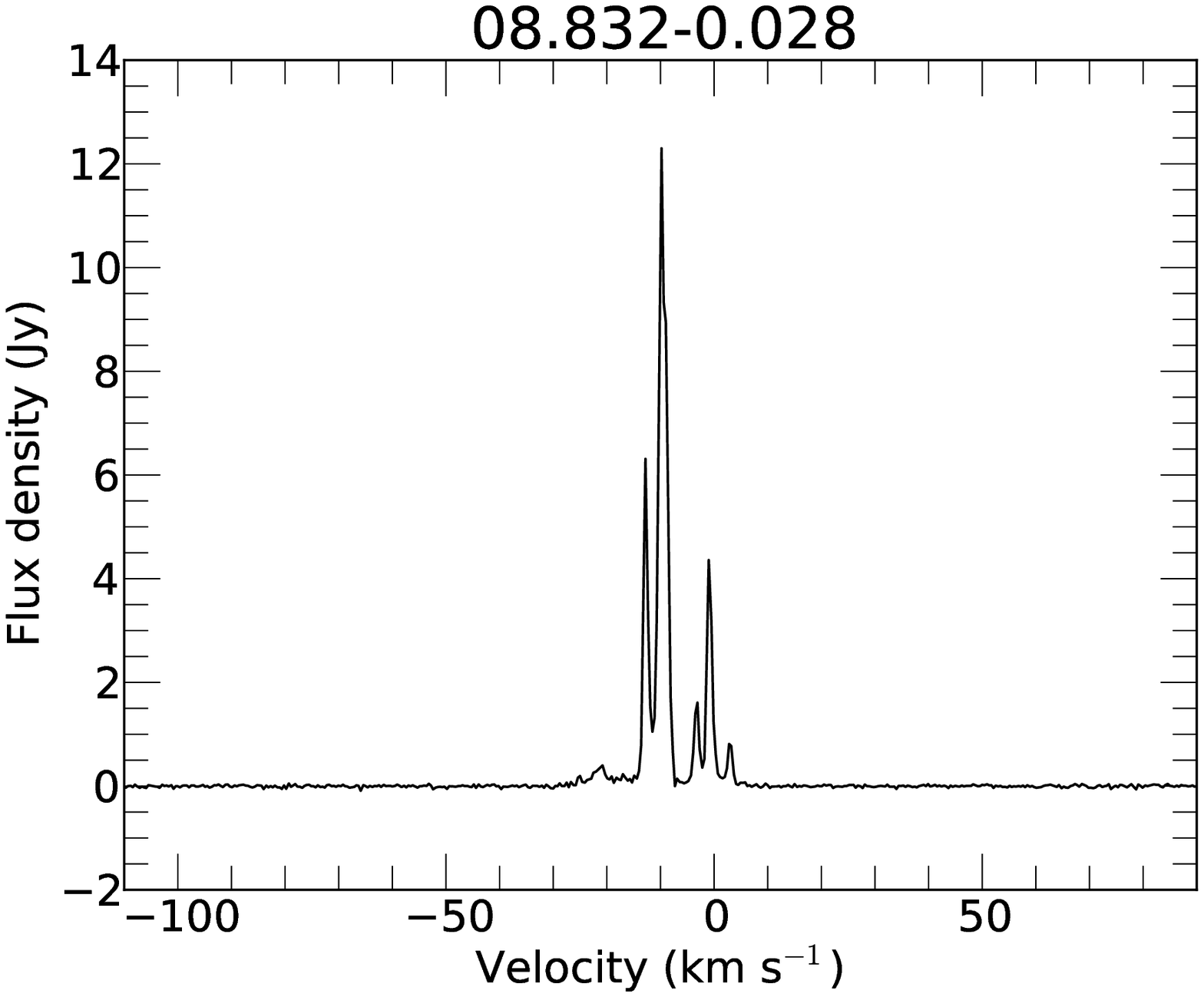}
\includegraphics[width=2.2in]{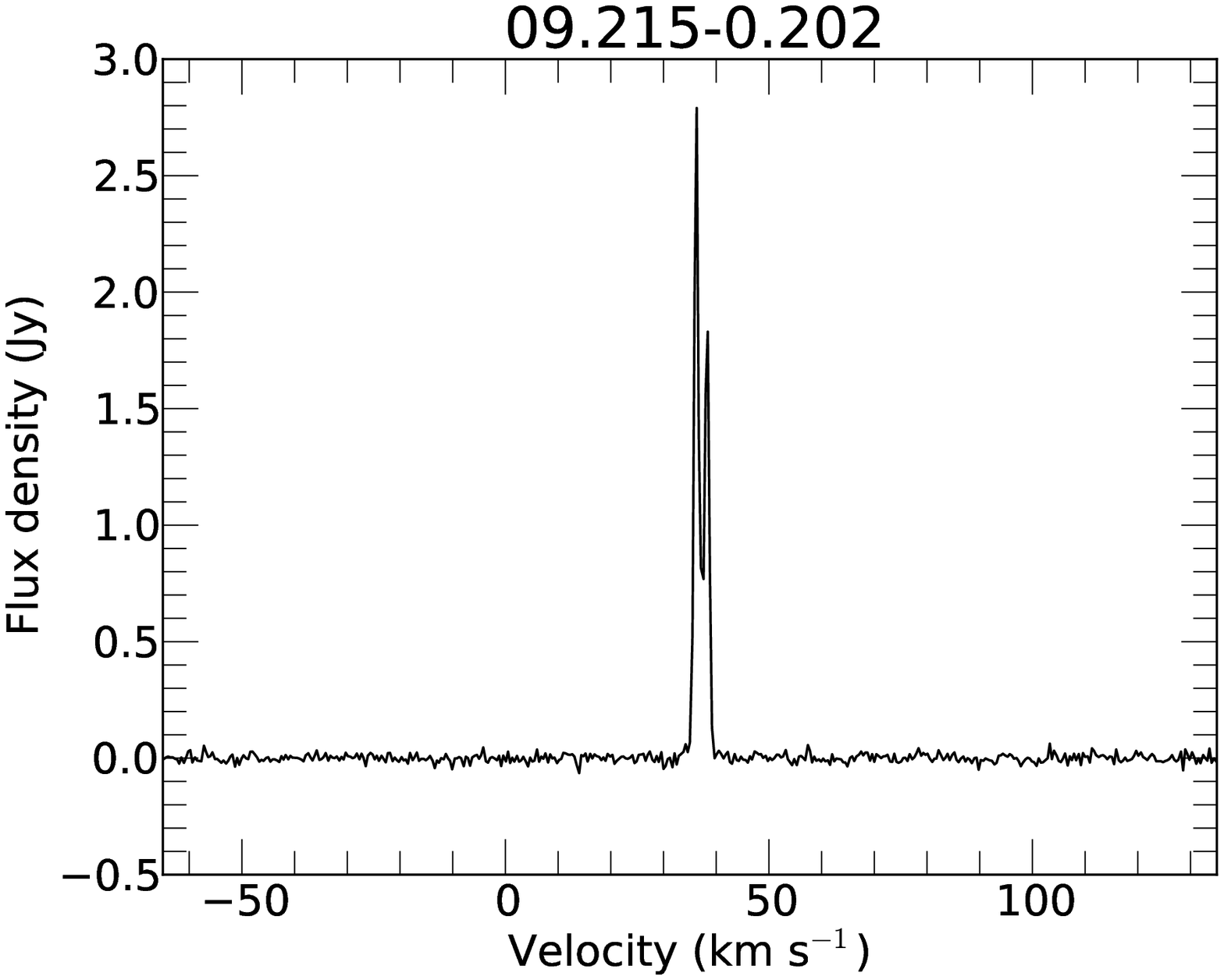}
\includegraphics[width=2.2in]{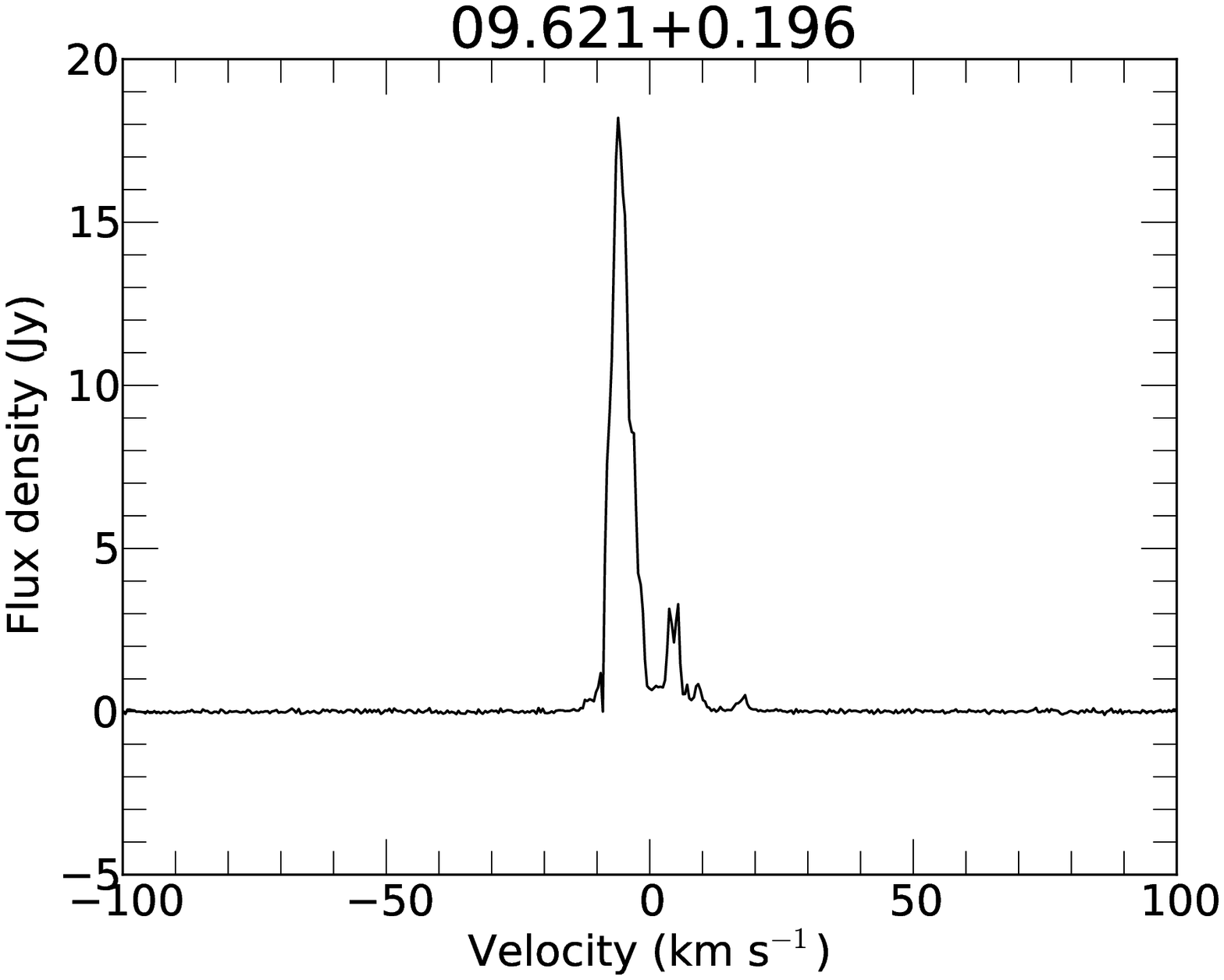}
\includegraphics[width=2.2in]{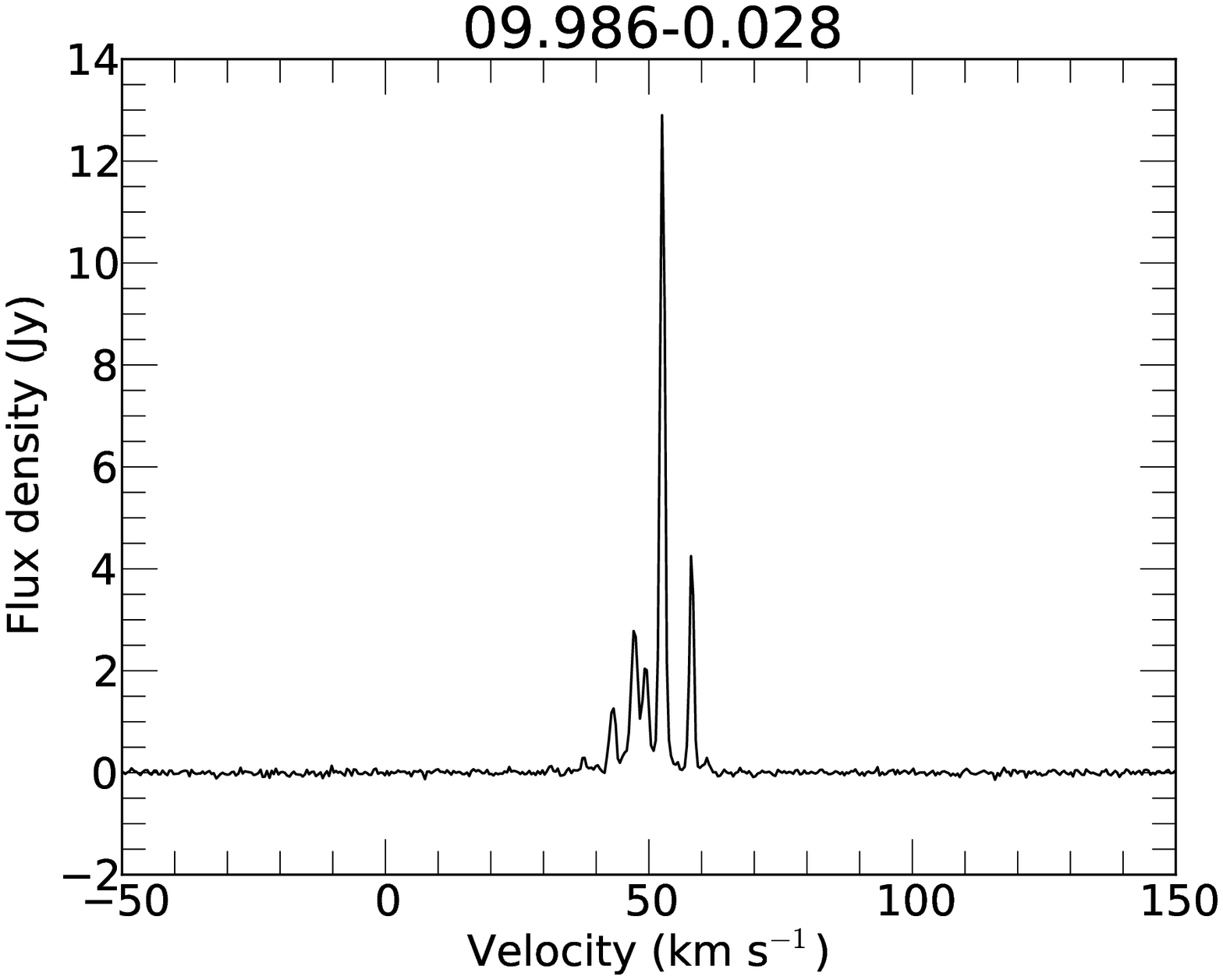}
\includegraphics[width=2.2in]{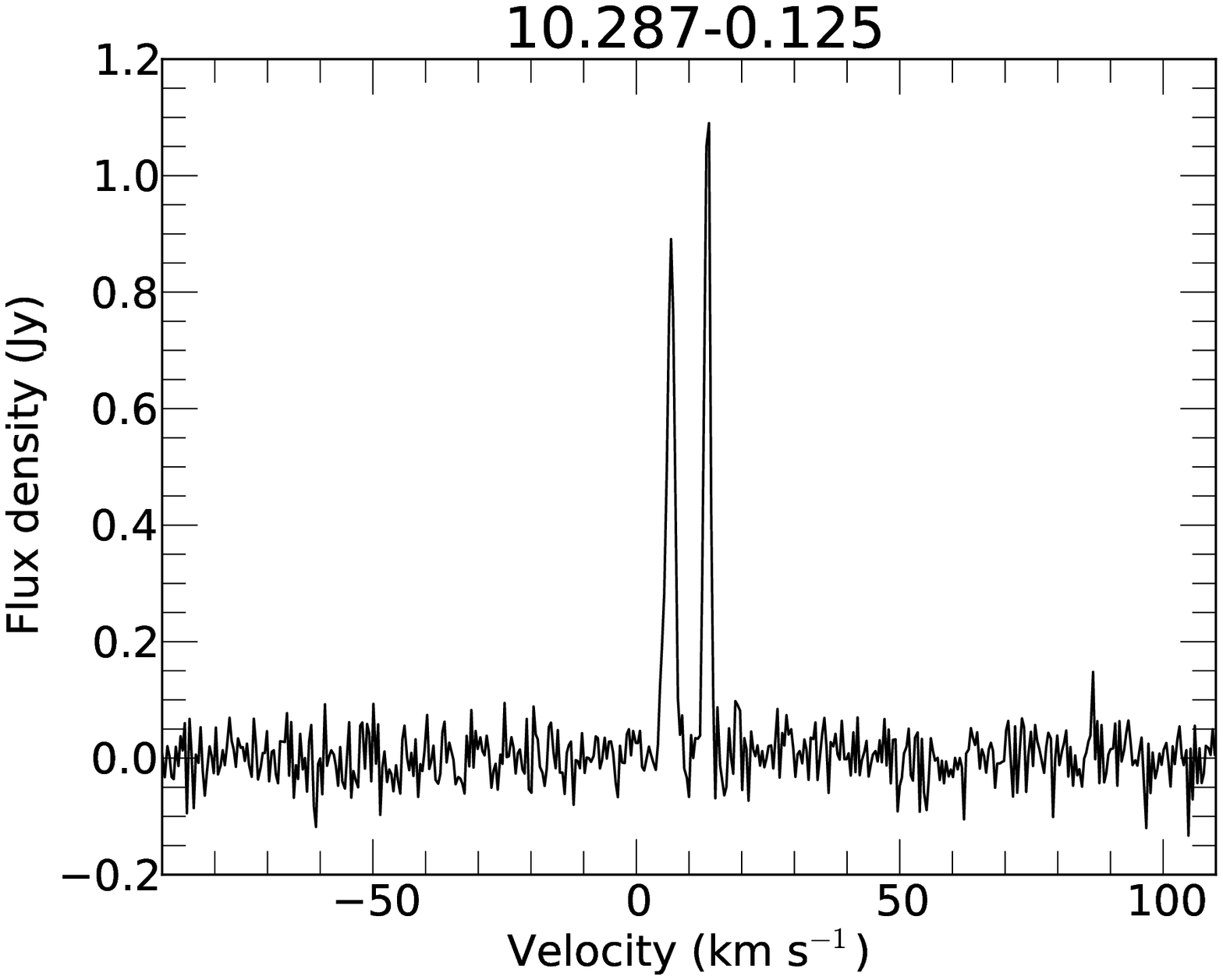}
\caption{Spectra obtained with the ATCA of water masers associated with 6.7-GHz methanol masers.}
\label{fig:assoc}
\end{figure*}

\begin{figure*}
\includegraphics[width=2.2in]{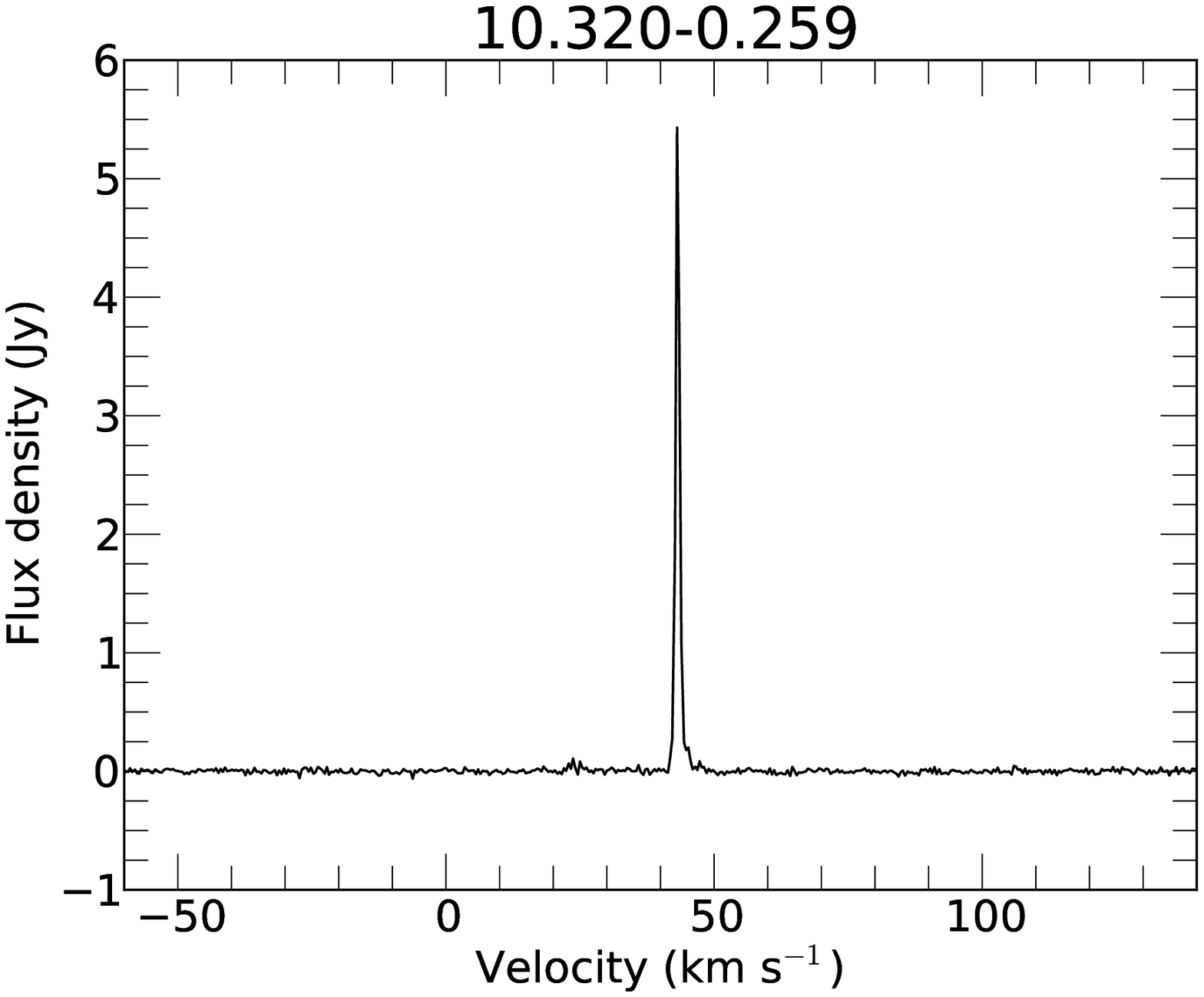}
\includegraphics[width=2.2in]{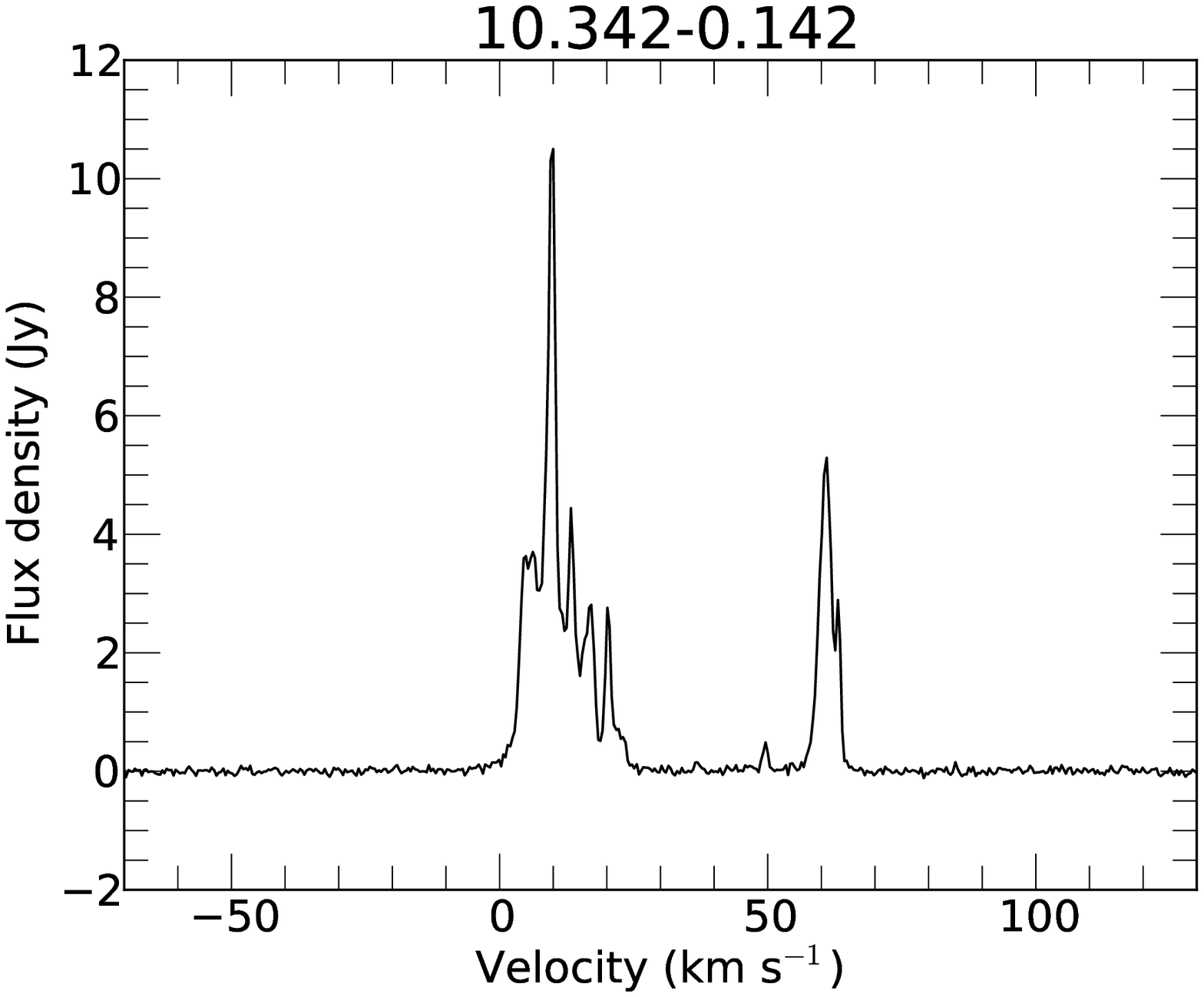}
\includegraphics[width=2.2in]{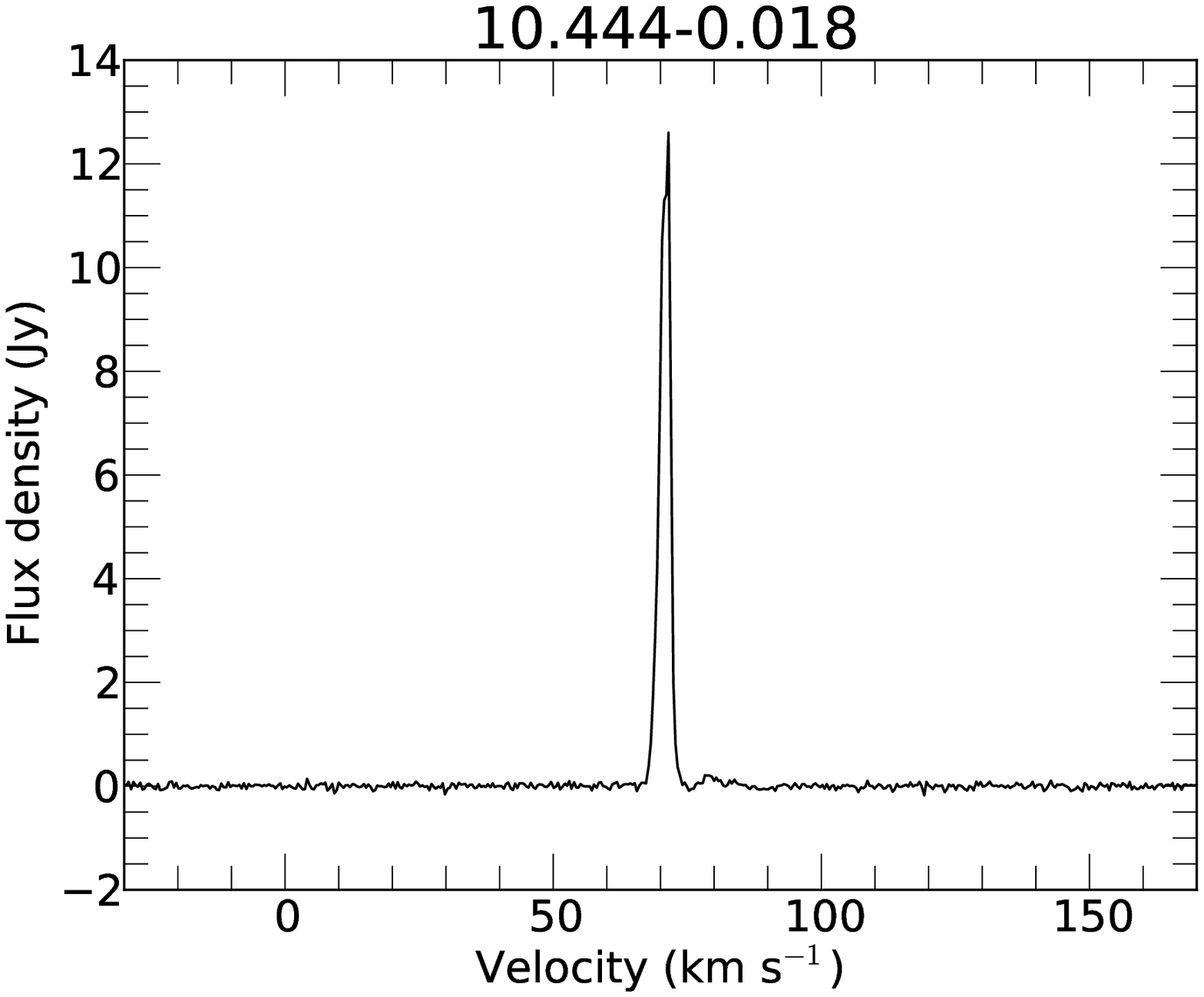}
\includegraphics[width=2.2in]{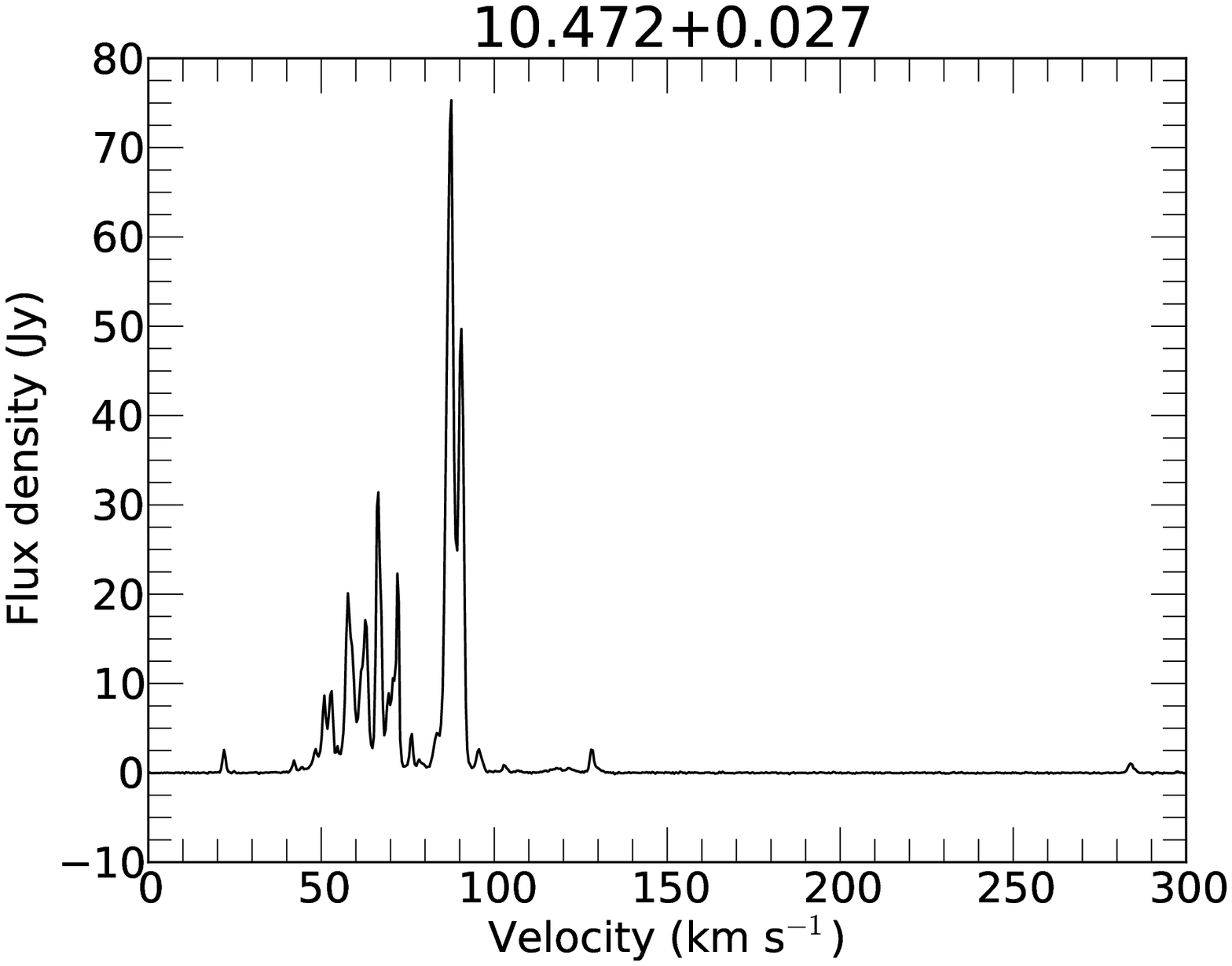}
\includegraphics[width=2.2in]{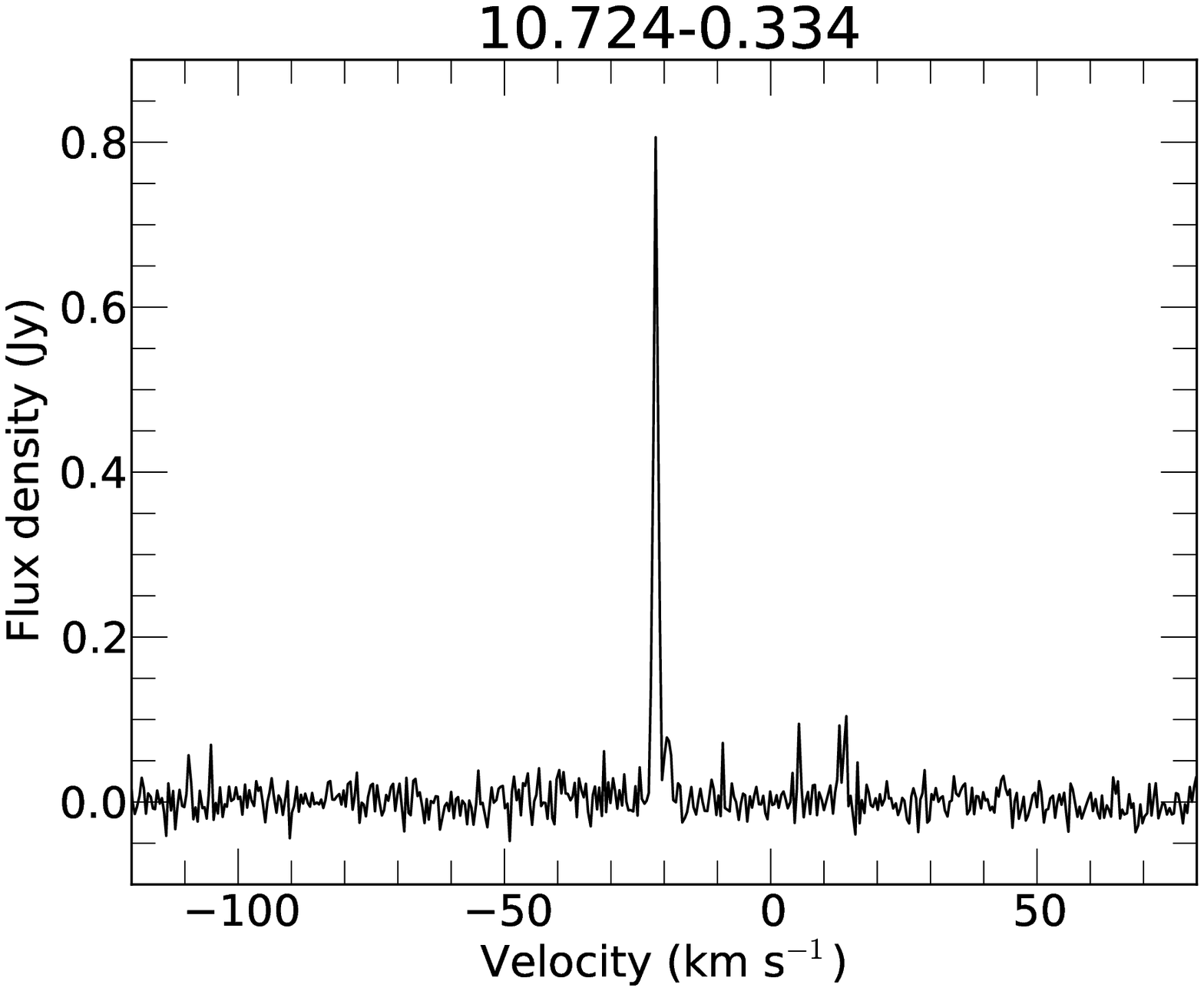}
\includegraphics[width=2.2in]{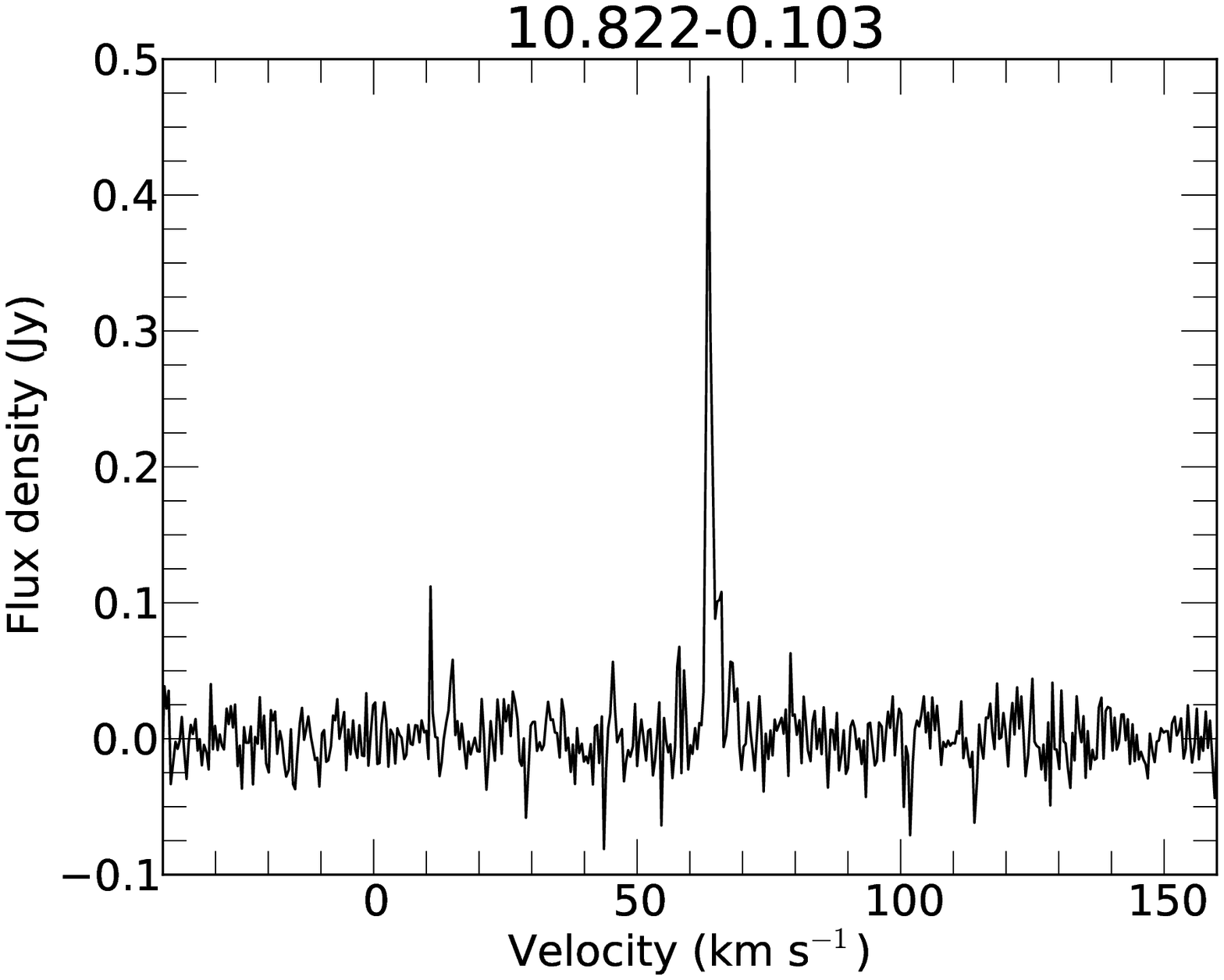}
\includegraphics[width=2.2in]{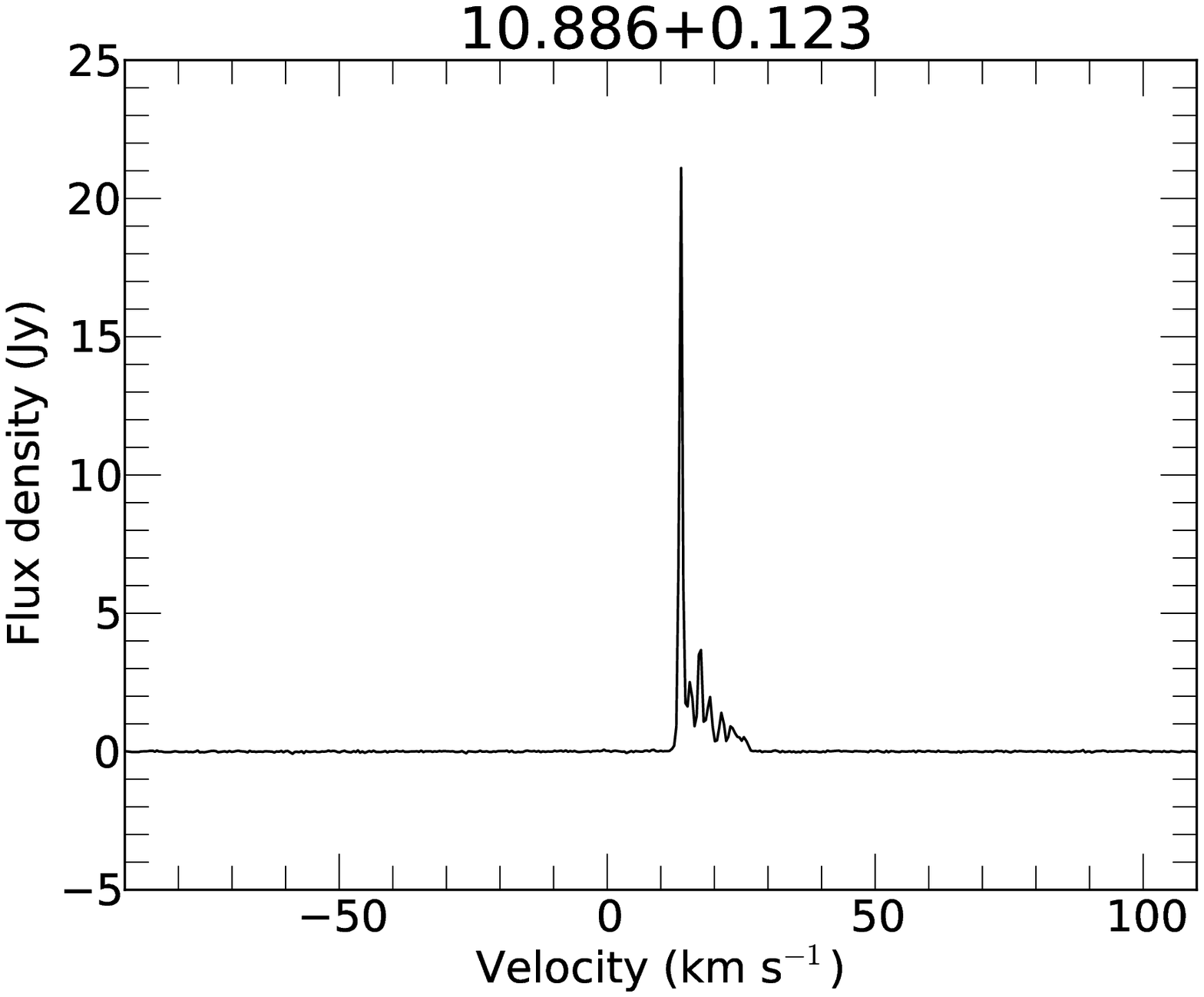}
\includegraphics[width=2.2in]{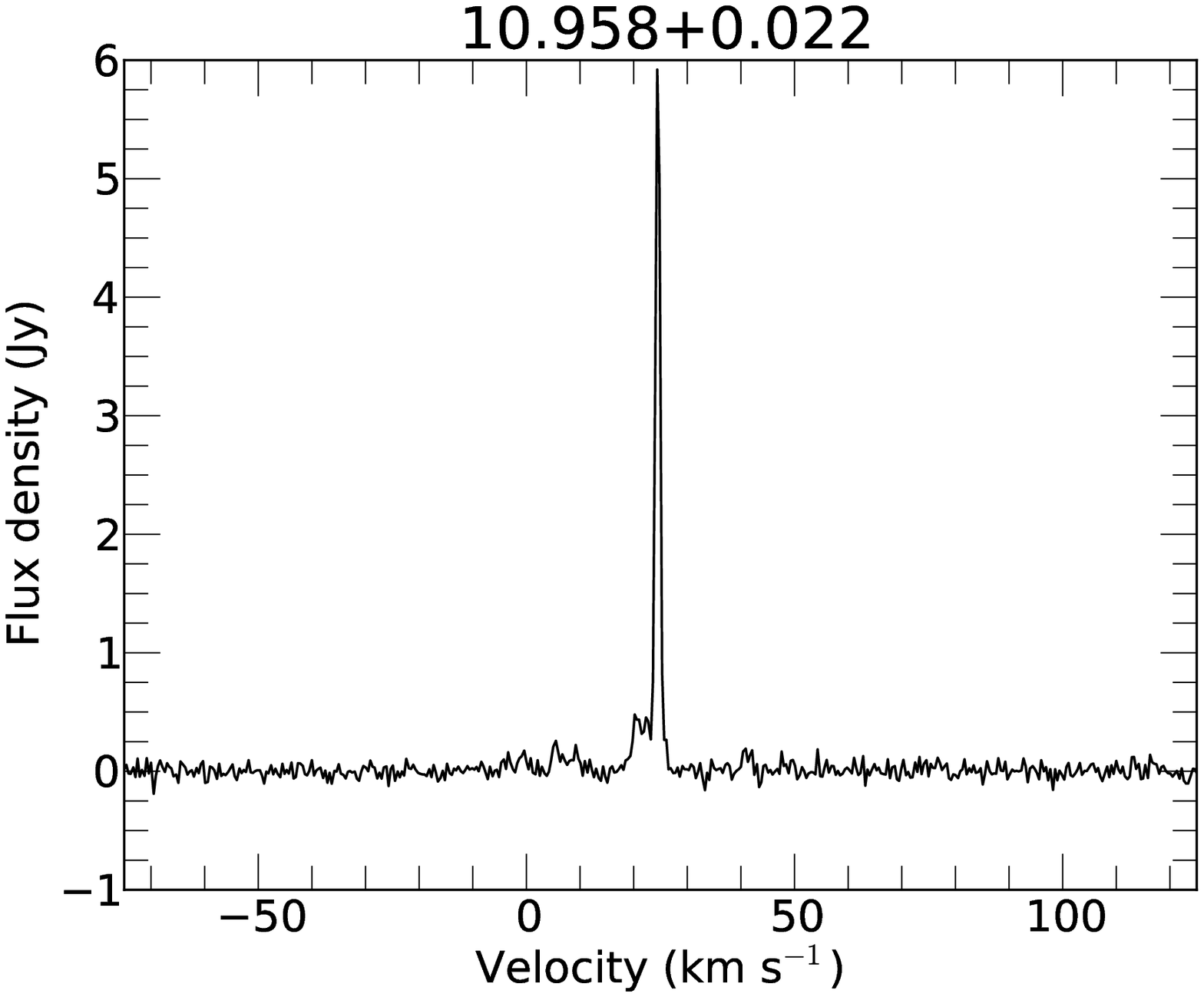}
\includegraphics[width=2.2in]{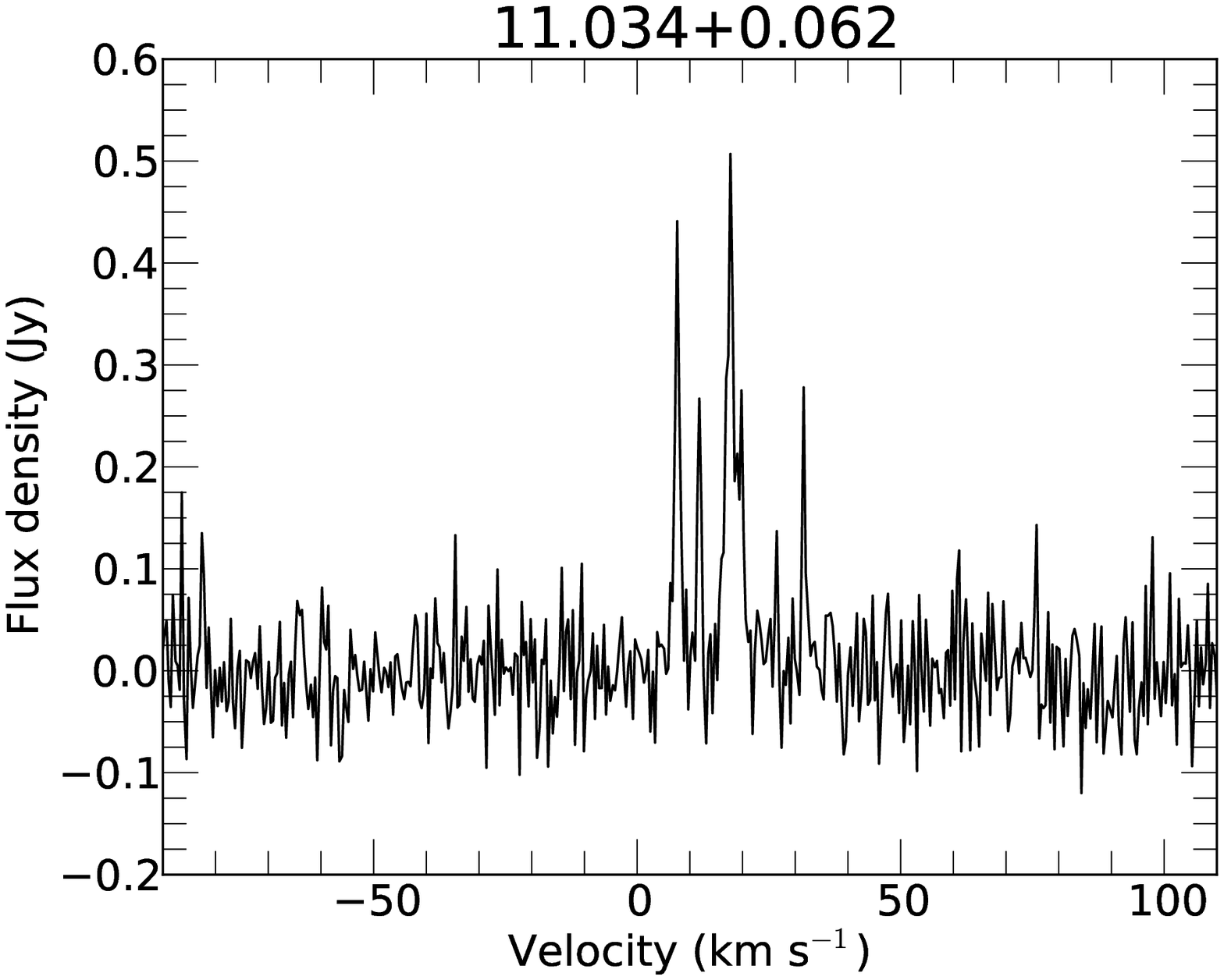}
\includegraphics[width=2.2in]{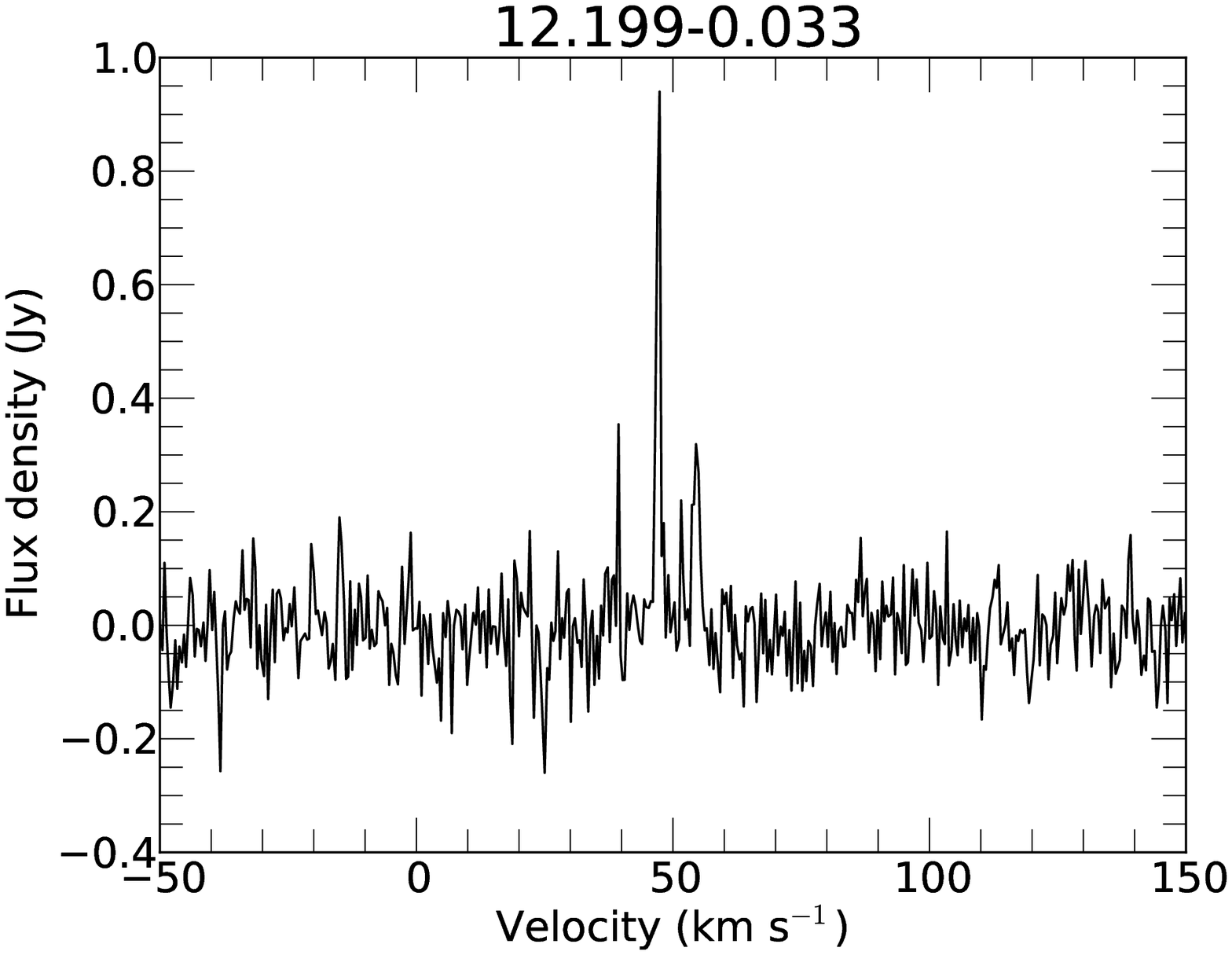}
\includegraphics[width=2.2in]{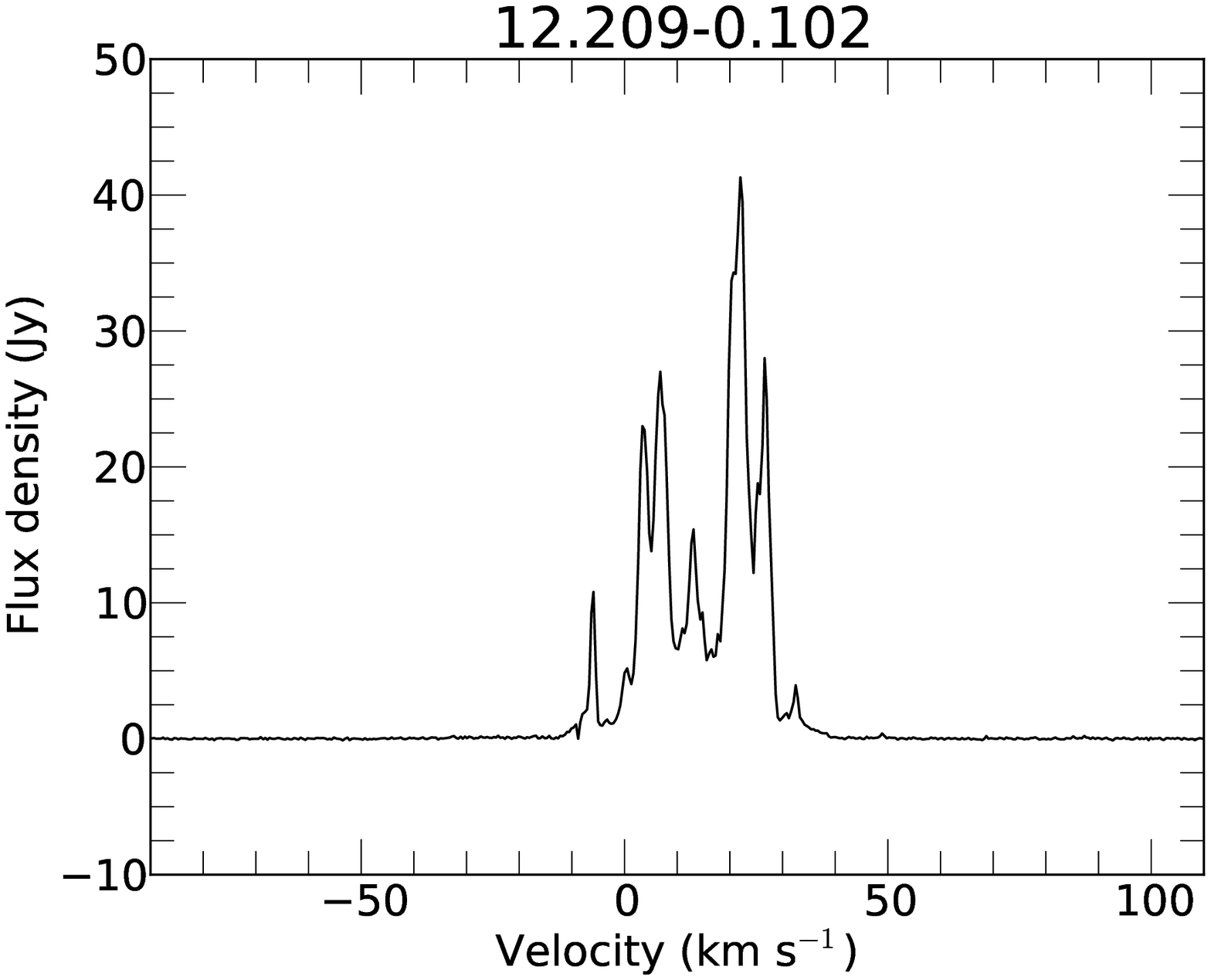}
\includegraphics[width=2.2in]{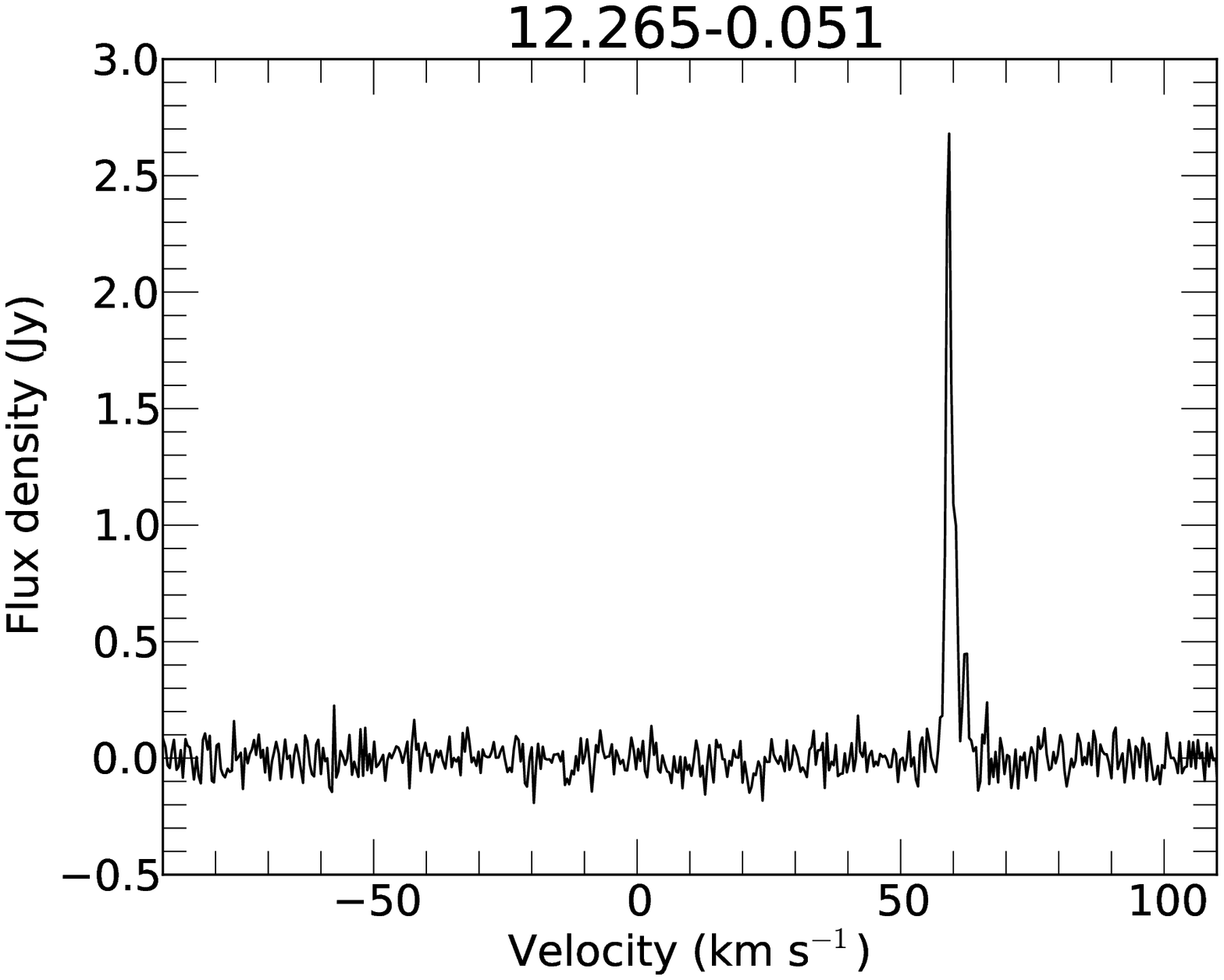}
\includegraphics[width=2.2in]{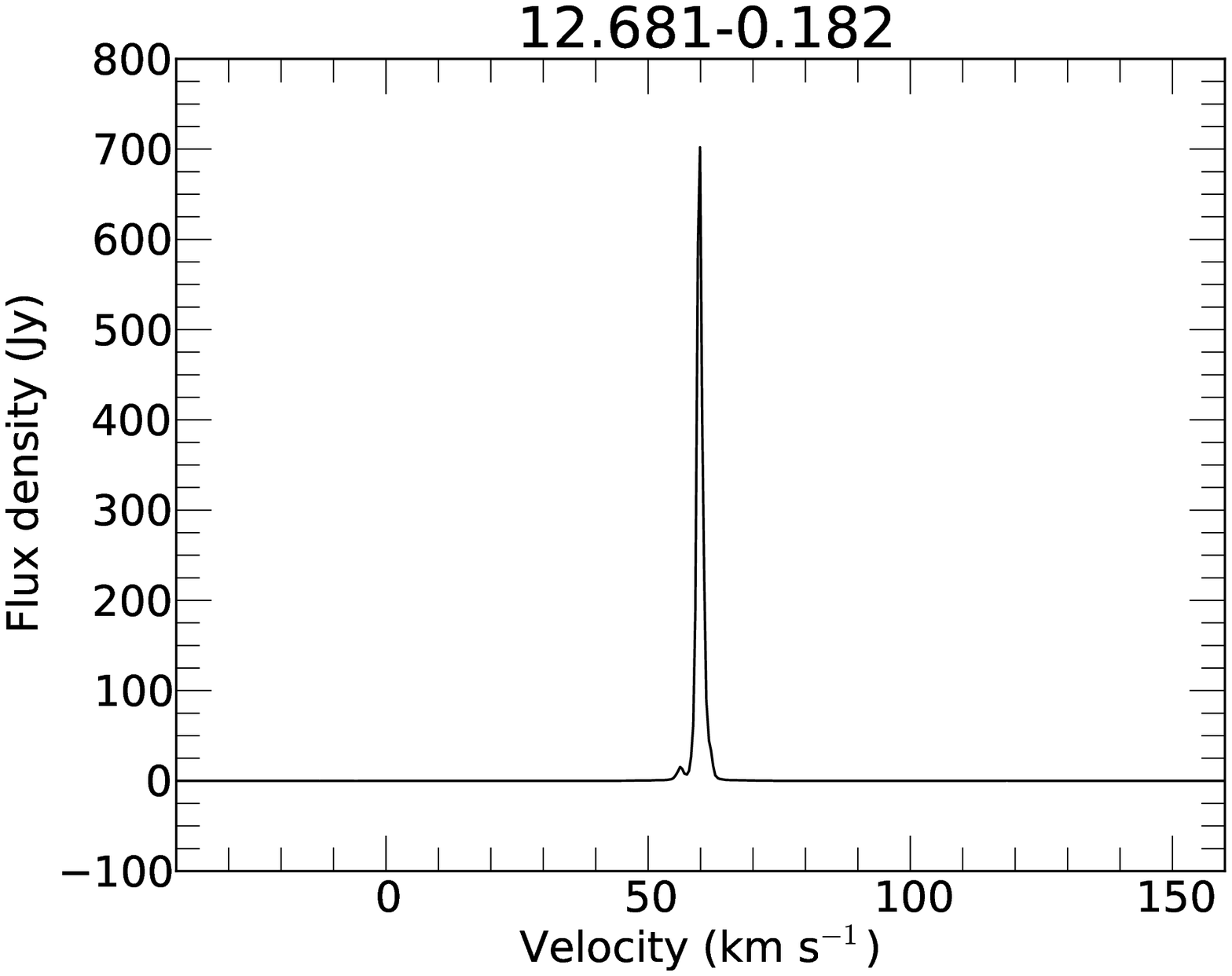}
\includegraphics[width=2.2in]{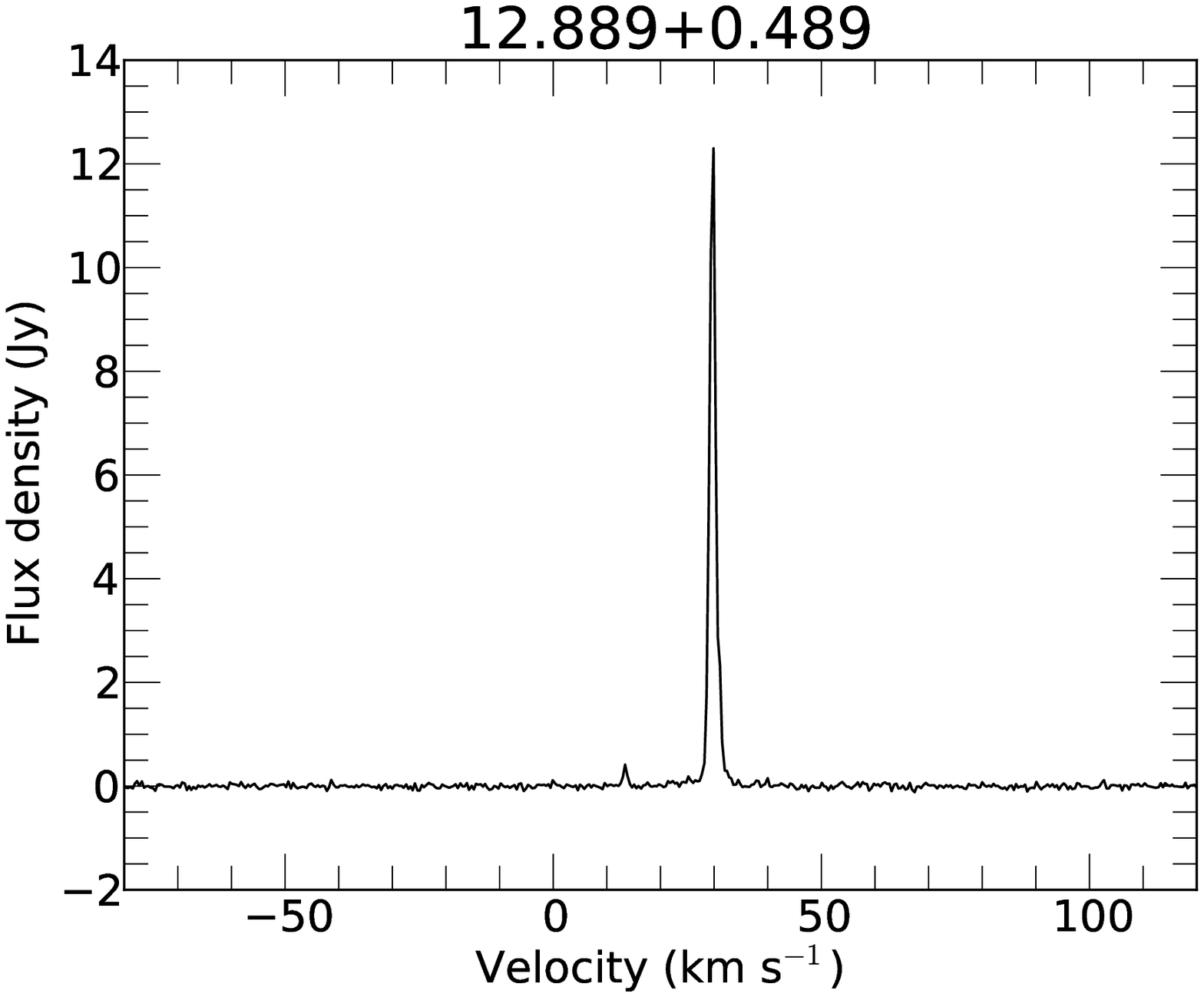}
\includegraphics[width=2.2in]{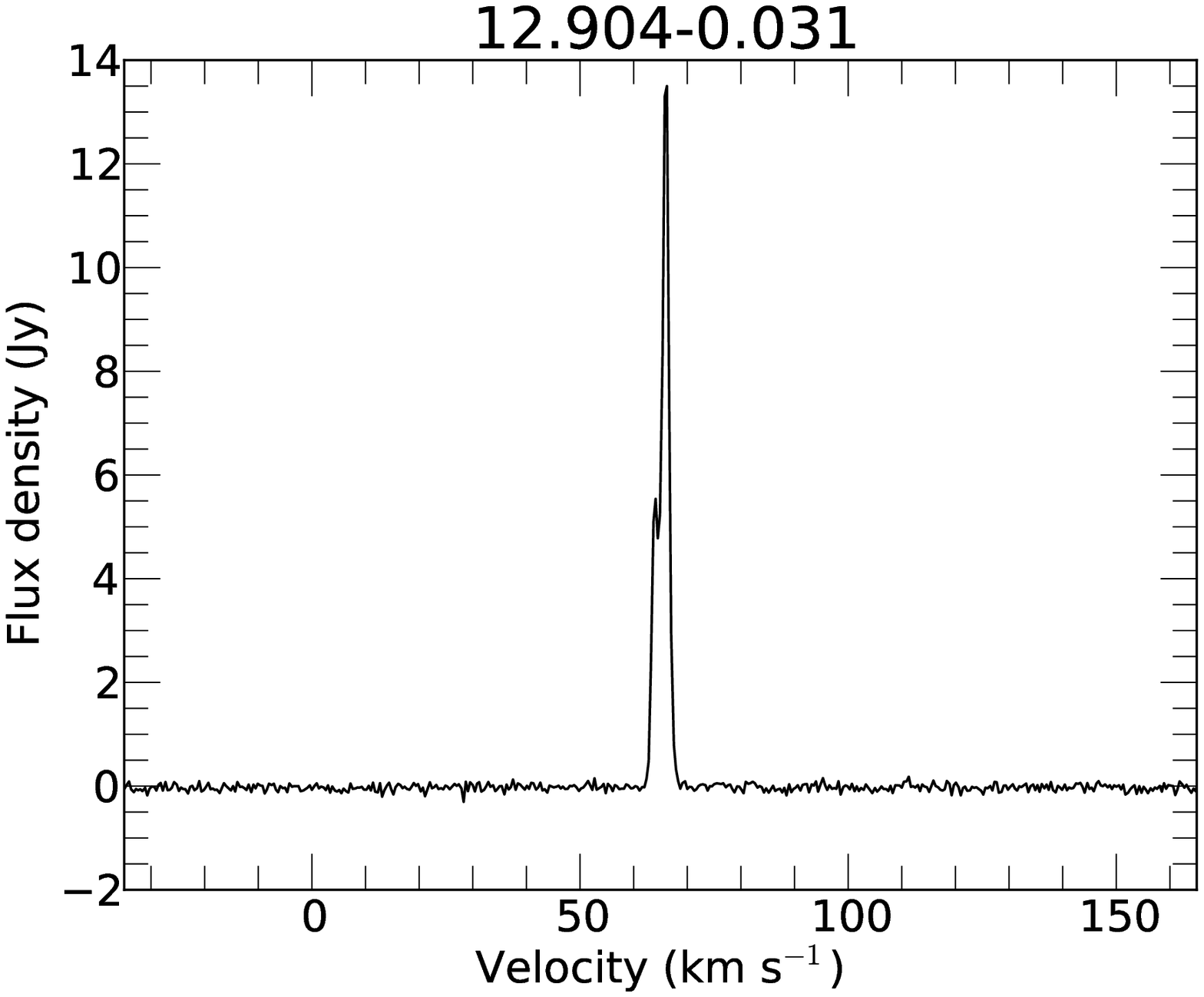}
\\
\contcaption{Spectra obtained with the ATCA of water masers associated with 6.7-GHz methanol masers.}
\end{figure*}

\begin{figure*}
\includegraphics[width=2.2in]{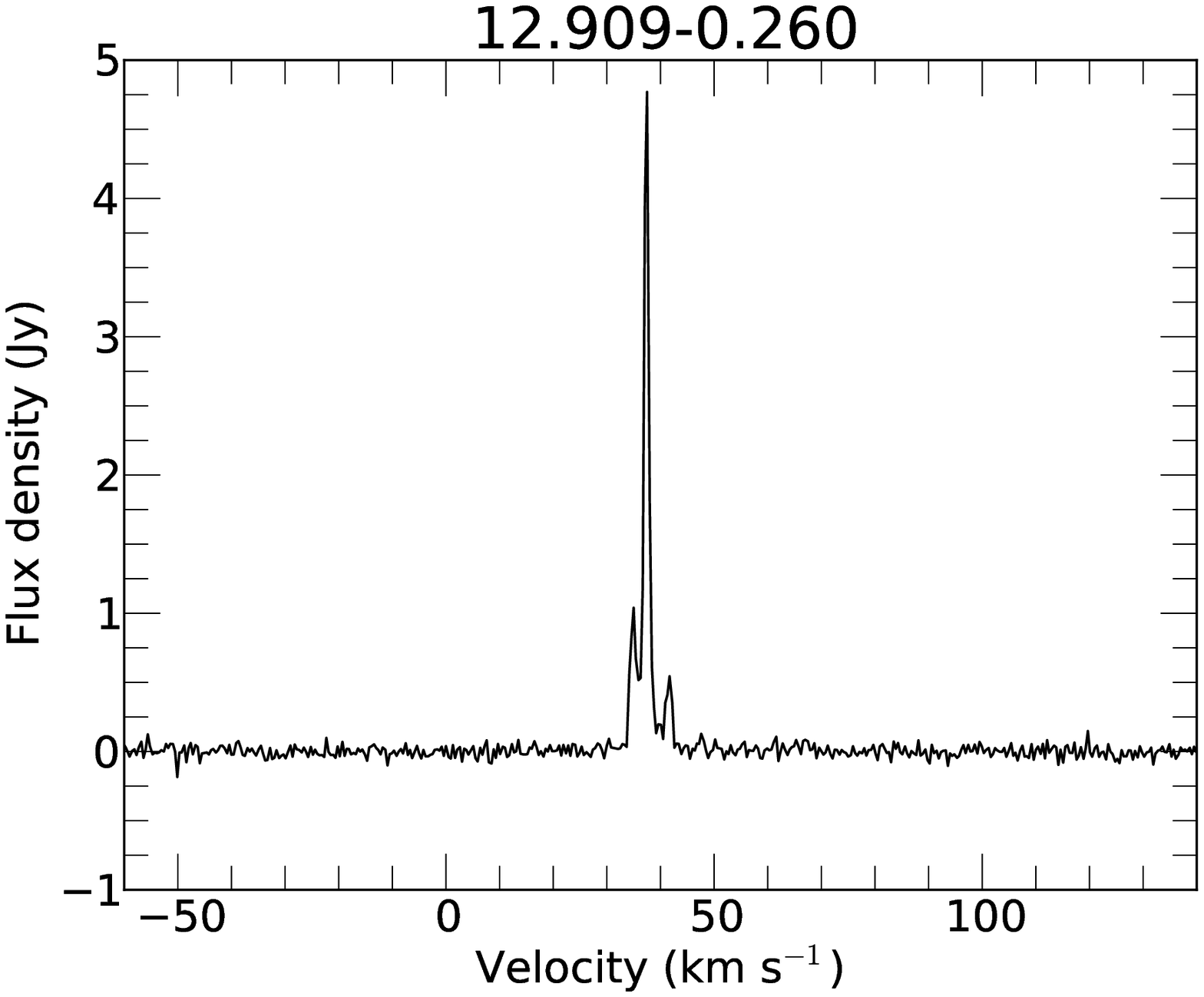}
\includegraphics[width=2.2in]{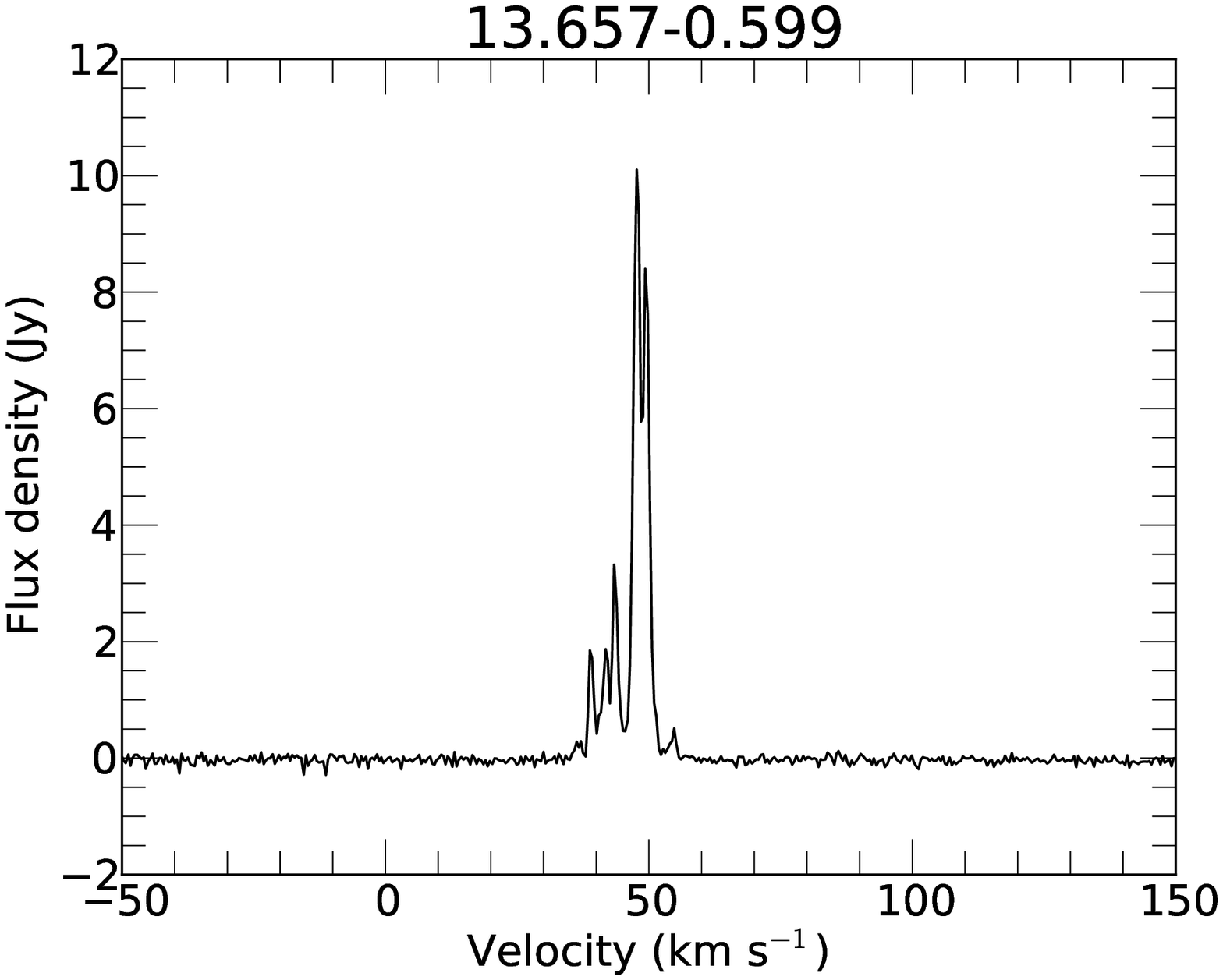}
\includegraphics[width=2.2in]{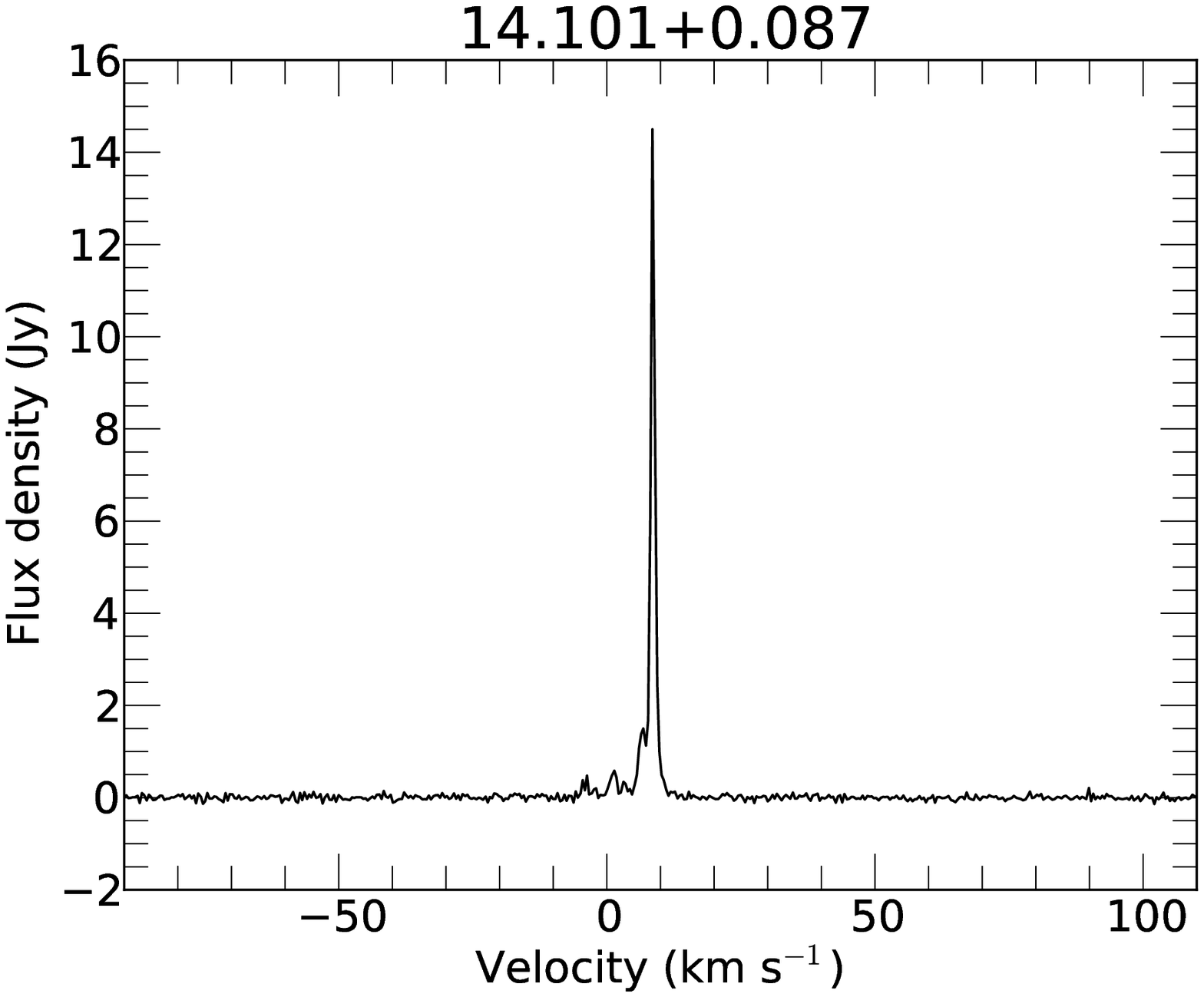}
\includegraphics[width=2.2in]{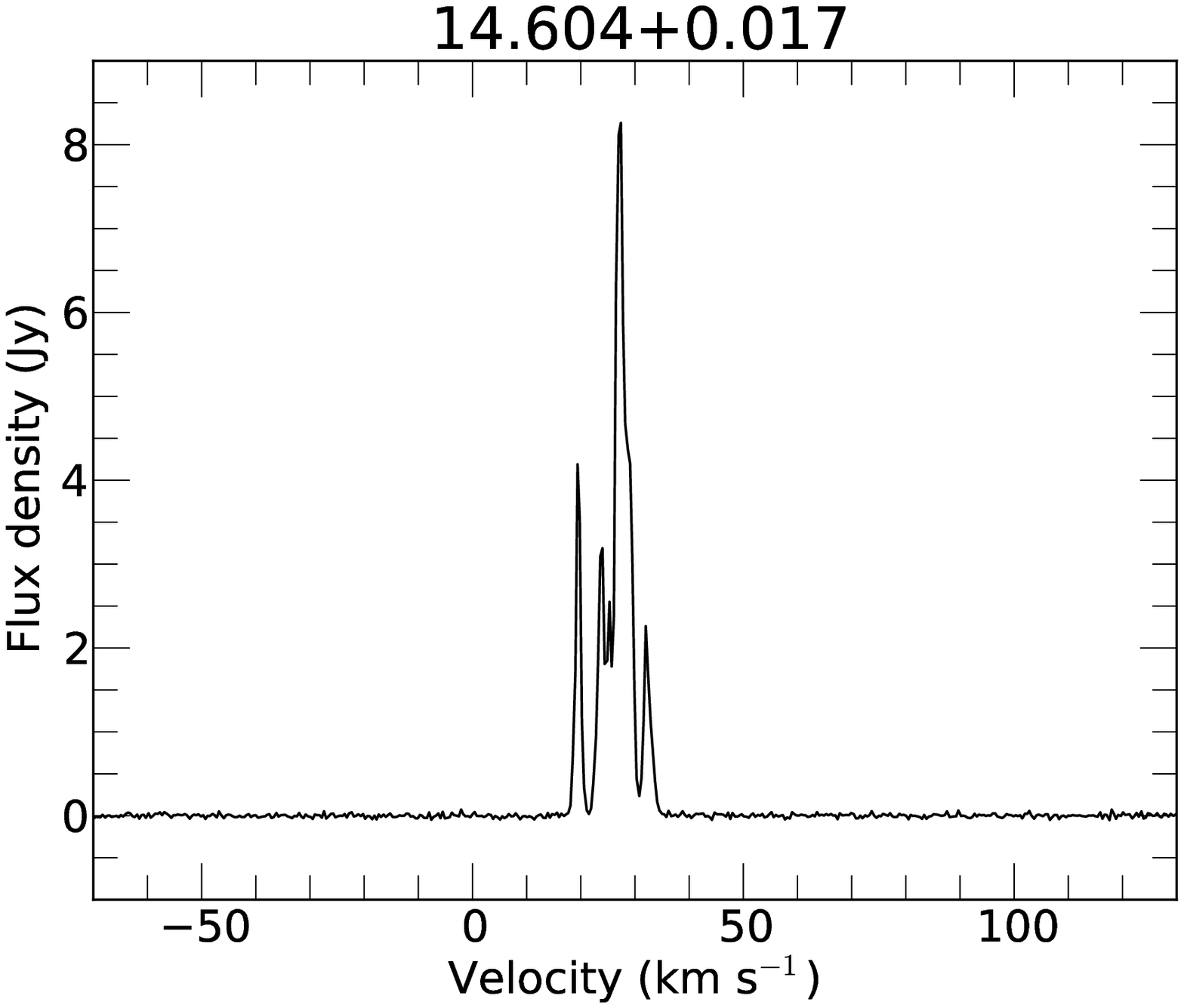}
\includegraphics[width=2.2in]{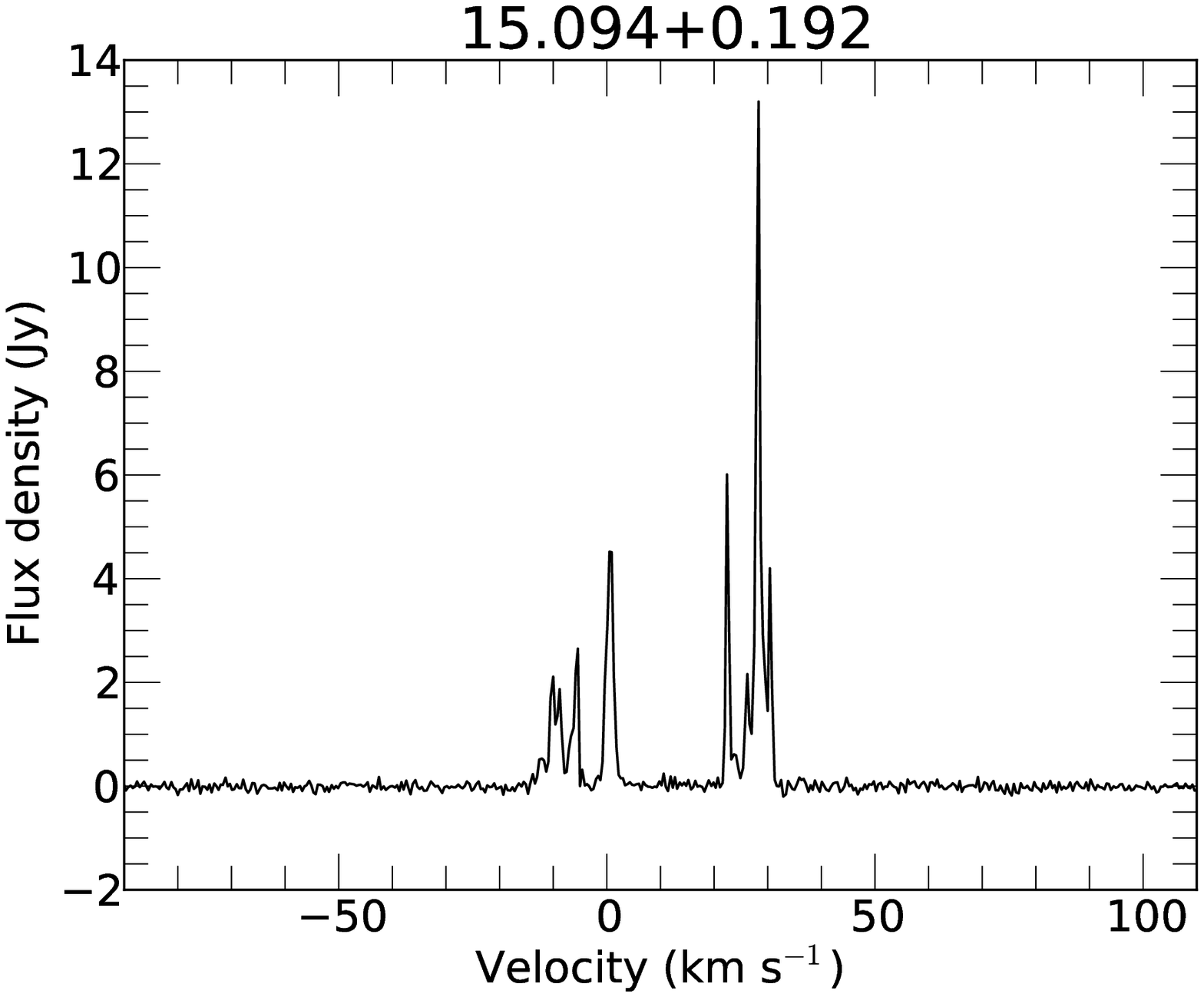}
\includegraphics[width=2.2in]{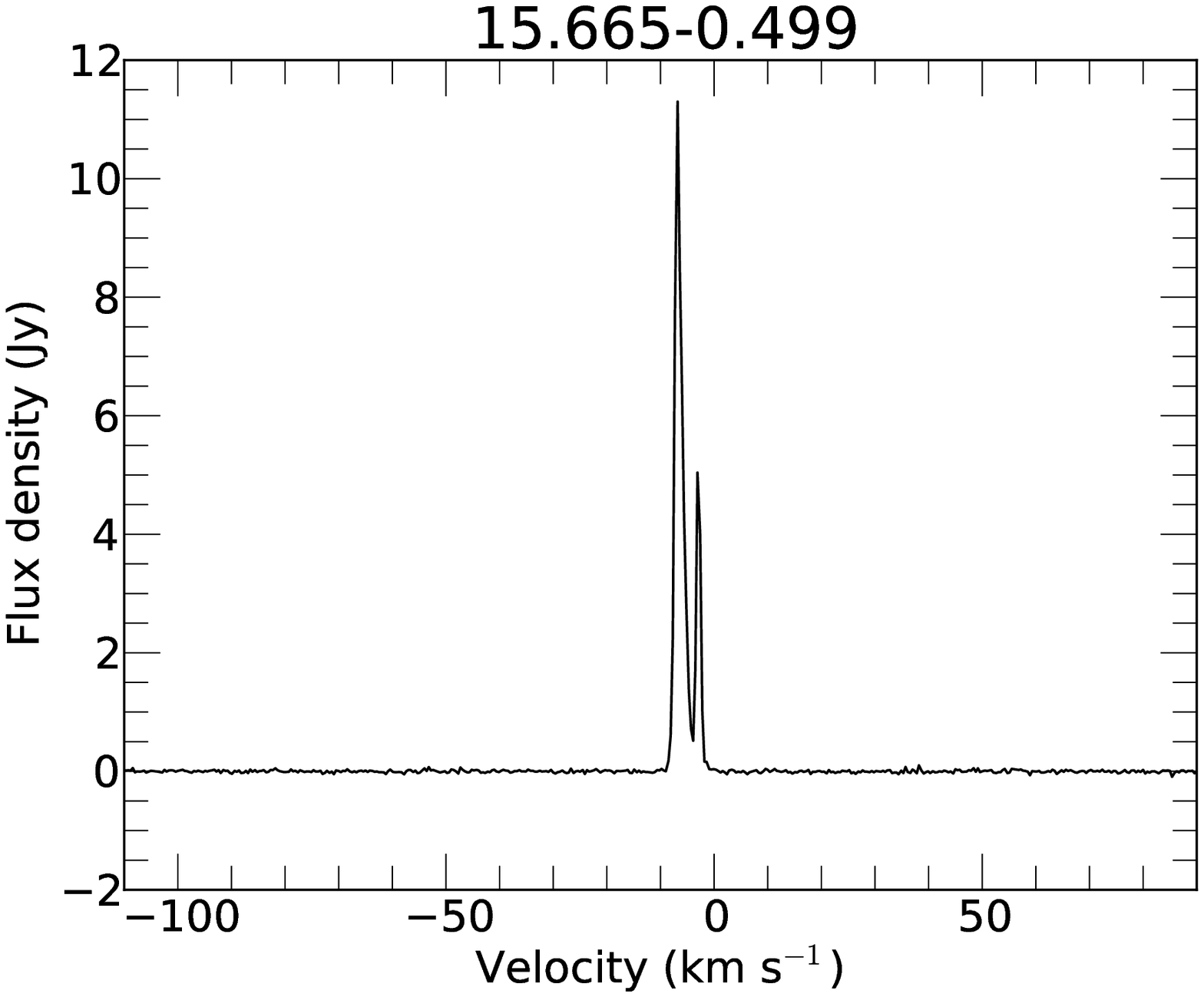}
\includegraphics[width=2.2in]{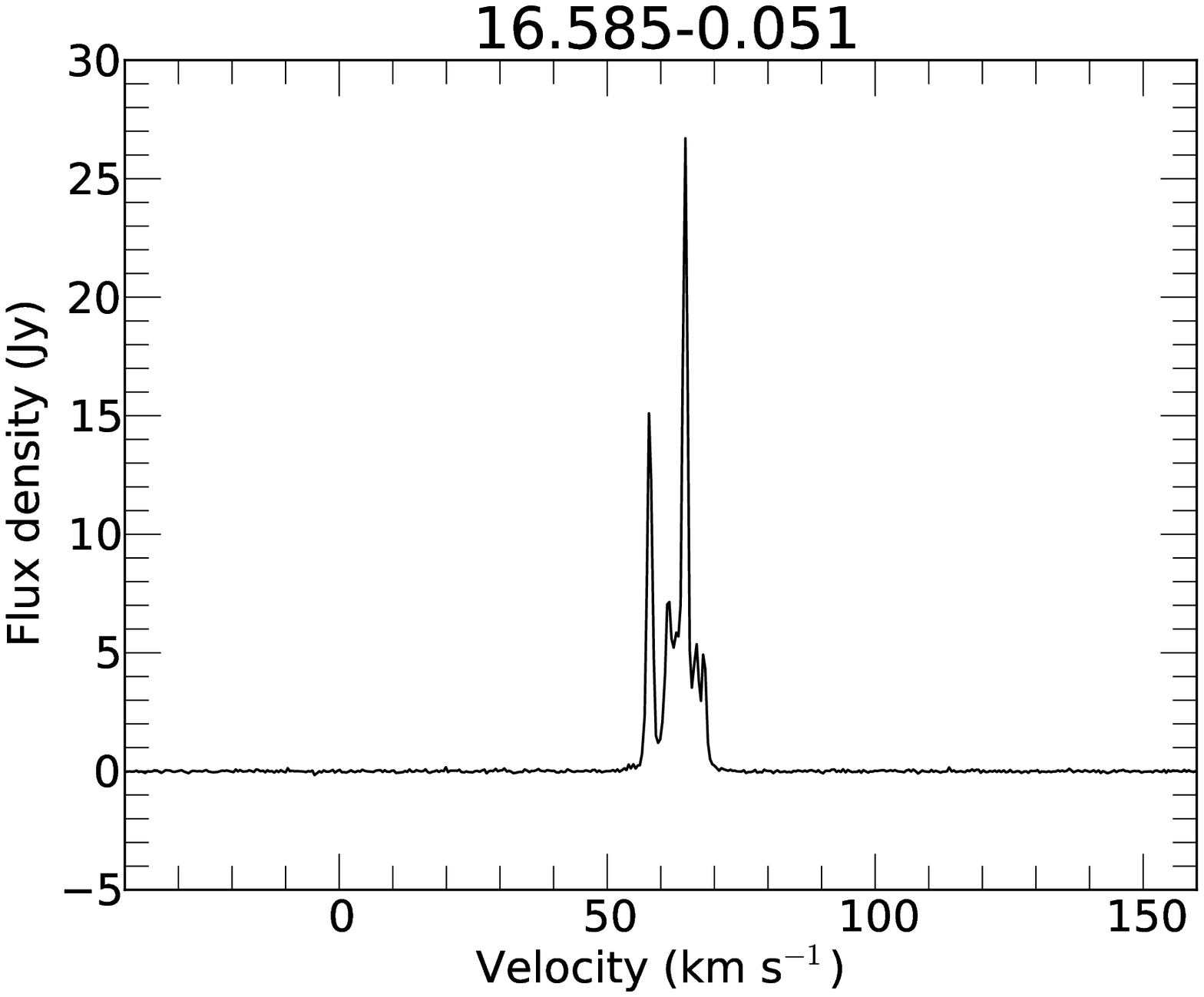}
\includegraphics[width=2.2in]{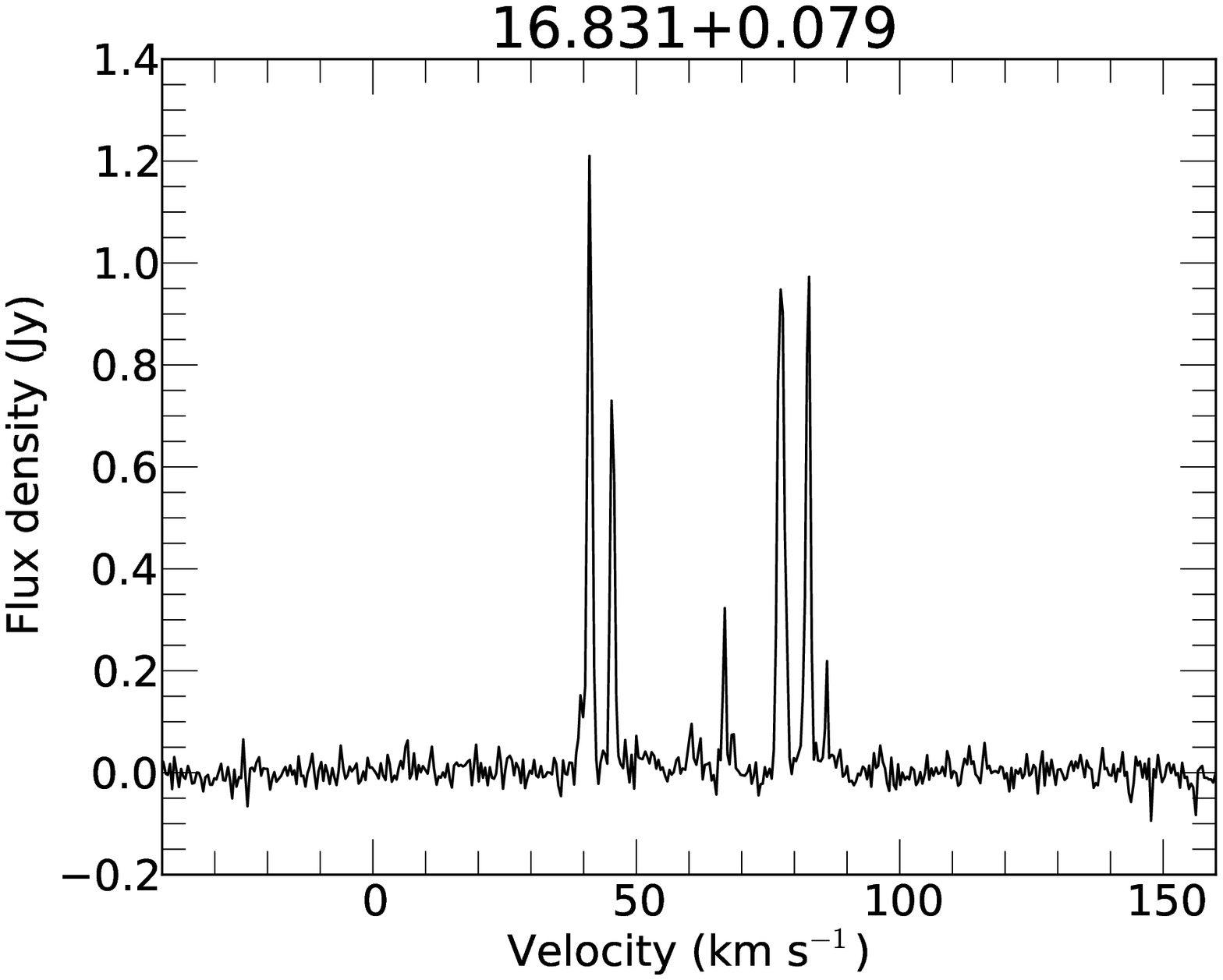}
\includegraphics[width=2.2in]{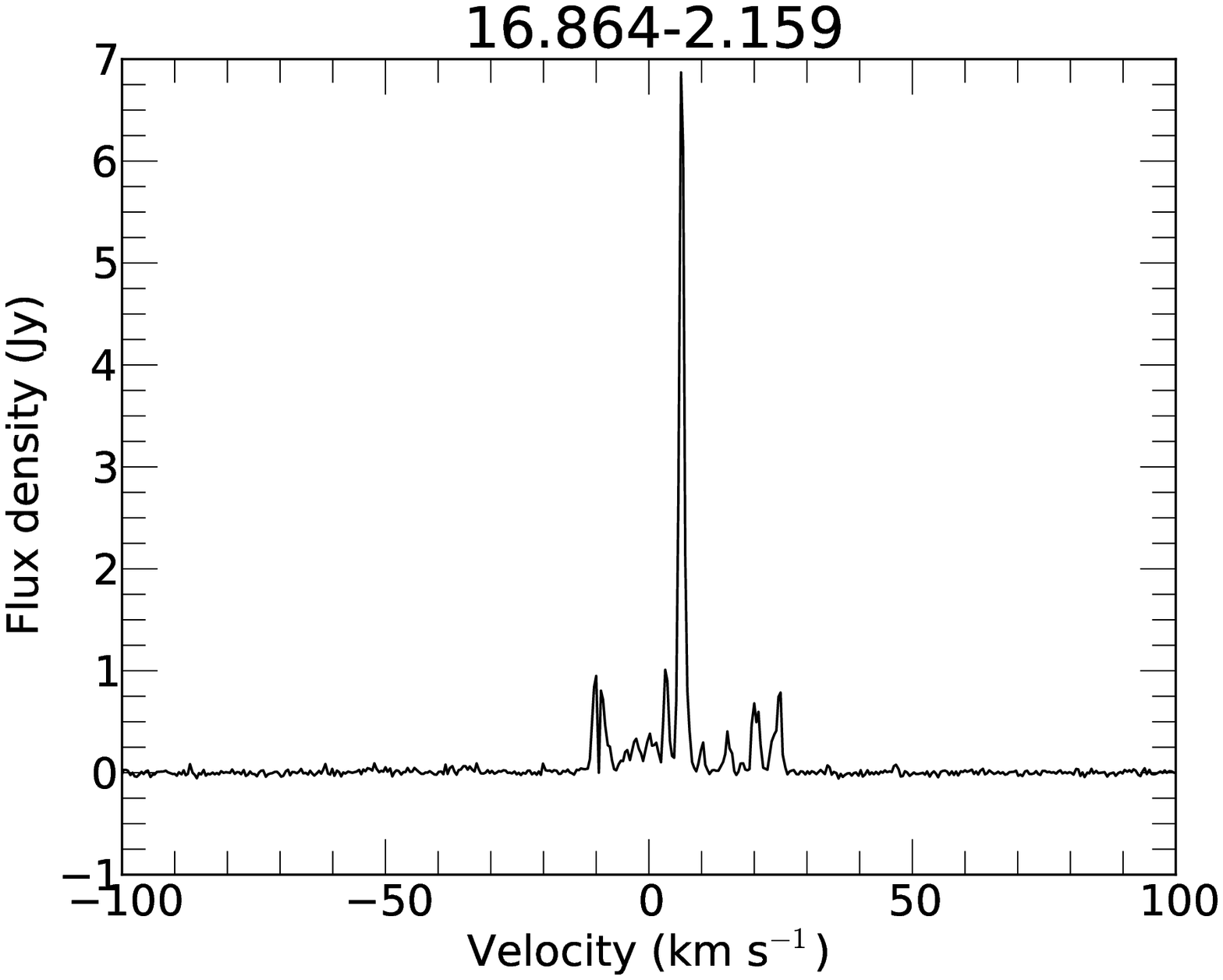}
\includegraphics[width=2.2in]{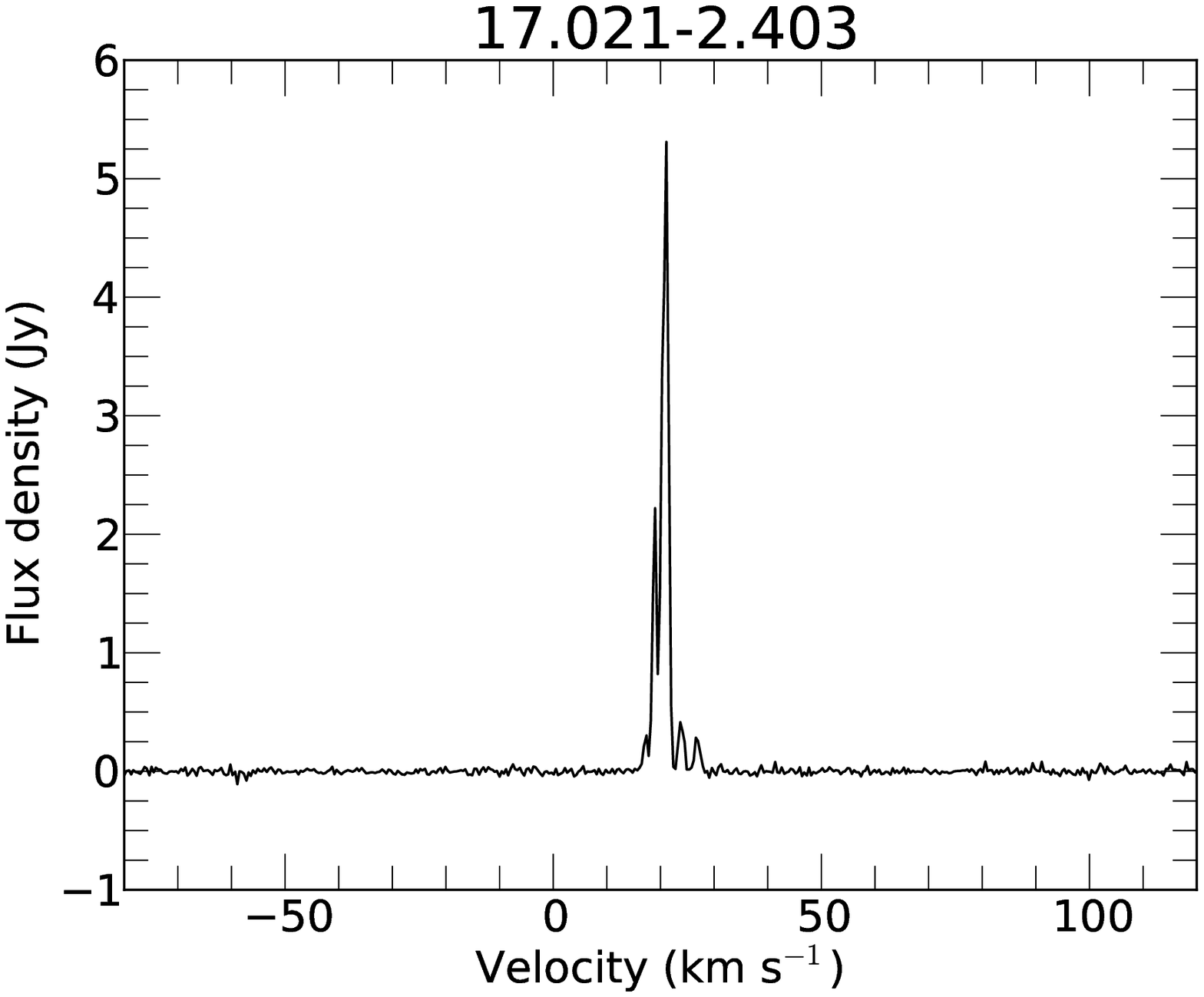}
\includegraphics[width=2.2in]{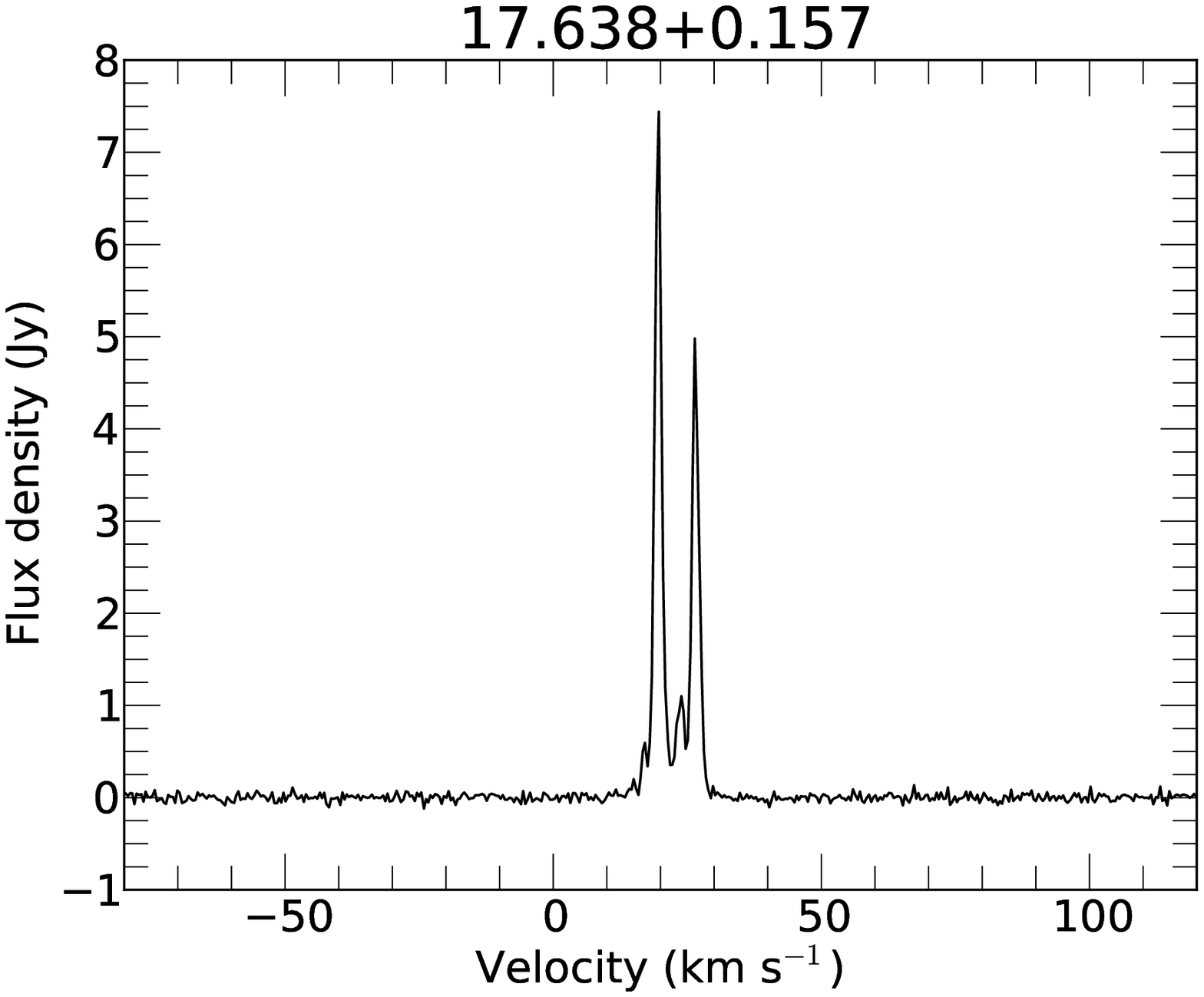}
\includegraphics[width=2.2in]{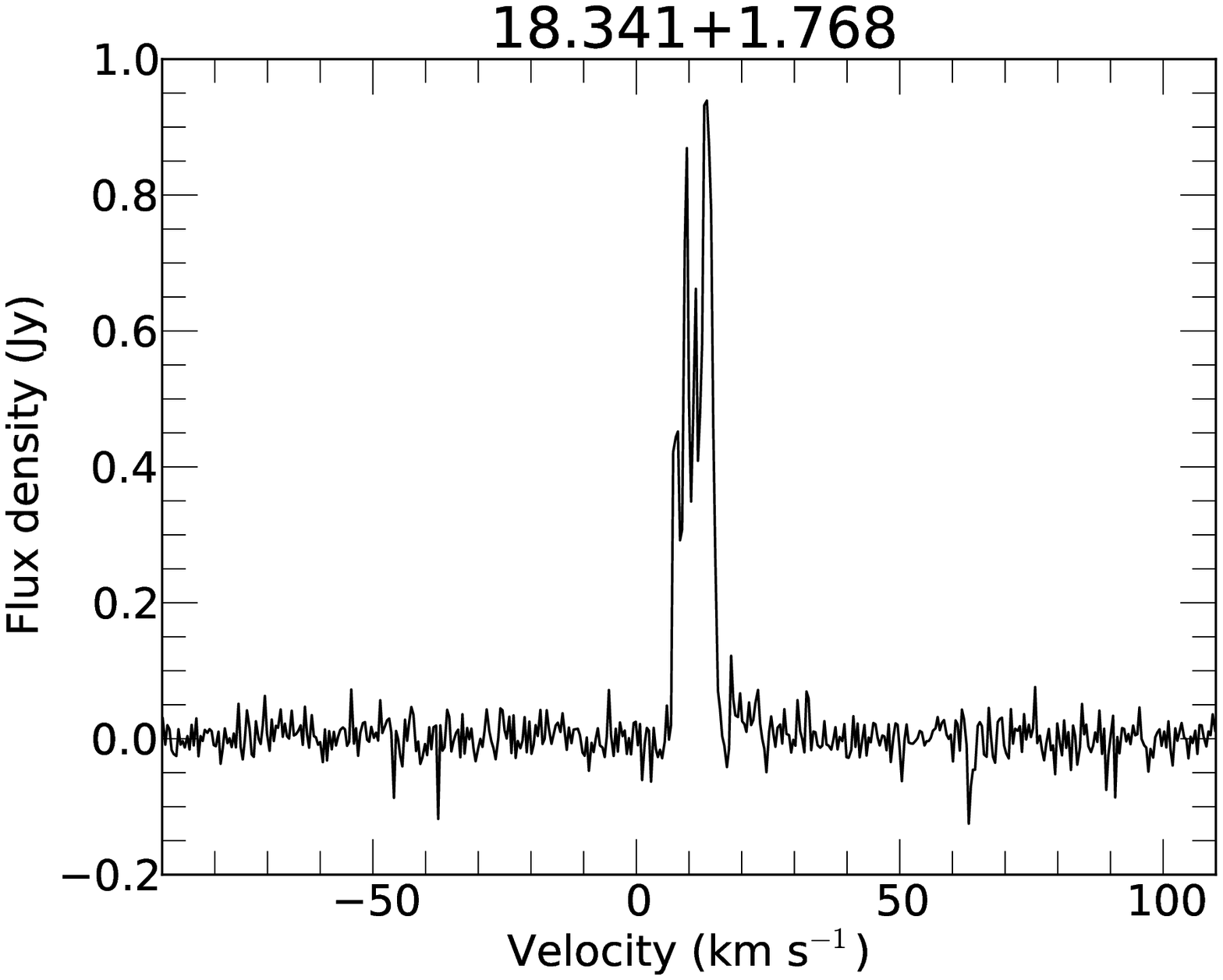}
\includegraphics[width=2.2in]{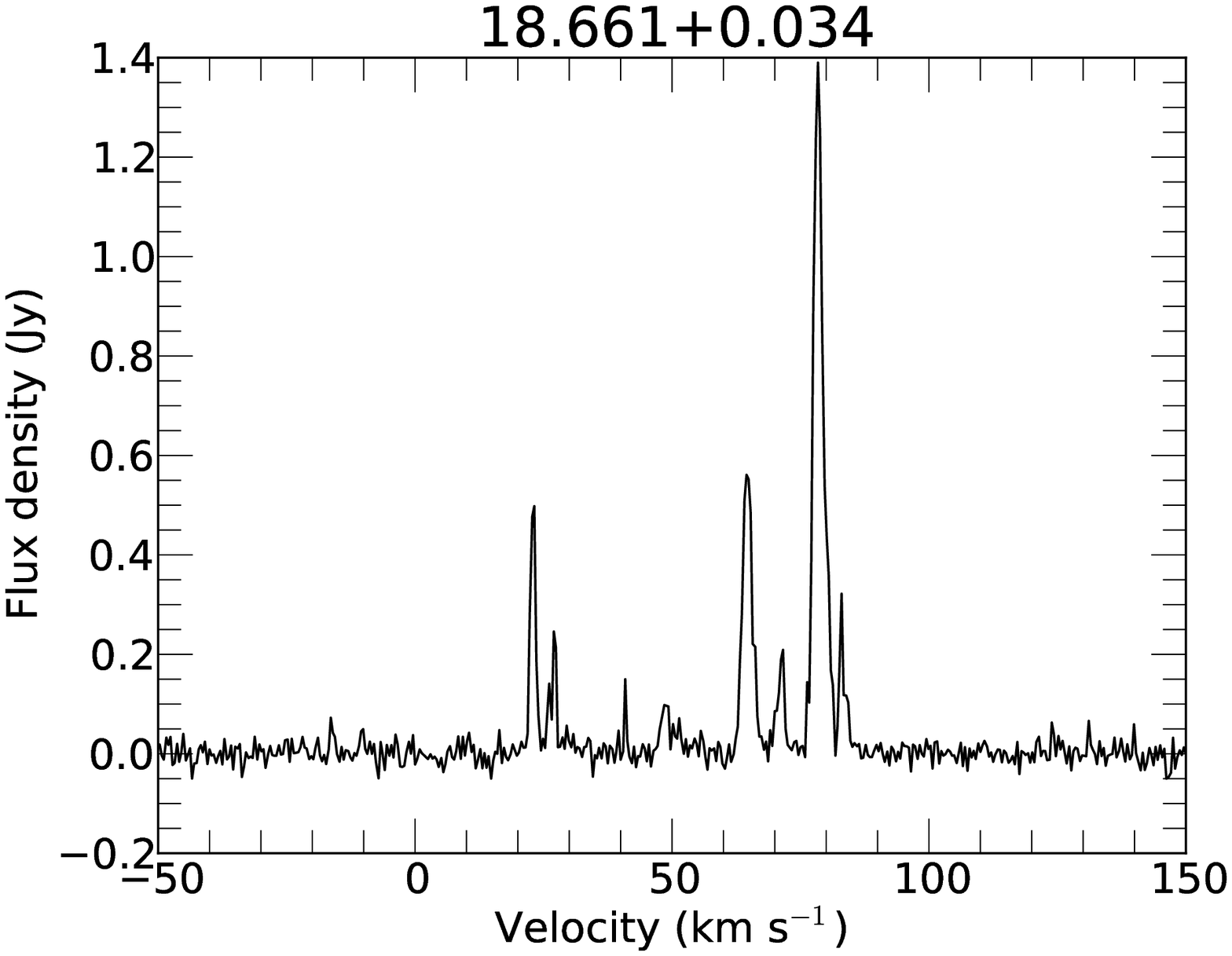}
\includegraphics[width=2.2in]{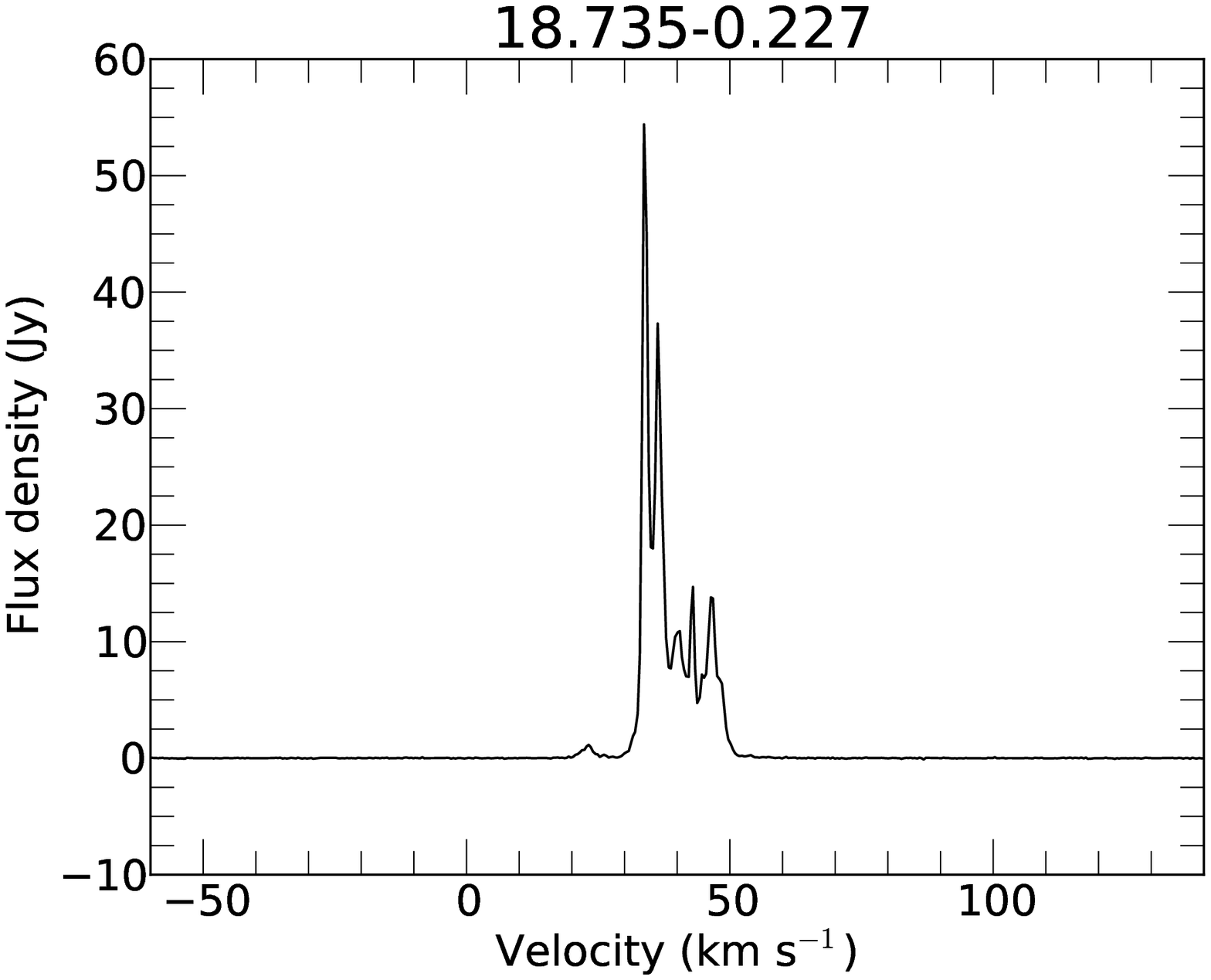}
\includegraphics[width=2.2in]{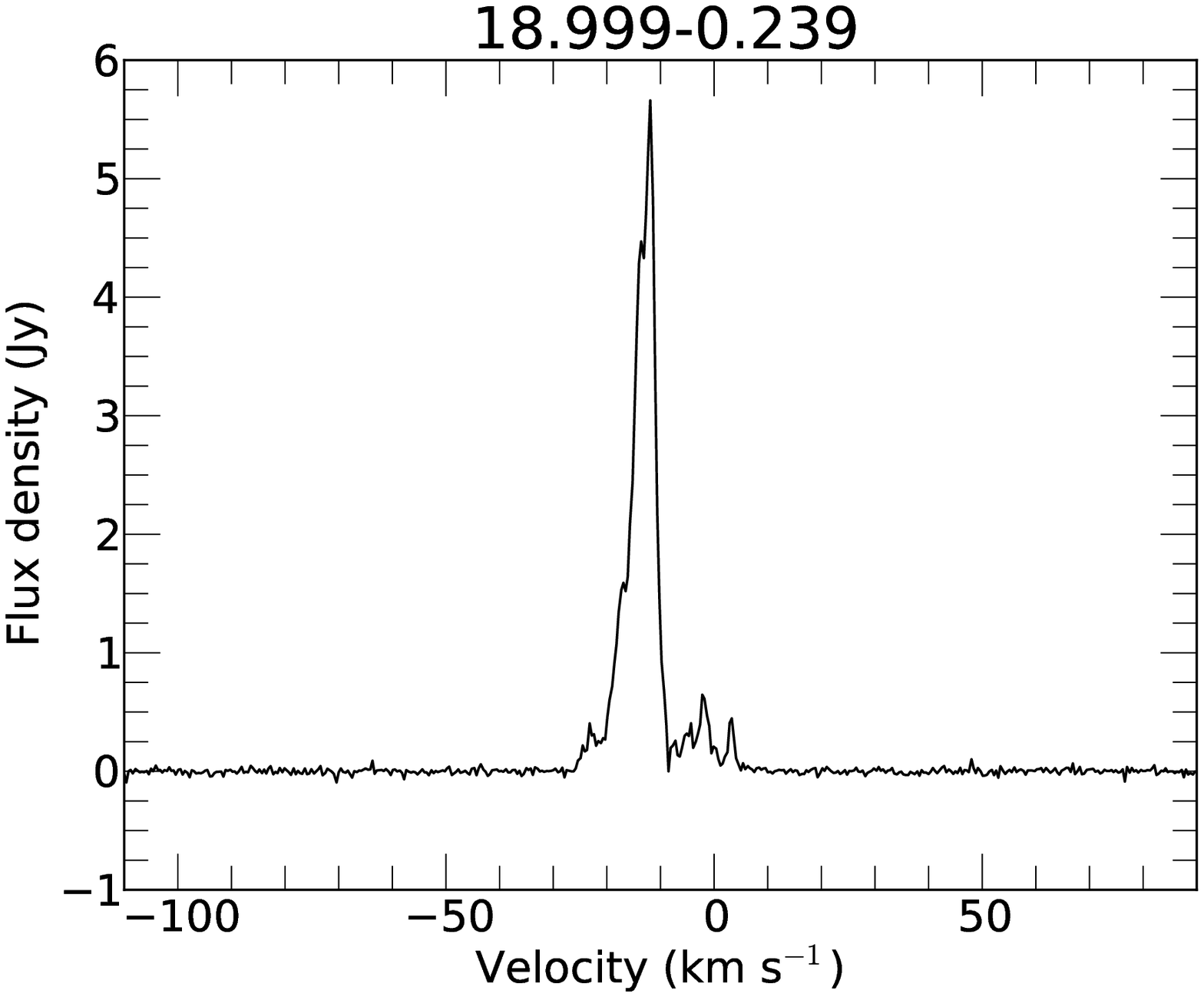}
\\
\contcaption{Spectra obtained with the ATCA of water masers associated with 6.7-GHz methanol masers.}
\end{figure*}

\begin{figure*}
\includegraphics[width=2.2in]{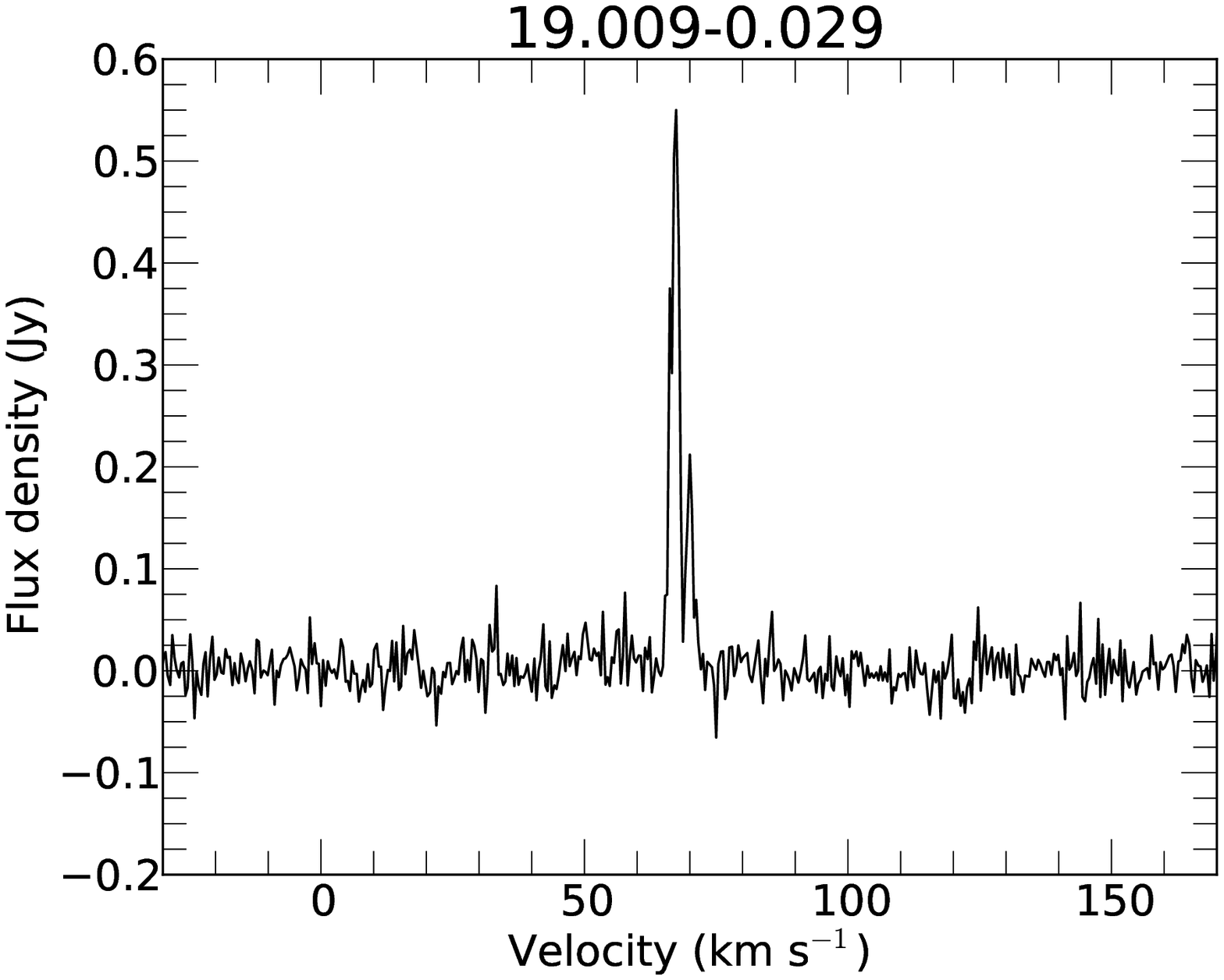}
\includegraphics[width=2.2in]{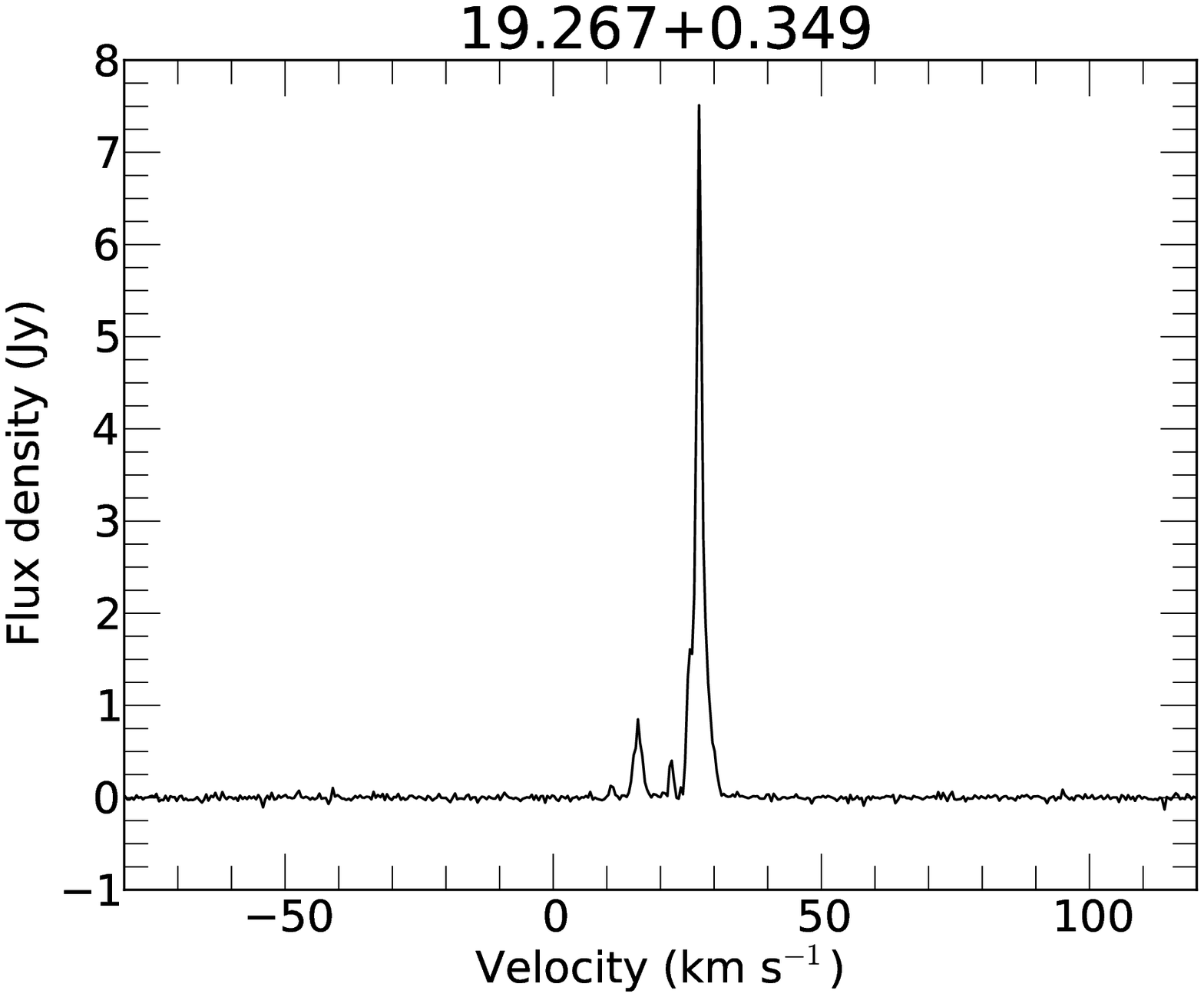}
\includegraphics[width=2.2in]{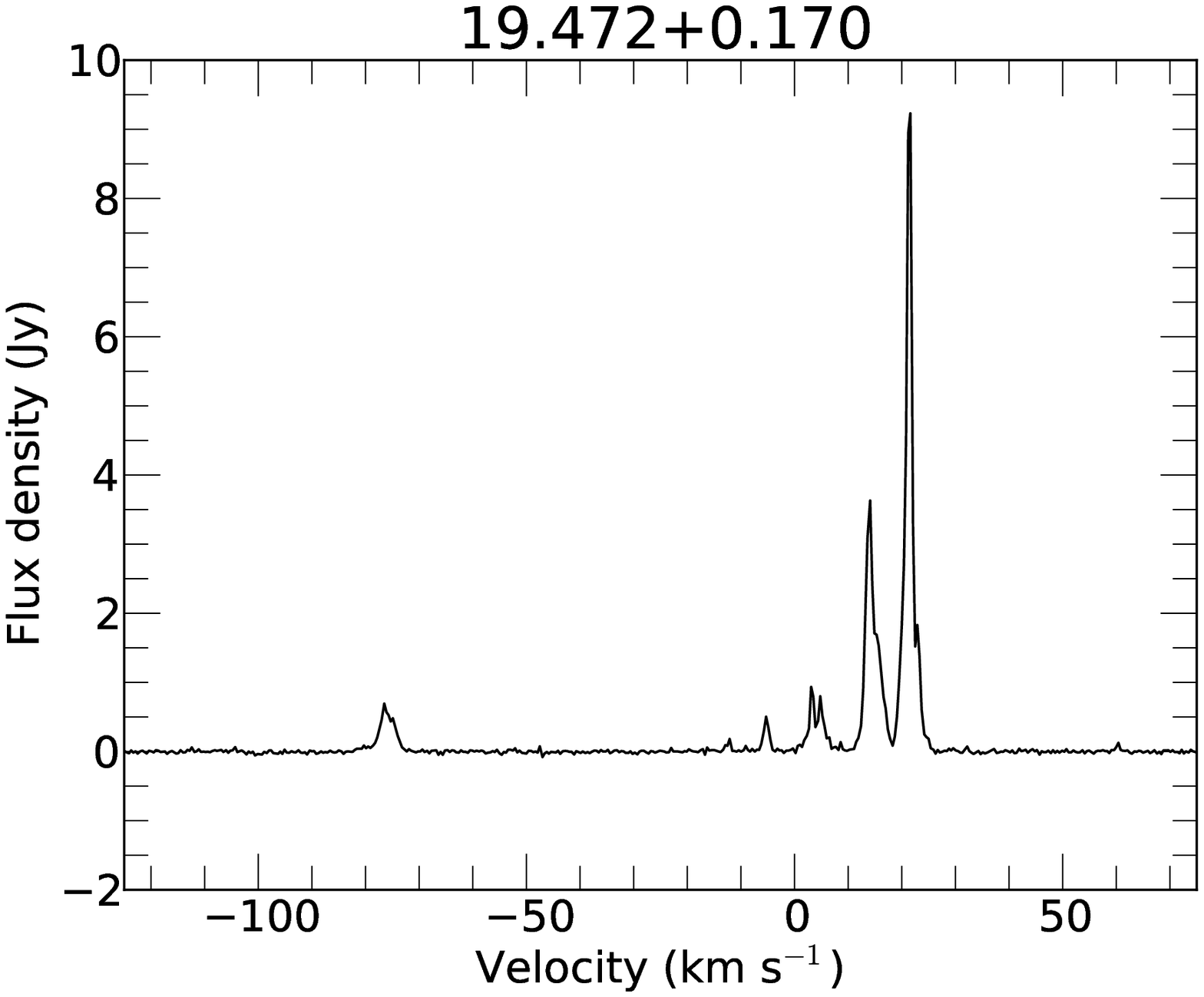}
\includegraphics[width=2.2in]{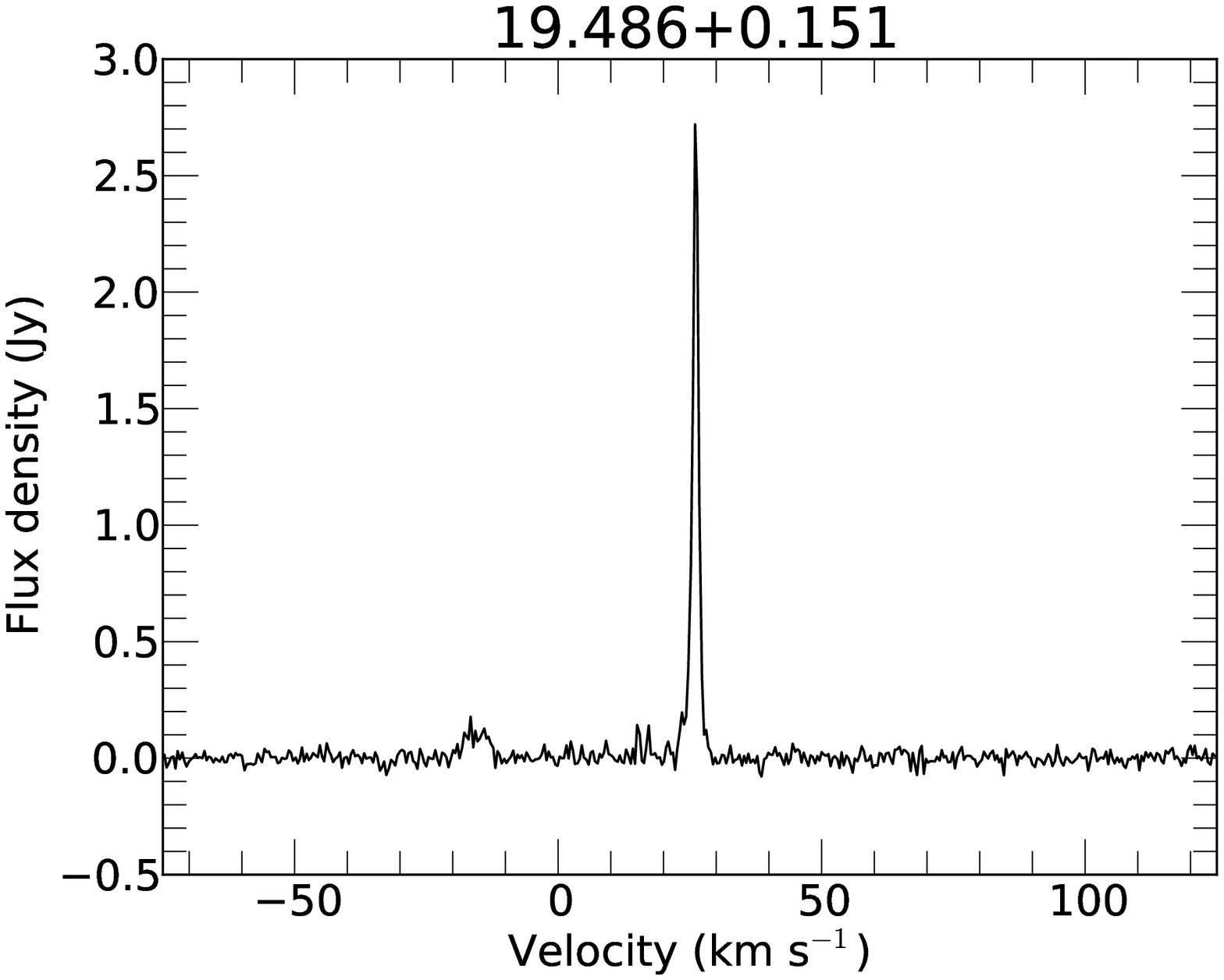}
\includegraphics[width=2.2in]{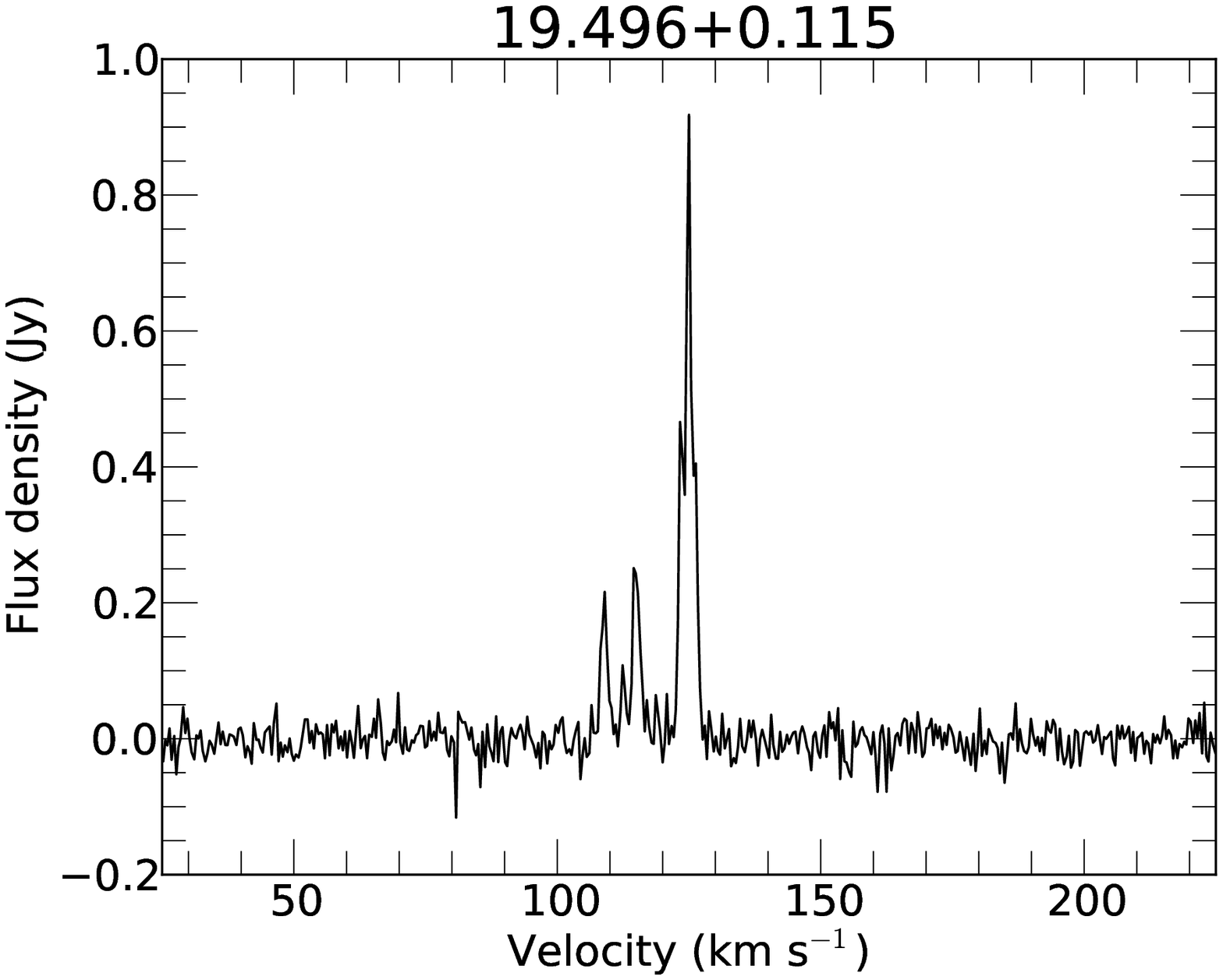}
\includegraphics[width=2.2in]{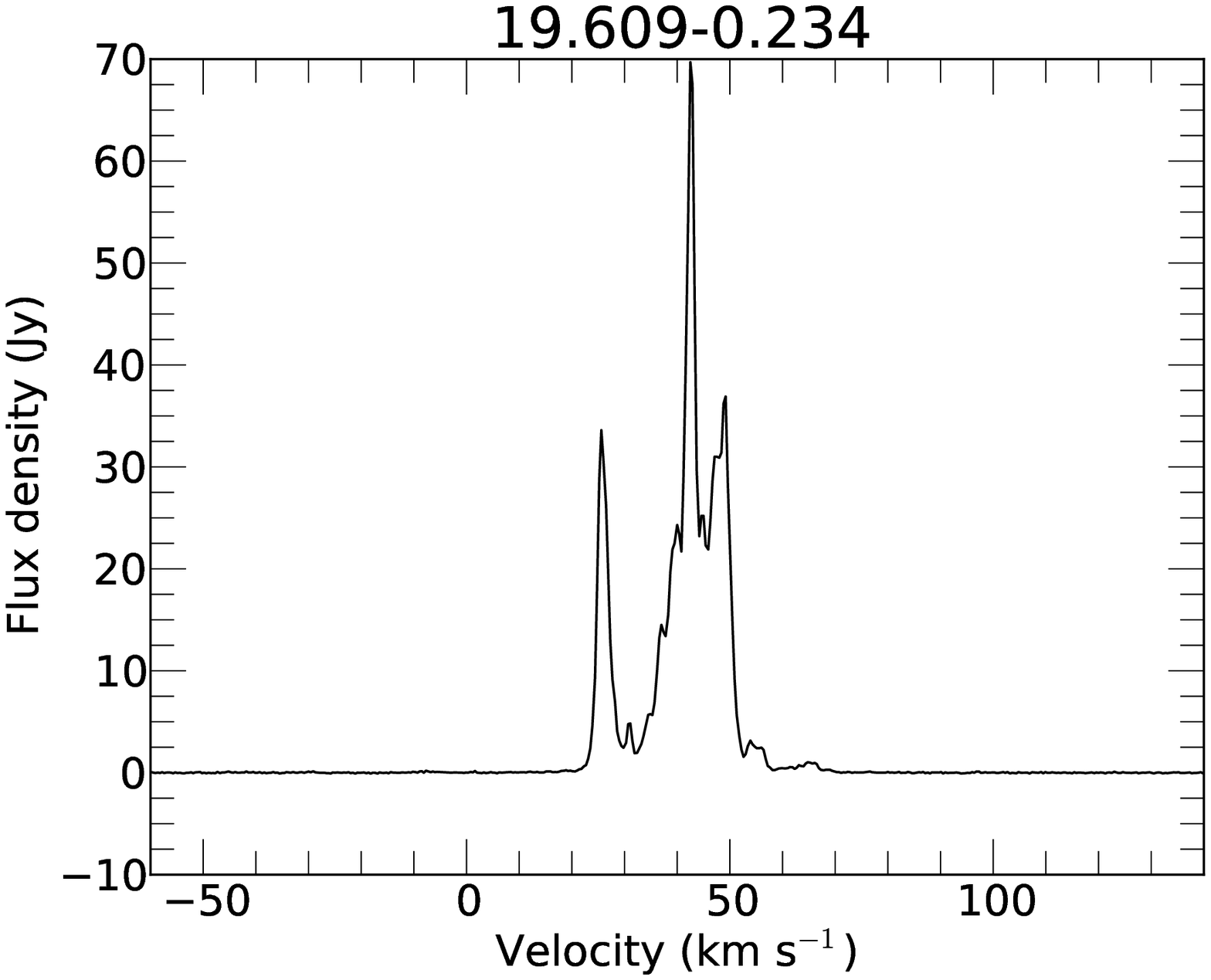}
\includegraphics[width=2.2in]{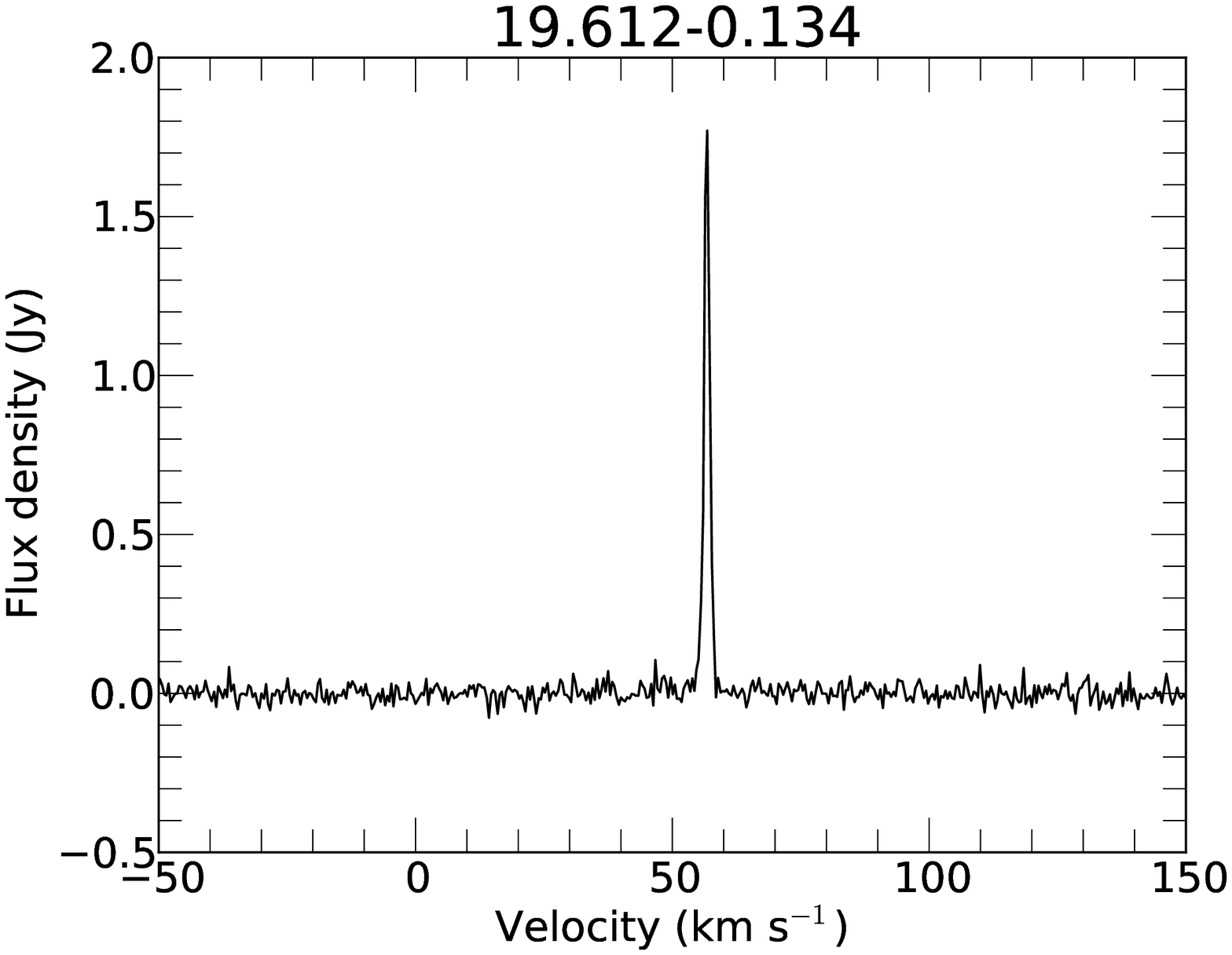}
\includegraphics[width=2.2in]{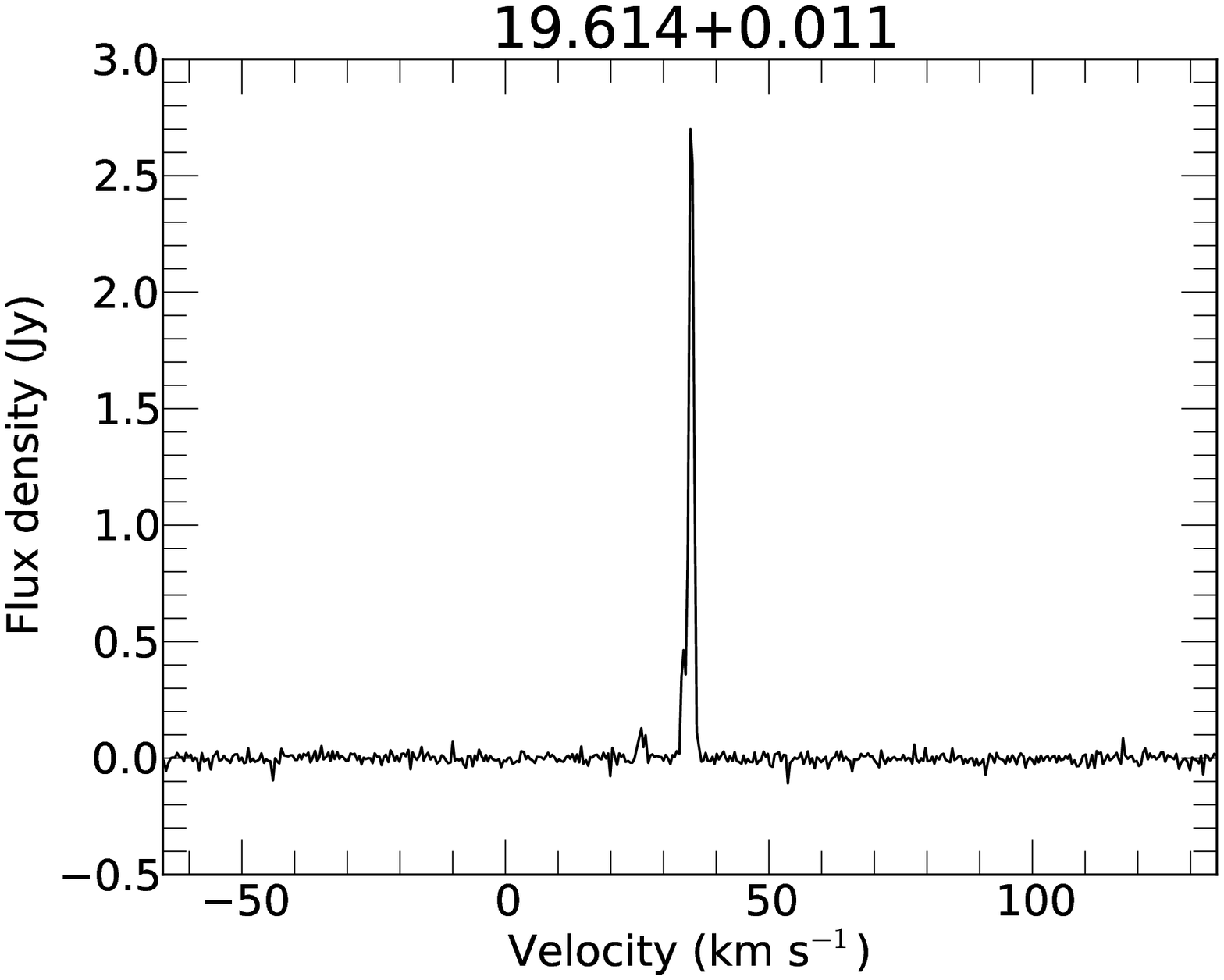}
\includegraphics[width=2.2in]{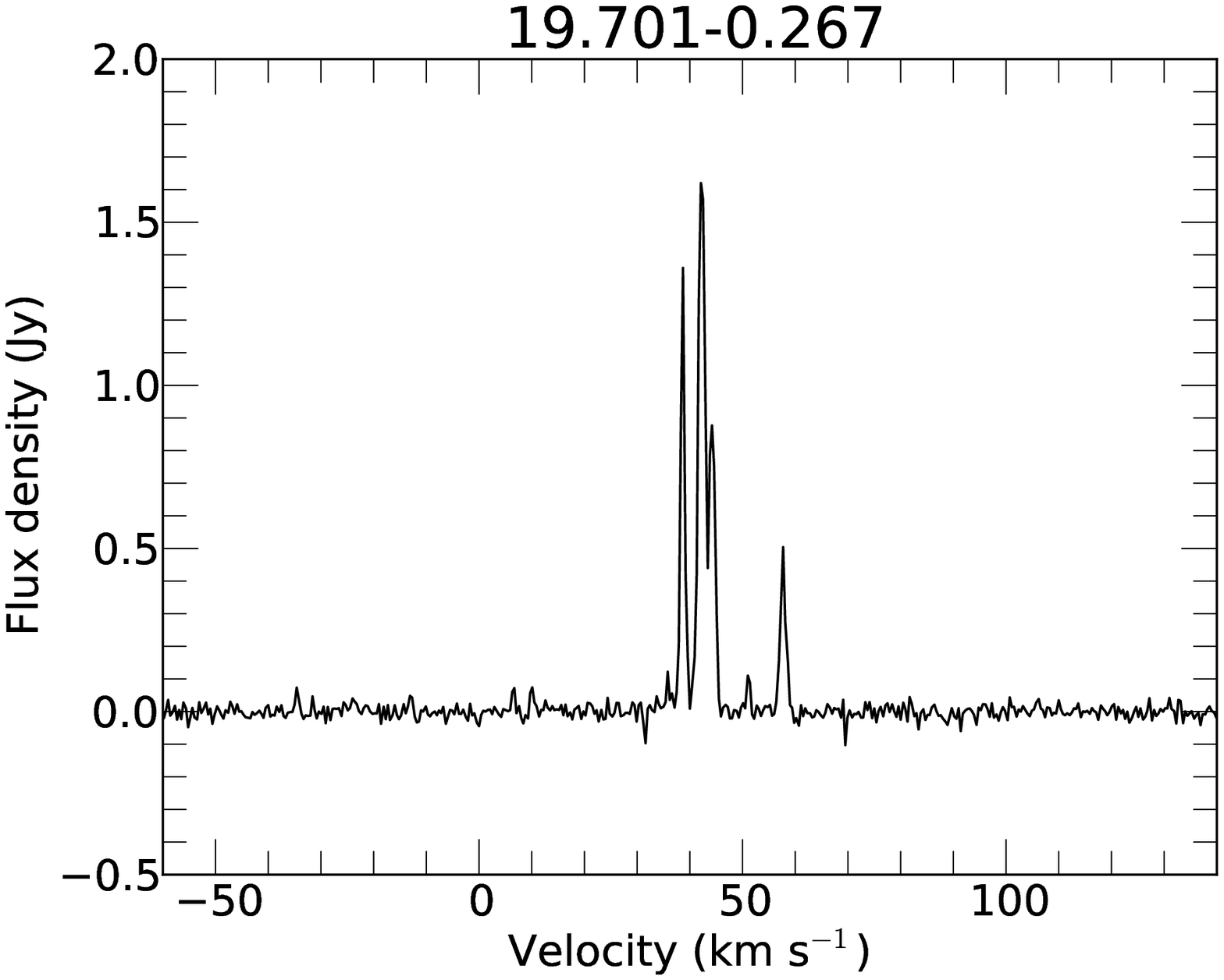}
\includegraphics[width=2.2in]{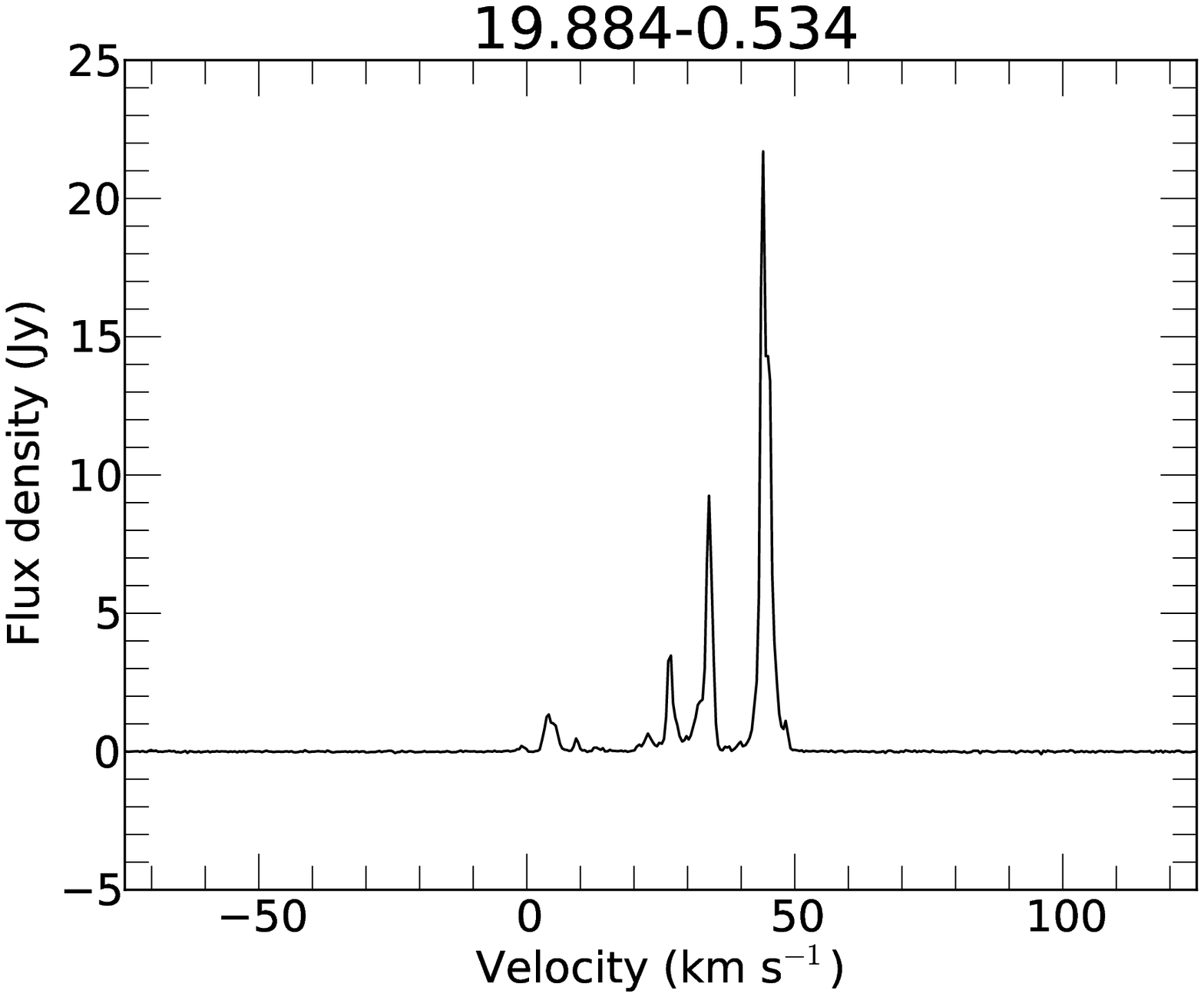}
\\
\contcaption{Spectra obtained with the ATCA of water masers associated with 6.7-GHz methanol masers.}
\end{figure*}

\subsection{Comments on individual sites of maser emission.}
\label{sec:indiv}

\textit{06.610--0.082, 06.795--0.257, 08.139+0.226, 08.669--0.356, 09.621+0.196, 09.986--0.028, 10.287--0.125, 10.342--0.142, 10.958+0.022, 12.209--0.102, 12.681--0.182, 12.889+0.489 and 17.638+0.157}. These water masers were also observed by \cite{breen10oh} and all have shown some variation in their spectra.

\textit{10.287--0.125, 10.886+0.123, 12.909--0.260, 15.665--0.499, 19.496+0.115,19.884--0.534, 14.631--0.577, 18.667+0.025, 18.888--0.475, 19.365--0.030}.  These sources we identified as having extended emission in the GLIMPSE 4.5~$\mu$m band that appear similar to the EGOs identified by \cite{chen13a,chen13b,cyganowski08} but are not in their catalogues. 

\textit{06.189--0.358}. This source has a strong methanol maser (221.6~Jy) and a weaker water maser (13.1~Jy). Inspection of the GLIMPSE three colour images shows emission in the 4.5~$\mu$m band in excess of the other bands around the object the masers are associated with and this object is embedded within an Infrared Dark Cloud.

\textit{10.323--0.160}. No water maser emission was detected associated with this methanol maser in our observations (5 sigma detection limit of 75~mJy), however \cite{breen10oh} detected a water maser with peak flux density of 3.2~Jy here in their 2004 observations (not observed in 2003). 

\textit{10.444--0.018}. The GLIMPSE three colour image of this source does not show anything at the location of these masers, although it is associated with a 1.1-mm dust clump detected with the Bolocam Galactic Plane Survey. This is a distant source at 11.0~kpc so this may account for the lack of mid-infrared emission. Also associated with this source is an OH and 12.2-GHz methanol maser \citep{breen14,caswell98} the presence of which suggests that this is not a particularly young maser region. 

\textit{10.472+0.027}. This water maser has a high velocity feature offset 250~km~s$^{-1}$ from the systemic velocity of the region (assuming the 6.7-GHz methanol maser peak velocity to be at the systemic velocity) and covers a total velocity range of 296~km~s$^{-1}$. It is offset 0.9~arcsec from the 6.7-GHz methanol maser and is the highest velocity water maser feature known in any high mass star formation region. This may be because previous studies have not had the velocity coverage of this survey ($>$~800~km~s$^{-1}$). Further discussion of this source is presented in \cite{titmarsh13}.  This source was also detected in the \cite{breen10oh} observations, although they did not have sufficient velocity coverage to detect the extreme high velocity features seen in our observations.

\textit{10.480+0.033}. We did not detect any water maser emission at this site, however, it was observed by \cite{breen10oh} in 2003 with a peak flux density of 12~Jy but not detected in their 2004 observations (detection limit of 0.2~Jy).

\textit{11.034+0.062}. This water maser had a peak flux density of 0.507~Jy in our observations and was also observed by \cite{breen10oh} with a peak flux density of 0.6~Jy in 2003 but was not detected in their 2004 observations (detection limit of 0.2~Jy).

\textit{11.109--0.114}.  In our search we found no water maser emission associated (5 sigma detection limit of 35~mJy), however, \cite{pillai06b} found weak ($\sim$ 0.3~Jy) water maser emission in their observations in 2004.

\textit{11.497--1.485}. No water maser emission was detected associated with this methanol maser in our observations (5 sigma detection limit of 75~mJy), however \cite{breen10oh} detected a water maser with peak flux density of 112~Jy here in their 2004 observations (not observed in 2003).

\textit{11.904--0.141}. We did not detect any water maser emission at this site, however, it was observed by \cite{breen10oh} in 2003 with a peak flux density of 0.3~Jy but not detected in their 2004 observations (detection limit of 0.2~Jy).

\textit{12.203--0.107}. No water maser emission was detected associated with this methanol maser in our observations (5 sigma detection limit of 55~mJy), however \cite{breen10oh} detected a water maser with a peak flux density of 6~Jy in 2003 and 8~Jy in 2004.

\textit{12.909--0.260}. This water maser had a peak flux density of 4.77~Jy in our observations. It was also detected by \cite{breen10oh} in 2003, with peak of 0.8~Jy, but undetectable (less than 0.2~Jy) in 2004.

\textit{14.457--0.143} we note that there is a typographical error in the MMB paper \cite{green10}. The declination given is --$16^{\circ}27'57''.5$, however, the correct declination is --$16^{\circ}26'57''.5$.

\textit{18.999--0.239}. The 6.7-GHz methanol maser has one narrow feature with a peak velocity of 69.4~km~s$^{-1}$, yet the associated water maser peaks at -11.86~km~s$^{-1}$ with no emission anywhere near the velocity of the methanol. \cite{caswell08} suggested that the water masers dominated by blueshifted emission such as this one generally occur at an early stage in the evolution of a YSO as they generally do not have an associated OH maser \citep[this source does not have an OH maser detected above 0.2~Jy;]{caswell13}. This source was previously unknown, and inspection of the GLIMPSE three colour images shows an EGO associated with this YSO, suggesting that this water maser may be powered by an outflow or shock from this source (see Section \ref{sec:intro}).

\section{Discussion}
\label{sec:discussion}

There have been a number of previous studies which have compared relative detection rates and properties of water and methanol masers based on various sample selection criteria \citep{beuther02,szymczak05,xu08,breen10oh}. \citet{beuther02} made arcsecond resolution observations of water and 6.7-GHz methanol maser emission towards a sample of 24 star formation regions selected on the basis of infrared ({\em IRAS}) colours. They found approximately 60 per cent of the sources with 6.7-GHz methanol masers had an associated water maser (within 1.5 arcseconds) and approximately 65 per cent of the sources with a water maser had an associated 6.7-GHz methanol maser.  

\citet{xu08} made a sensitive (rms noise $\sim$0.1 - 0.2~Jy) though lower resolution search (half-power beam width of $\sim$2 arcminutes) for 6.7-GHz methanol masers towards a sample of 89 water masers thought to be associated with star formation regions, selected from the Arcetri catalogue \citep{comoretto90,brand94}. 81 of the target sources had infrared luminosities which suggest that they are high-mass star formation regions.  \citet{xu08} detected 10 new 6.7-GHz methanol masers. The \citet{xu08} selection criteria initially selected 178 sources, (161  thought to be high-mass star formation regions) of which  47 sources were already known to have an associated 6.7-GHz methanol maser and so were not targeted.  Thus the combined detection rate of 6.7-GHz methanol masers towards their sample of 22-GHz water masers associated with high-mass star formation was at least 57 out of 161 (35 per cent).  

Of all the previous investigations in the literature, the two which have the greatest similarity to the current study are those undertaken by \citet{breen10oh} and \citet{szymczak05}. \citet{szymczak05} used the Effelsberg 100m telescope to search for water masers towards a flux-limited (peak intensity greater than 1.6 Jy), but statistically complete sample of 6.7-GHz methanol masers.  They detected 41 water masers towards their sample of 79 sources, a detection rate of 52 $\pm$ 8 per cent for a single epoch search (with an rms noise of 0.45~Jy). Differences between our observations and those undertaken by \citet{szymczak05} are the significantly higher sensitivity and angular resolution and somewhat larger  size of our target sample.  The first two of these enable us to unambiguously determine when the two maser species are associated, to determine their association with infrared and other sources and also to better investigate if the detection rate for water masers changes with properties such as the 6.7-GHz maser luminosity. 

\citet{breen10oh} used the ATCA to make a sensitive search (5-$\sigma$ detection limit in a 1~km~s$^{-1}$ channel typically less than 0.2~Jy) for water masers towards a large sample of OH and 6.7-GHz methanol masers.  In total, 270 6.7-GHz methanol masers were observed, resulting in water maser detections towards 198 sources, a rate of 73 $\pm$ 5 per cent for a two epoch search. Here and below we use $\sqrt{N_{det}}$, where $N_{det}$ is the number of detections, as an estimate of the uncertainty in the detection rate (this method assumes that the non-detections are due to purely stochastic effects, and considering source evolution there is not a more appropriate method to estimate the errors). The chief difference between our observations and those undertaken by \citet{breen10oh} is that we have a statistically complete target sample. The 6.7-GHz methanol masers observed by \citet{breen10oh} were primarily those with associated OH maser emission, or those without OH maser emission which had an accurate position determined by \citet{caswell09}. The \citet{caswell09} sample in particular is biased towards sources with a higher 6.7-GHz peak flux density.

The detection rate we have achieved towards a statistically complete sample of 6.7-GHz methanol masers (46 $\pm$ 6 per cent) is for a single epoch search. It is slightly lower than that obtained by either \citet{szymczak05} in a single epoch search or \citet{breen10oh} in a two epoch search. Comparing our results with those of \cite{szymczak05}, we can see that to within the uncertainty of the two studies the detection rates are the same.  Furthermore, the \citet{szymczak05} study had significantly lower positional accuracy (telescope half power beam width of 40 arcseconds at 22-GHz) and had a less sensitive limit for their target sample (approximately a factor of 2 higher than for the MMB). Both of these factors may influence the detection rate and when we restrict our MMB sample to only sources with a peak flux density greater than 1.6 Jy and relax our association criteria to 40 arcseconds we find a detection rate for water masers towards these sources of 52 percent, the same as \citet{szymczak05}. 

We are also able to compare our results with those of with those of HOPS, the H$_2$O Southern Galactic Plane Survey \citep[HOPS; ][]{walsh11}. HOPS is the first large-scale water maser survey that is not biased to pre-existing known likely associations.  The HOPs rms noise level is 1 - 2 Jy over most of the survey and hence is likely to have detected the majority of sources with a peak intensity greater than 10~Jy (subject to variability), and some of the sources in the 5 -- 10~Jy range. HOPS was performed with a 22-m single dish and so has significantly poorer angular resolution, and thus positional uncertainties as great as 1 arcminute or more. 

\begin{figure}
\includegraphics[width=3.5in]{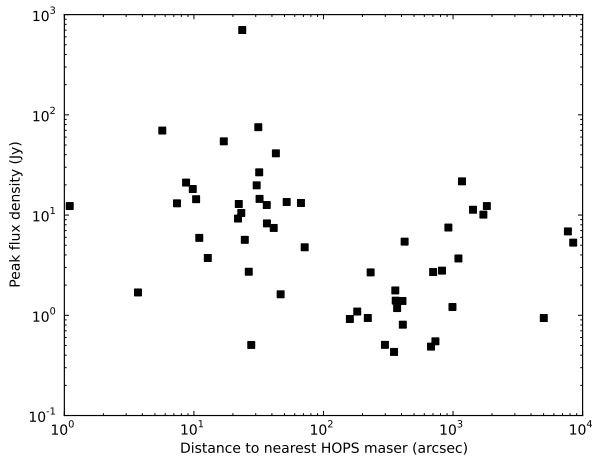}
\caption{Water maser peak flux density from our observations versus the distance to the nearest HOPS water maser position.}
\label{fig:hops}
\end{figure}

Figure \ref{fig:hops} plots the peak flux densities of the water masers that we detected against the distance to the nearest maser in the HOPS catalogue. This shows a clear break at separations of around 100~arcseconds, and strongly suggests  that the water masers from our observations with separations greater than 100~arcseconds represent sources with no HOPS counterpart, and conversely, many HOPS sources not associated with methanol masers. They tend to be our weaker detections, consistent with their non-detection in the lower sensitivity HOPS survey. Only 29  ATCA water masers have a possible HOPS counterpart closer than 100~arcseconds, and individual inspection suggests that 20 of these are true counterparts. Their positions were mostly within 1 arcminute of those observed with the ATCA, and the median offset was 20 arcseconds. There were no additional HOPS sources with a methanol counterpart.  

The many HOPS sources further than 100 arcsec from a methanol site will be a fruitful field of investigation when high resolution HOPS follow-up data is available and the water maser follow-up of MMB masers is fully complete.  

\subsection{Luminosities}

Work by \cite{szymczak05} found no statistically significant difference between the luminosities of 6.7-GHz methanol masers with and without associated water masers. Comparing the 6.7-GHz methanol maser luminosities may be a useful probe of the relative evolutionary timeline of objects with and without water masers as \cite{breen10ev,breen11b} found that methanol masers increase in luminosity as they evolve. 

We compared the mean and median integrated luminosities of the methanol masers with and without associated water masers in our sample (given in Table \ref{table:lummeans}). These luminosities were calculated using the \cite{green11} H{\sc i} self-absorption resolved kinematic distances where available, and the remainder with near distances from the \cite{reid09} rotation curve with the same solar motion parameters as \cite{green11}. Using a t-test we find no statistically significant difference between the 6.7-GHz  methanol maser integrated luminosities between those with and without water (p-value of 0.07). Note that when we exclude the extremely luminous source G09.621+0.196 of $\sim$ 60000~Jy~km~s$^{-1}$~kpc$^2$ the p-value becomes 0.03 because of the reduction of variance in the population that have associated water masers. This suggests that perhaps 6.7-GHz methanol masers with associated water masers generally have greater luminosities. The next paper in this catalogue series will explore this further with a larger sample size.

 \cite{breen10oh}, in their primarily OH targeted sample of water masers, investigated the flux densities of solitary water masers and those associated with other masers species and continuum emission. They found the water maser flux densities increased through the association categories of solitary, to sources associated with 6.7-GHz methanol masers, to sources associated with OH masers and 22-GHz continuum. They suggested that this may be evidence that water maser flux densities increase with age, however, they caution that some solitary water masers maybe associated with low mass stars.

\begin{figure}
\includegraphics[width=3.5in]{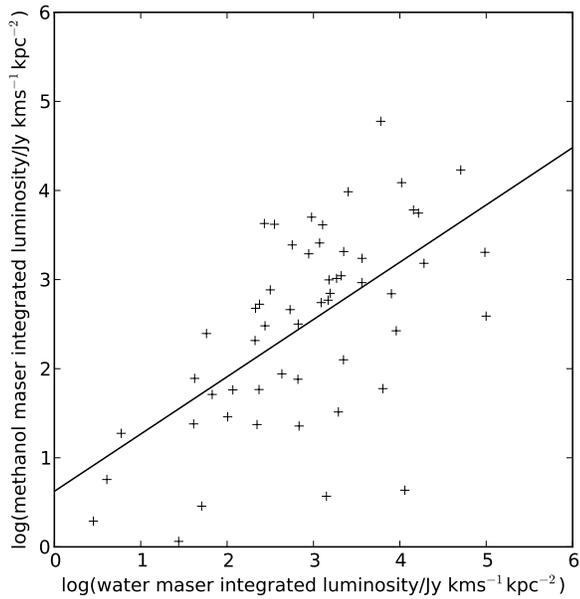}
\caption{Log 6.7-GHz methanol maser integrated luminosity vs. log water maser integrated luminosity.}
\label{fig:lum}
\end{figure}

\begin{figure}
\includegraphics[width=3.5in]{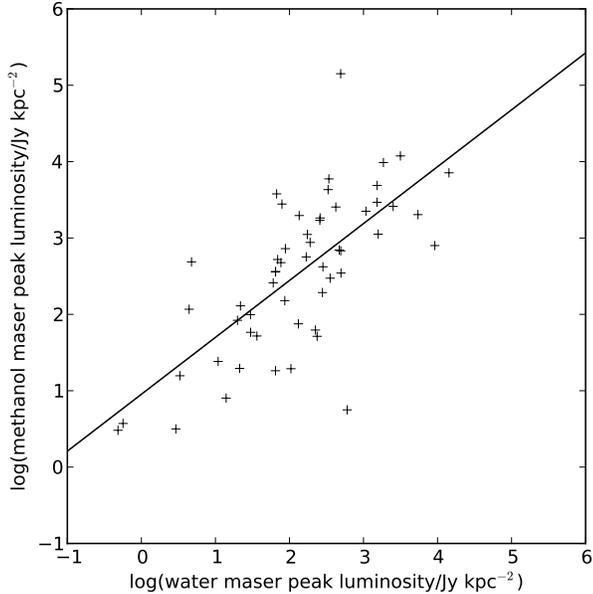}
\caption{Log 6.7-GHz methanol maser peak luminosity vs. log water maser peak luminosity.}
\label{fig:peaklum}
\end{figure}

If water masers do increase in luminosity with age, then since there is evidence that the methanol maser luminosities increase with age \citep{breen10ev}, we might expect the water and methanol maser luminosities to be correlated. Figures \ref{fig:lum} and \ref{fig:peaklum} show the water maser versus the methanol maser integrated and peak luminosities respectively. We find the integrated luminosity shows a weak correlation (Pearson's correlation coefficient of 0.59) and a linear least squares fit to this data gives a statistically significant slope of 0.64 (p-value of 1.53e-06). The peak luminosities show a tighter correlation (Pearson's correlation coefficient of 0.70) and a linear least squares fit gives a statistically significant slope of 0.74 (p-value of 2.42e-09). We note that the peak and integrated flux densities of water and methanol masers are also correlated, though the correlations are better for luminosities. The luminosities should not produce a spurious correlation as the distances come from three different sources (astrometric distances, kinematic distances from H{\sc i} self-absorption and kinematic distances from the methanol masers).

\begin{figure}
\begin{center}
 \includegraphics[width=3.0in]{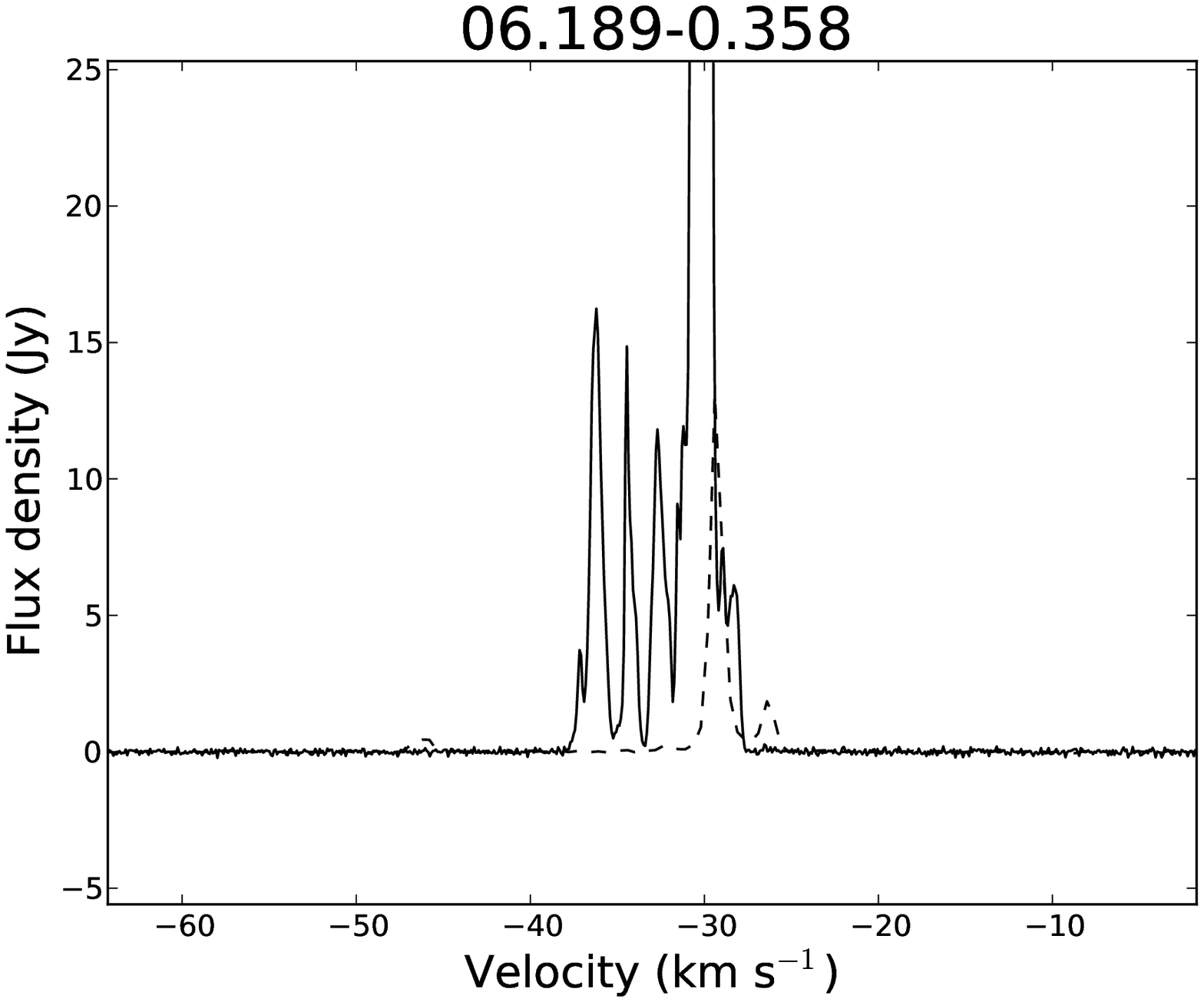} 
 \includegraphics[width=3.0in]{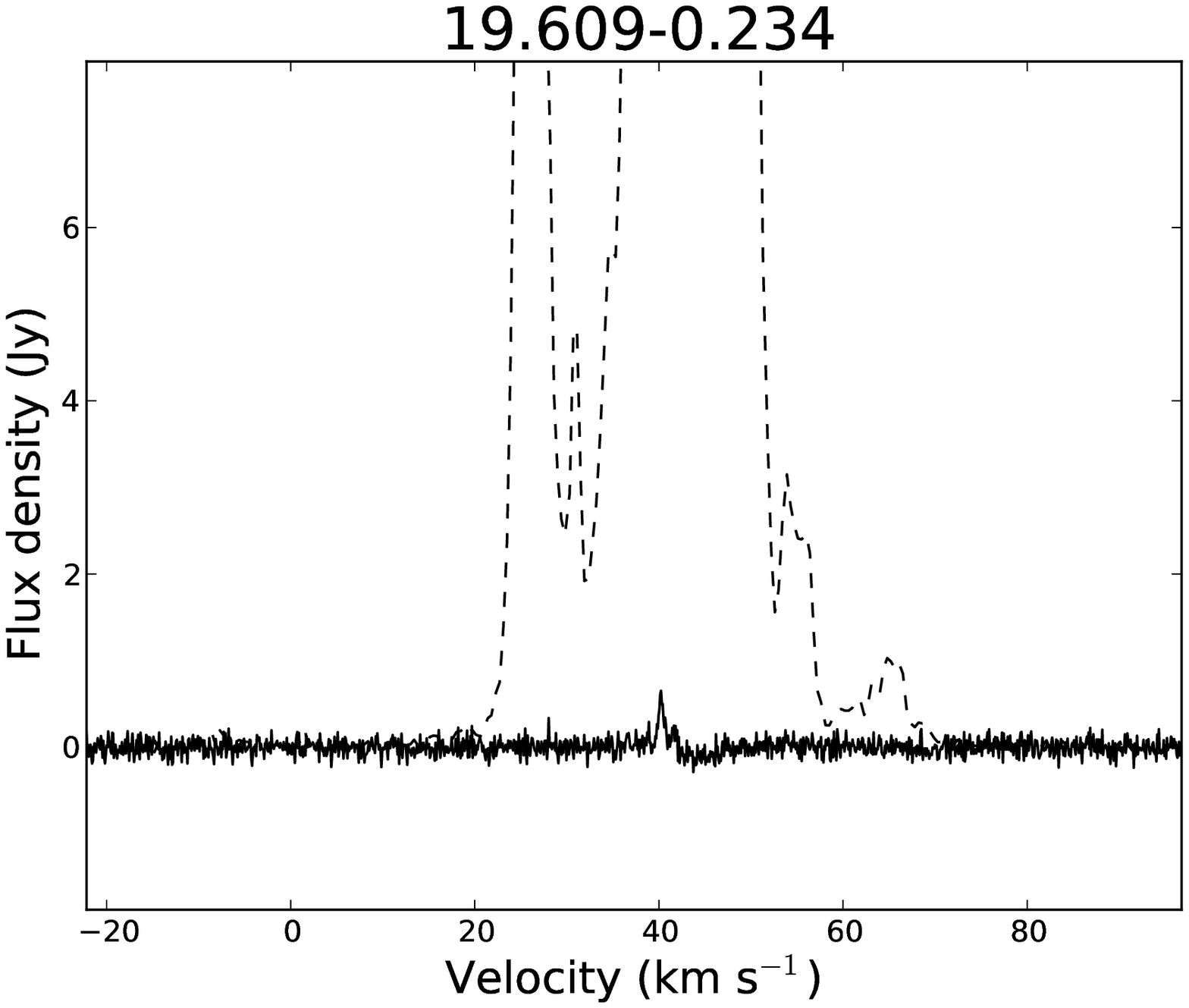}
 \caption{Spectra of associated 22-GHz water masers taken with the ATCA (shown with dashed lines) and 6.7-GHz methanol masers from the MMB survey (shown with solid lines).}
   \label{egmasers}
\end{center}
\end{figure}

This is a general trend and some individual maser sources do not show this correlation between the luminosities of the two maser species. The sources in this sample with the most extreme differences between the 6.7-GHz methanol and 22-GHz water maser intensities are shown in Figure~\ref{egmasers}. The source G06.189-0.358 has 6.7-GHz methanol maser emission with a peak flux density of 221.6~Jy, while the water maser emission has a peak of only 13.1~Jy. In contrast G19.609-0.234 has a water maser peak flux density of 69.7~Jy associated with a weak methanol maser with a peak of 1~Jy.

 \begin{table}
\caption{Luminosities of the 6.7-GHz methanol masers with and without associated water masers and including/excluding the highly luminous source G09.621+0.196 of $\sim$~60000~Jy~km~s$^{-1}$~kpc$^2$.}
\begin{tabular}{l l l r}
\hline
 & Mean int. & Median int. & Std. dev. \\
 & luminosity & luminosity & \\
 & \multicolumn{2}{c}{(Jy km s$^{-1}$ kpc$^2$)} &  \\
 \hline
 Methanol with water & 2791. & 476. & 14667. \\
 Methanol with water & 1736. & 468. & 3366. \\
 (without 09.621+0.196) & & & \\
 Methanol only & 908. & 476. & 631. \\
 \hline
\end{tabular}
\label{table:lummeans}
\end{table}

\subsection{Velocities of water maser emission}

Water masers are well known for having velocity features offset from the systemic velocity of the region. The most extreme known water maser in a high-mass star forming region is G10.472+0.027 and has features redshifted up to 250~km~s$^{-1}$ from the systemic velocity of the region and covers a total velocity range of nearly 300~km~s$^{-1}$ \citep{titmarsh13}. Methanol masers at 6.7-GHz typically show emission over a much smaller range of velocities \citep[usually less than 16~km~s$^{-1}$;][]{caswell09} and the central velocities are typically within $\pm$ 3 km~s$^{-1}$ of the systemic velocity of the region \citep[]{szymczak07,caswell09,pandian09,green11}. In the investigations below we have used the velocities at the peak flux densities rather than the central velocities, but this will make negligible difference to our results. 

\begin{figure}
\includegraphics[width=3.5in]{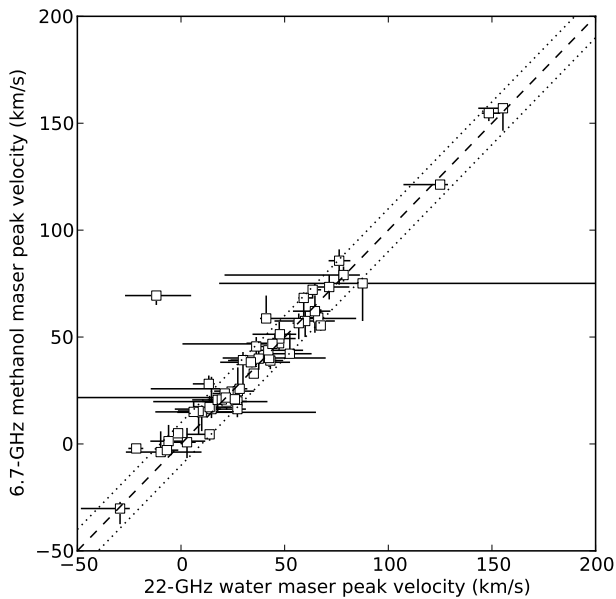}
\caption{Water maser peak velocities vs. associated 6.7-GHz methanol maser peak velocities are shown with squares. The horizontal and vertical bars represent the total velocity ranges of the water and methanol masers respectively. Also plotted is a dashed line with a slope of 1 and two dotted lines showing a deviation of $\pm$10~km~s$^{-1}$ from the dashed line.}
\label{fig:velocities}
\end{figure}

Figure \ref{fig:velocities} shows the velocities of the peak emission of the methanol versus the water masers. Our sample shows water and methanol velocities lie within 10~km~s$^{-1}$ of each other for 46 of our 55 detections i.e. 84$\pm$ 13 per cent, taking the uncertainty as $\sqrt{N_{samp}}$ (the notable outlier in Figure \ref{fig:velocities} is 18.999-0.239 which is discussed in Section \ref{sec:indiv}). The slightly smaller fraction, 78 per cent, that was found by \cite{breen10oh} in their similar study, is consistent with our result to within the uncertainties. We note that the \cite{breen10oh} sample is biased more towards sources with OH masers. \cite{szymczak05} found an even smaller fraction, 61 per cent, displaying velocities differences less than 10~km~s$^{-1}$. This might be because the \citeauthor{szymczak05} water masers were observed with a larger beam size, and so not all the water masers may be associated with the same powering object as the methanol masers.

\begin{figure}
\includegraphics[width=3.5in]{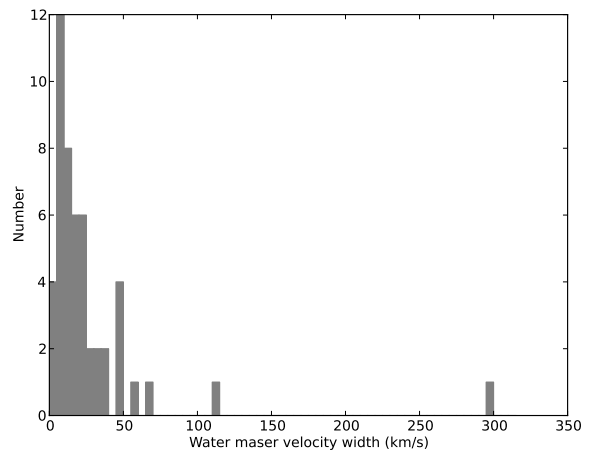}
\caption{Velocity ranges of the water masers associated with a methanol maser.}
\label{fig:velwidths}
\end{figure}

In Figure \ref{fig:velwidths} we show the distribution of the water maser velocity ranges in our survey. The total velocity range is typically less than $\sim$~50~km~s$^{-1}$, the median velocity range is 17~km~s$^{-1}$ and the average is 27~km~s$^{-1}$. This is close to previous work by \cite{breen10oh} who found a median velocity range of 15~km~s$^{-1}$ in their sample of 379 masers.  \cite{szymczak05} found that 33 per cent of their water masers had velocity ranges greater than 20 km~s$^{-1}$ and 22 per cent greater than 40 km~s$^{-1}$. We find 38 and 15 per cent $\pm$ 13 per cent greater than 20 km~s$^{-1}$ and 40 km~s$^{-1}$ respectively which is consistent within the uncertainties of the \citeauthor{szymczak05} study.

Comparing our velocities with the \cite{caswellbreen10} sample of 32 water masers from their unbiased survey, we find the water masers in our sample typically have smaller velocity ranges. \cite{caswellbreen10} looked at the fraction of water masers with emission spread greater than 30~km~s$^{-1}$ from the systemic velocity. They found that 34 per cent of water masers associated with both methanol and OH masers displayed emission over such a wide range, and 31 per cent for water masers with only methanol masers associated. In contrast, in our sample we found only 11 per cent of the water masers had velocity ranges greater than 30~km~s$^{-1}$ from the systemic velocity. 

Since high velocity emission in water masers is an indicator of outflow activity in the powering source, we wish to see if water maser velocity ranges increase with age. \cite{breen10ev} found evidence that 6.7-GHz methanol masers increase in luminosity with age, so we have compared the methanol maser luminosity to the water maser velocity range. If more outflow activity occurs with age, then the water maser total velocity range may be correlated with the methanol maser luminosity. However, no correlation was found, so we can not draw any conclusions about different evolutionary phases playing any role in the difference between the sample of \cite{caswellbreen10} and ours.

\subsection{Water maser variability}
\label{sec:var}

It is common for water masers to show variability in their flux densities, with studies such as the Arcetri project \citep{brand03,felli07} covering a wide range of timescales and luminosities of the powering YSO.Ê The Arcetri project continued for several decades, but with a fairly small sample size (43 in \cite{felli07}, and a subset of 14, with more detailed analysis, in \cite{brand03}).

\cite{breen10oh} observed a larger sample of 253 water masers over two epochs (in 2003 and 2004), 17 per cent of which were only detected on one epoch. Those detected only once tended to have simple spectra with only a few velocity features and one-third of these had peak flux densities less than 2 Jy at the epoch they were detected. About a quarter of these were not associated with other maser emission. They also found no statistically significant difference between the proportions of variable water masers associated with methanol and OH masers than from their entire sample. 

\begin{figure}
\includegraphics[width=3.5in]{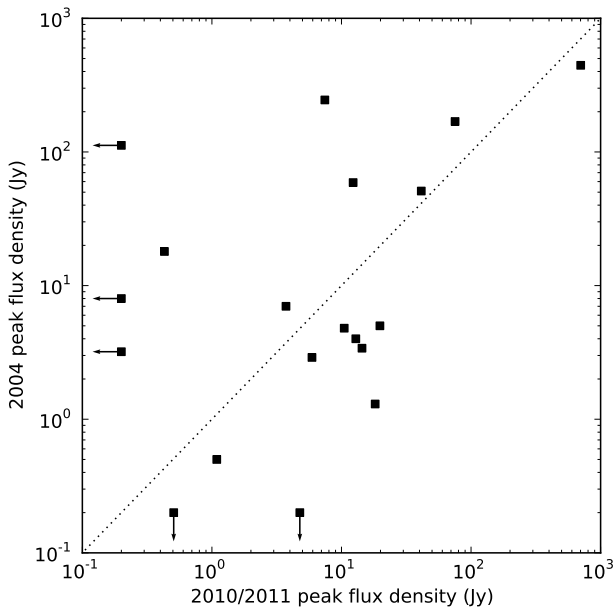}
\caption{Comparison of peak flux densities of the water masers that were observed in both the MMB follow-up (observed in 2010/2011) and the Breen et al., (2010a) sample. Since many of the Breen et al., (2010a) water masers were observed over a few epochs the flux densities used here are from only the 2004 epoch. Where masers were detected in one epoch and not the other, 3 $\sigma$ upper limits on the flux densities are shown with arrows. Also plotted is a dotted line with a slope of 1.}
\label{fig:variability}
\end{figure}

Some of our sample (observed in 2010 and 2011) overlap with the water masers observed by \citet{breen10oh} in 2004, with many having very different spectra. Figure \ref{fig:variability} compares (on a log-log scale) the peak flux densities of the water masers from the current sample which were also observed in 2004 by \cite{breen10oh}. All of them show some variation in peak flux density, although some of these may be due different spectral features changing their relative intensities causing features at different velocities to be the peak. There was no overall trend for the masers to increase or decrease in peak flux density with time, with nine getting brighter and nine becoming weaker. Also, masers with both simple and complex spectra were seen to vary. The most extreme case of variability is G11.497--1.485. This maser was 112~Jy in the 2004 observations by \cite{breen10oh} and not detected (5$\sigma$ detection limit of 75 mJy) when we observed it in 2010. 

Figure~\ref{fig:variability} also allows us to make an assessment of the possible effect of variability on the completeness of our sample. The observations of \citet{breen10oh} have comparable sensitivity to our observations. There are 19 sources which were observed in both the 2004 observations of \cite{breen10oh} and in the current observations and of these, 5 (25 per cent) were detectable in only one of the two epochs. (We have used the 2004 data from \cite{breen10oh} for comparison as eight of the overlapping masers were not observed in their 2003 observations. Section \ref{sec:indiv} gives details on the overlapping maser sample.) Although the sample size is small, this suggests that a complete sample of 6.7-GHz methanol masers observed six to seven years apart will differ by at least 10 per cent of the population due to variability.

\begin{figure}
\includegraphics[width=3.5in]{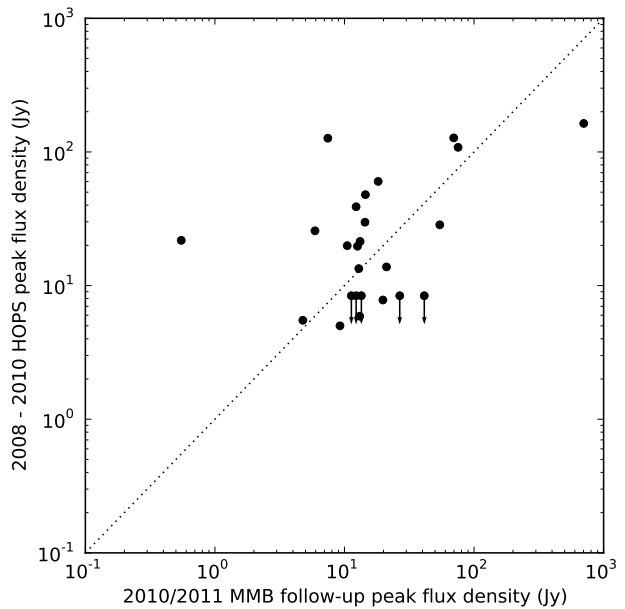}
\caption{Peak flux densities of water masers detected in both the MMB follow-up (observed in 2010/2011) and HOPS (observed between 2008 - 2010). Sources detected in our survey but not HOPS are shown with arrows indicating an upper limit for the HOPS non-detection. The dotted line is a slope of 1.}
\label{fig:hops_var}
\end{figure}

Another survey useful to study variability in our sample is HOPS, an unbiased but less sensitive search for water masers. HOPS is estimated to be 98\% complete down to 8.4 Jy. Figure \ref{fig:hops_var} shows the masers detected in both our survey and HOPS. Also included are masers detected in our survey that had peak flux densities greater than 8.4 Jy that were undetected in HOPS (these are shown with arrows indicating the upper limits of their detection). There were no water masers associated with methanol masers in HOPS that went undetected in our survey. We had more masers in common between our sample and the HOPS survey than with the \cite{breen10oh} survey, since the latter did not cover the whole longitude range, and was conducted 6-7 years earlier (compared to 1-3 years earlier for HOPS). Furthermore, the HOPS survey was limited to stronger masers which are thought to be less variable than weaker masers.

\subsection{GLIMPSE associations}

The \textit{Spitzer} Galactic Legacy Infrared Midplane Survey Extraordinaire \citep[GLIMPSE]{benjamin03} images have been used to examine the mid-infrared environments of the regions where the masers occur (images used are available online). 115 of the 119 6.7-GHz methanol masers in the $l=6^{\circ} - 20^{\circ}$ range are within the range covered by the GLIMPSE I and II data sets. 

\cite{cyganowski08} have identified many EGOs from the GLIMPSE ~I survey region and \citet{chen13a} have recently identified 98 new EGO sources from the GLIMPSE~II region. \cite{chen13a} investigated both the infrared and 3-mm molecular line properties of these sources and found them to be consistent with the EGOs being associated with high-mass protostellar objects undergoing active accretion, but at a stage prior to the formation of a UCHII region \citep{chen13a,chen13b}. An earlier study has explored associations of EGOs with solitary water masers (no methanol or OH), and water masers accompanied by methanol and/or OH \citep{breen10oh}.  We are now able to expand these results with the statistics of association with methanol masers without water masers.

We first assemble an extensive sample of EGOs for this comparison. The \cite{cyganowski08} sample overlaps with our survey in the $l=10^{\circ} - 20^{\circ}$ range and only two of our masers have an EGO match within 30~arcseconds of any EGO in their catalogue (12.904-0.031 and 19.009-0.029). 30~arcseconds was chosen because it is approximately the largest angular extent of the EGOs in that sample and visual inspection of three colour images of these maser sites confirms that they are associated with the \cite{cyganowski08} EGOs. Classification of EGOs is a somewhat subjective process and other authors have defined their own samples of EGOs \citep[e.g.][]{chen13a,chambers09}. The sample that has the most matches with ours is \cite{chen13a} which has seven matches with our masers in the $l=6^{\circ} - 10^{\circ}$ range. So for this range we used the \cite{chen13a} sample and for the rest we decided to search our sample for EGO candidates rather than match with an existing catalogue. Three colour images of each maser site were made using the standard three colour scheme, 3.6~$\mu$m - blue, 4.5~$\mu$m - green, 8.0~$\mu$m - red and visual inspection of the three colour images reveals several other sources that are EGO candidates (see Tables \ref{table:assoc_sources} and \ref{table:notass_sources}). There are more EGOs associated with both water and methanol masers than there are with methanol masers only regions (the difference is statistically significant using a Pearson's chi-squared test, p-value of 0.03, see Table \ref{table:glimpse}). This may be because water masers are more dependent on the physical conditions of the surrounding environment interacting with the protostar (e.g. outflows interacting with gas the protostar is embedded in) than class II methanol masers. It could also be that since EGOs may be preferentially associated with the later stages of the methanol maser phase, that the water masers switch on later than the methanol masers. \cite{breen10oh} studied associations of EGOs with solitary water masers, water with associated methanol, water with OH and water with both methanol and OH masers. They found water masers with associated methanol and OH masers were most likely to have an EGO present. They suggest that EGOs occur late enough in the evolution of a protostar for an OH maser to have formed, but not so late that it has developed an UCH{\sc ii} region causing the methanol maser emission to cease.

Another indicator of the early stages of star formation that can be found in GLIMPSE images are Infrared Dark Clouds (IRDCs). IRDCs are regions of cold and dense gas and dust which are seen in absorption against the diffuse mid-infrared background from PAH emission, particularly in the 8~$\mu$m band \citep[e.g.][]{rathborne2006,pillai2006}. Although IRDCs do not trace a specific evolutionary phase of the high-mass star formation process, it is well established that very young protostellar objects, and perhaps even some prestellar cores are embedded within some IRDCs \citep[e.g.][]{ellingsen06}. 

The IRDC catalogue of \cite{peretto09} covers our $l=10^{\circ} - 20^{\circ}$ range and cross matching with them  gives 19 matches over our whole sample within 1~arcminute and no matches within 3~arcseconds. Thus we decided to inspect our infrared images for IRDCs (see Table \ref{table:glimpse}). The classification of objects associated with IRDCs is also somewhat subjective, as IRDCs can only be observed where there is sufficient background diffuse emission and not too much diffuse foreground emission between us and the cold dense gas. The sources we have classified as being associated with an IRDC include both sources which are infrared dark at 8~$\mu$m, as well as those associated with infrared bright objects which are surrounded by some infrared darkening. We found no statistically significant difference in the number of IRDCs associated with sources with both methanol and water masers and those with only methanol masers. 

\cite{gallaway13} investigated at the GLIMPSE images of the MMB sources in the longitudes $186^{\circ} \leq l \leq 20^{\circ}$ and found 26 per cent were associated with IRDCs and 12 per cent were IR-dark and do not have an IRDC. In this study we only found the number of IRDCs to be consistent with their study, however, we found only one source to be IR-dark ($<$ 1 per cent).

\begin{table}
\caption{Numbers of Extended Green Objects and Infrared Dark Clouds found in GLIMPSE associated with 6.7-GHz methanol masers with and without associated water masers. (There are 63 methanol only sources and 52 with water as well.)}
\begin{tabular}{l c c}
\hline
 & Methanol & Methanol \\
 & only sources & with water  sources \\
 \hline
EGO & 5 & 10 \\
IRDC & 11 & 10 \\
\hline
\end{tabular}
\label{table:glimpse}
\end{table}

\subsection{Associations with 1.1-mm emission from dust clumps}

The Bolocam Galactic Plane Survey \citep[BGPS;][]{rosolowsky10} used the Bolocam instrument on the Caltech Submillimeter Observatory to undertake a continuum survey of the Milky Way at 1.1-mm. This wavelength is a good tracer of thermal dust emission from the coldest, densest gas and dust cores, the locations believed to be the sites of the earliest stages of high-mass star formation. We have used the second data release of the BGPS \citep{ginsburg13} to compare with our maser sample. The BGPS covered the longitude range ($6^{\circ} - 20^{\circ}$) of our water maser observations with a latitude coverage of $\pm1.5^{\circ}$ (which covers the range of the vast majority of 6.7-GHz methanol maser detections, all but three are in this longitude range). 92 of the 116 6.7-GHz methanol masers in this region had a BGPS counterpart within 33~arcseconds (the effective resolution of the BGPS), and of these 49 had a water maser also. The resolution of the BGPS is much coarser than the size of a star forming core associated with a maser which means that the BGPS flux densities will potentially include emission from many other surrounding sources from the clustered environments where high mass stars are formed. 

Previous work by \cite{chen12} used BGPS dust clumps as targets for a search for class I methanol masers. They compared the BGPS flux densities, gas masses and beam averaged column densities of the clumps with and without masers associated. They found that BGPS sources with a class I methanol maser had higher BGPS flux densities and beam averaged column densities than those without a maser. Details of how the gas mass and beam averaged column density were derived are given in equations 1 and 2 of \cite{chen12}. 

\begin{figure}
\includegraphics[width=3.5in]{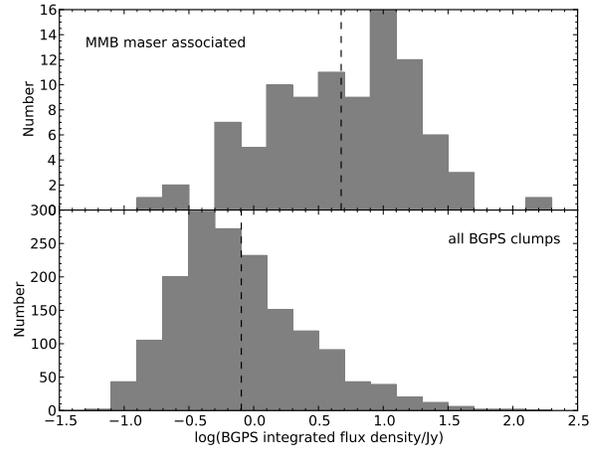}
\caption{Number of sources as a function of BGPS integrated flux density.  The top panel shows the dust clumps with associated 6.7-GHz methanol masers and the bottom panel are all the BGPS dust clumps in the $l$ = 6 -- 20$^{\circ}$ region. The dashed lines represent the means.}
\label{fig:hist_sint}
\end{figure}

\begin{figure}
\includegraphics[width=3.5in]{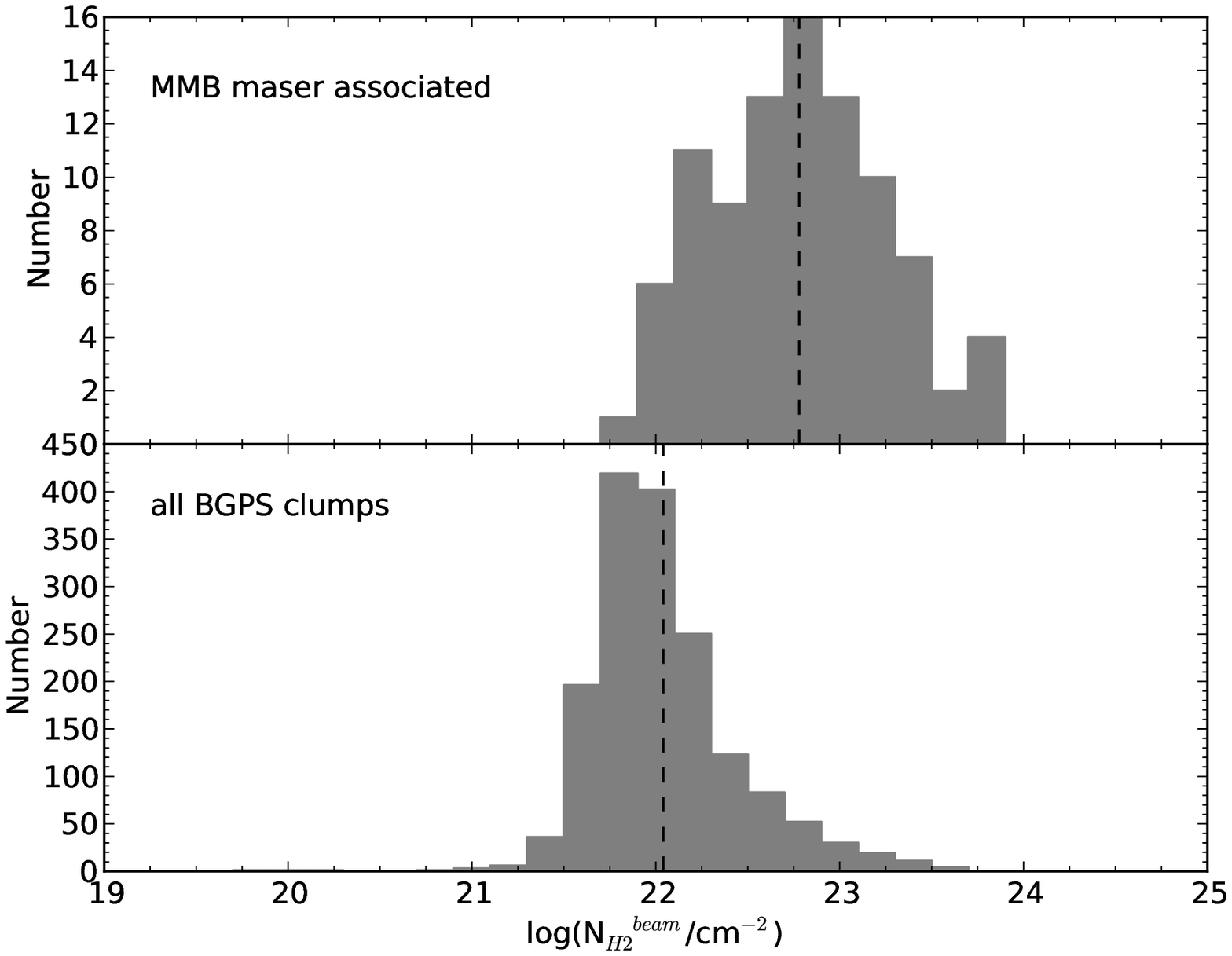}
\caption{Number of sources as a function of beam averaged column density.  The top panel shows the dust clumps with associated 6.7-GHz methanol masers and the bottom panel are all the BGPS dust clumps in the $l$ = 6 -- 20$^{\circ}$ region. The dashed lines represent the means.}
\label{fig:hist_nh2}
\end{figure}

Similar to \cite{chen12} we found BGPS sources with either class II methanol or water masers had higher BGPS integrated flux densities and beam averaged column densities than the general population (see Figures \ref{fig:hist_sint} and \ref{fig:hist_nh2}). The beam averaged column density is only dependent on the 40 arcsecond flux density as in our calculations we assumed a temperature of 20~K for all sources. No comparison of the gas masses was made due to the lack of distance estimates for the sources without masers. No statistically significant differences in integrated flux density and beam-averaged column density were found between BGPS sources with both and methanol and water masers and the methanol only sources.

\begin{figure}
\includegraphics[width=3.5in]{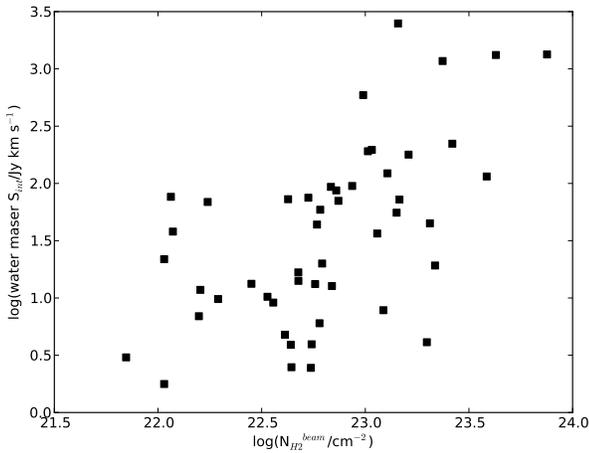}
\caption{Water maser integrated flux density vs. beam averaged column density.}
\label{fig:bol_nh2}
\end{figure}

\cite{chen12} found that intensity of the class I methanol maser emission is correlated with the mass and beam averaged column densities of the BGPS sources. Similar to \cite{chen12}, we found a correlation between the water maser integrated flux density and the beam averaged column density (see Figure \ref{fig:bol_nh2}). However, no correlations were apparent when doing similar comparisons with the 6.7-GHz methanol maser sample. It is not surprising that these water masers show similar correlations to the class I methanol masers as they are also collisionally pumped. Sources with higher integrated column densities tend to be more massive and so this trend is likely a result of there being a larger volume of gas where the conditions are conducive to masing in these sources.

\begin{figure}
\includegraphics[width=3.5in]{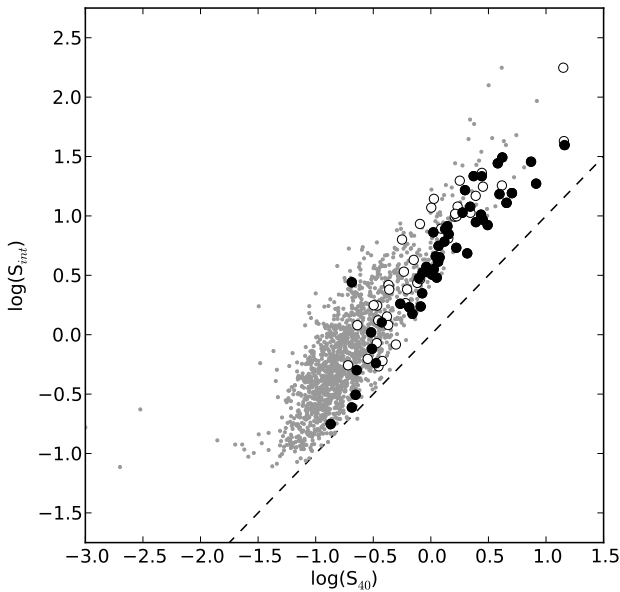}
\caption{BGPS integrated flux density vs. BGPS 40 arcsecond flux density.  All the BGPS dust clumps in the $l$ = 6 -- 20$^{\circ}$ region are plotted with grey dots, clumps with only 6.7-GHz methanol masers associated are open circles and clumps with both water and methanol masers associated are black dots. The dashed line has a slope of 1.}
\label{fig:sint_vs_s40}
\end{figure}

\begin{figure}
\includegraphics[width=3.5in]{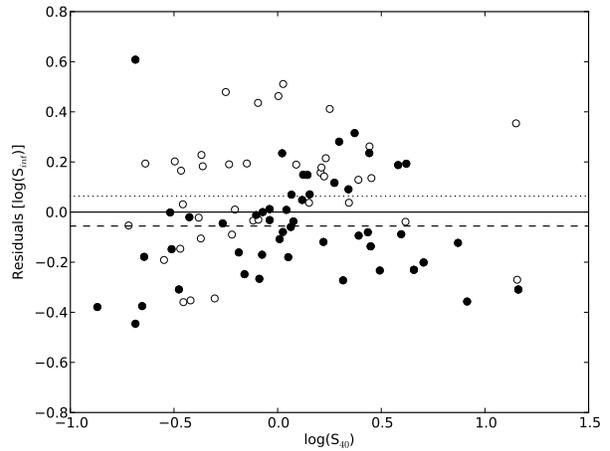}
\caption{Residuals from a linear fit to all the BGPS clumps with an associated methanol maser in the previous figure. Clumps with 6.7-GHz methanol masers associated are open circles and clumps which also have water associated are black dots. The fit to all the clumps with masers is the solid line, the mean of the residuals for methanol-only clumps is the top dashed line and the mean of the water and methanol associated residuals is the lower dashed line.}
\label{fig:res}
\end{figure}

In Figure \ref{fig:sint_vs_s40} we have plotted the BGPS integrated flux densities against the BGPS 40 arcsecond flux densities. BGPS sources with associated masers are typically the brighter sources, implying that they are higher mass clumps. BGPS clumps with masers associated tend to be the most compact sources having a greater fraction of their integrated flux densities within the 40 arcsecond beam than the majority of the non-maser BGPS sources. It also appears that clumps associated with both water and methanol masers are more compact than those associated with methanol only, although there is a large degree of overlap in these two samples. We have done a linear least squares fit to all the dust clumps in Figure \ref{fig:sint_vs_s40} that have any maser emission and the differences in the residuals between the maser association categories are shown in Figure \ref{fig:res}. The mean of the residuals of the clumps that have only methanol masers associated is greater than that of the clumps with both water and methanol masers. This difference is statistically significant; a t-test gives a p-value of 0.01. Dust is accreted onto the protostar as it evolves which implies that the more compact dust clumps may be older. Hence, the dust clumps with both water and methanol masers may be older than the methanol only clumps. This suggests that both class II methanol and water maser emission occurs during the phase of high-mass star formation where large-scale infall is still in progress, consistent with recent observations of molecular gas \citep{peretto13}.

While protostars with both 6.7-GHz methanol and water masers may generally be older than those with just 6.7-GHz methanol masers, we suggest that the evolutionary phase traced by water masers is less well-defined than that traced by 6.7-GHz methanol. Our findings support the argument of \cite{breen14} that class II methanol maser transitions are better tied to the evolutionary phase of the protostar as they are radiatively pumped, existing close to the protostar and so are more closely linked to the protostar's properties. In contrast, water masers, being collisionally pumped and occurring at the interaction of outflows and the surrounding environment could exist over a longer time scale and a wider variety of conditions and so are not well tied to a specific evolutionary phase of the protostar.

\section{Conclusions}

We have performed targeted observations for water masers towards all sources within the complete 6.7-GHz methanol maser sample from the MMB in the longitude range $6^{\circ}$ to $20^{\circ}$. We found 55 water masers associated within $\sim$3~arcseconds of their 6.7-GHz methanol maser targets ($\sim$~46 per cent). This is consistent with previous studies when we consider the angular resolution and flux density limitations of those studies. All the 6.7-GHz methanol masers from the MMB have also been observed for water masers in the longitude range $310^{\circ}$ through $0^{\circ}$ to $6^{\circ}$, with results to be reported in subsequent papers.

We have found that 6.7-GHz methanol masers with associated water masers generally have greater integrated luminosities. Since 6.7-GHz methanol masers get brighter with age this may be evidence that water masers are associated with older methanol masers.

The peak velocities of the water and methanol masers are very well correlated in this sample, with 84 per cent having peak velocities within $\pm$ 10 km~s$^{-1}$. We found the median velocity range of our water masers is 17 km~s$^{-1}$ and the average to be 27 km~s$^{-1}$. This is consistent with other methanol-selected samples of water masers, but is smaller than that found in an unbiased survey of water masers by \cite{caswell10}.

From the GLIMPSE mid-infrared images, we found that there was no statistically significant difference in the number of IRDCs associated with sources that had both water and methanol masers and the methanol-only sources. But we did find that there are more EGOs associated with water and methanol maser sources than with the methanol-only sources. 

The 1.1-mm data from the BGPS was used to study the properties of the dust emission at these maser sites. We found that dust clumps with masers associated had higher BGPS integrated flux densities than the general population of dust clumps in our survey region. The maser-associated clumps also had a greater fraction of their flux densities within the 40 arcsecond beam of BGPS, implying that they are the more compact sources. The sources with both water and methanol masers are generally more compact than those with only methanol masers associated. If these dust clumps collapse to form a protostar as they age, then this suggests that 6.7-GHz methanol masers with water masers associated are older than those without water masers.

Like \cite{breen14}, we conclude that water masers are unlikely to trace such a well defined evolutionary phase in the formation of a protostar as the 6.7-GHz methanol maser transition, which are radiatively pumped and exist close to the protostar and hence are very dependant on its evolutionary phase. Water masers are collisionally pumped, occurring at locations where the protostar is interacting with the surrounding environment (e.g. outflows) which may not exist alongside a specific evolutionary stage of the protostar. Since the methanol maser luminosities, associations with EGOs and 1.1-mm dust clump data suggest that water masers often occur later in the 6.7-GHz methanol maser phase, it is likely that the interactions required to produce a water maser increase as the protostar evolves, and therefore the presence of water maser emission alone does not provide a good evolutionary diagnostic. However, when water masers are compared with other diagnostics like 6.7-GHz methanol masers with positional accuracy of a few arcseconds, then this can help constrain the evolutionary state of the YSO.

\section*{Acknowledgments}

The Australia Telescope is funded by the Commonwealth of Australia for operation as a National Facility managed by CSIRO. This research has made use of NASA's Astrophysics Data System Abstract Service, the NASA/IPAC Infrared Science Archive (which is operated by the Jet Propulsion Laboratory, California Institute of Technology, under contract with the National Aeronautics and Space Administration) and data products from the GLIMPSE survey, which is a legacy science programme of the \textit{Spitzer Space Telescope}, funded by the National Aeronautics and Space Administration, and information from the BGPS survey database at http://irsa.ipac.caltech.edu/data/BOLOCAM\_GPS/. Shari Breen is the recipientÊof an Australian Research Council DECRA FellowshipÊ(project number DE130101270).

\bibliography{bib_paper1}

\begin{thebibliography}{65}
\expandafter\ifx\csname natexlab\endcsname\relax\def\natexlab#1{#1}\fi

\bibitem[{{Bartkiewicz} {et~al}\mbox{.}(2011){Bartkiewicz}, {Szymczak},
  {Pihlstr{\"o}m}, {van Langevelde}, {Brunthaler}, \& {Reid}}]{bartkiewicz11}
{Bartkiewicz} A., {Szymczak} M., {Pihlstr{\"o}m} Y.~M., {van Langevelde} H.~J.,
  {Brunthaler} A., {Reid} M.~J., 2011, \aap, 525, A120

\bibitem[{{Batrla} {et~al}\mbox{.}(1987){Batrla}, {Matthews}, {Menten}, \&
  {Walmsley}}]{batrla87}
{Batrla} W., {Matthews} H.~E., {Menten} K.~M., {Walmsley} C.~M., 1987, \nat,
  326, 49

\bibitem[{{Benjamin} {et~al}\mbox{.}(2003){Benjamin}, {Churchwell}, {Babler},
  {Bania}, {Clemens}, {Cohen}, {Dickey}, {Indebetouw}, {Jackson}, {Kobulnicky},
  {Lazarian}, {Marston}, {Mathis}, {Meade}, {Seager}, {Stolovy}, {Watson},
  {Whitney}, {Wolff}, \& {Wolfire}}]{benjamin03}
{Benjamin} R.~A. {et~al.}, 2003, \pasp, 115, 953

\bibitem[{{Beuther} {et~al}\mbox{.}(2002){Beuther}, {Walsh}, {Schilke},
  {Sridharan}, {Menten}, \& {Wyrowski}}]{beuther02}
{Beuther} H., {Walsh} A., {Schilke} P., {Sridharan} T.~K., {Menten} K.~M.,
  {Wyrowski} F., 2002, \aap, 390, 289

\bibitem[{{Brand} {et~al}\mbox{.}(1994){Brand}, {Cesaroni}, {Caselli},
  {Catarzi}, {Codella}, {Comoretto}, {Curioni}, {Curioni}, {Di Franco},
  {Felli}, {Giovanardi}, {Olmi}, {Palagi}, {Palla}, {Panella}, {Pareschi},
  {Rossi}, {Speroni}, \& {Tofani}}]{brand94}
{Brand} J. {et~al.}, 1994, \aaps, 103, 541

\bibitem[{{Brand} {et~al}\mbox{.}(2003){Brand}, {Cesaroni}, {Comoretto},
  {Felli}, {Palagi}, {Palla}, \& {Valdettaro}}]{brand03}
{Brand} J., {Cesaroni} R., {Comoretto} G., {Felli} M., {Palagi} F., {Palla} F.,
  {Valdettaro} R., 2003, \aap, 407, 573

\bibitem[{{Breen} {et~al}\mbox{.}(2010{\natexlab{a}}){Breen}, {Caswell},
  {Ellingsen}, \& {Phillips}}]{breen10oh}
{Breen} S.~L., {Caswell} J.~L., {Ellingsen} S.~P., {Phillips} C.~J.,
  2010{\natexlab{a}}, \mnras, 406, 1487

\bibitem[{{Breen} {et~al}\mbox{.}(2011){Breen}, {Ellingsen}, {Caswell},
  {Green}, {Fuller}, {Voronkov}, {Quinn}, \& {Avison}}]{breen11b}
{Breen} S.~L., {Ellingsen} S.~P., {Caswell} J.~L., {Green} J.~A., {Fuller}
  G.~A., {Voronkov} M.~A., {Quinn} L.~J., {Avison} A., 2011, \apj, 733, 80

\bibitem[{{Breen} {et~al}\mbox{.}(2014){Breen}, {Ellingsen}, {Caswell},
  {Green}, {Voronkov}, {Avison}, {Fuller}, {Quinn}, \& {Titmarsh}}]{breen14}
{Breen} S.~L. {et~al.}, 2014, \mnras, 438, 3368

\bibitem[{{Breen} {et~al}\mbox{.}(2012){Breen}, {Ellingsen}, {Caswell},
  {Green}, {Voronkov}, {Fuller}, {Quinn}, \& {Avison}}]{breen12}
{Breen} S.~L., {Ellingsen} S.~P., {Caswell} J.~L., {Green} J.~A., {Voronkov}
  M.~A., {Fuller} G.~A., {Quinn} L.~J., {Avison} A., 2012, \mnras, 421, 1703

\bibitem[{{Breen} {et~al}\mbox{.}(2010{\natexlab{b}}){Breen}, {Ellingsen},
  {Caswell}, \& {Lewis}}]{breen10ev}
{Breen} S.~L., {Ellingsen} S.~P., {Caswell} J.~L., {Lewis} B.~E.,
  2010{\natexlab{b}}, \mnras, 401, 2219

\bibitem[{{Breen} {et~al}\mbox{.}(2013){Breen}, {Ellingsen}, {Contreras},
  {Green}, {Caswell}, {Stevens}, {Dawson}, \& {Voronkov}}]{breen13}
{Breen} S.~L., {Ellingsen} S.~P., {Contreras} Y., {Green} J.~A., {Caswell}
  J.~L., {Stevens} J.~B., {Dawson} J.~R., {Voronkov} M.~A., 2013, \mnras, 435,
  524

\bibitem[{{Caswell}(1997)}]{caswell97}
{Caswell} J.~L., 1997, \mnras, 289, 203

\bibitem[{{Caswell}(1998)}]{caswell98}
{Caswell} J.~L., 1998, \mnras, 297, 215

\bibitem[{{Caswell}(2009)}]{caswell09}
{Caswell} J.~L., 2009, \pasa, 26, 454

\bibitem[{{Caswell} \& {Breen}(2010)}]{caswellbreen10}
{Caswell} J.~L., {Breen} S.~L., 2010, \mnras, 407, 2599

\bibitem[{{Caswell} {et~al}\mbox{.}(2010){Caswell}, {Fuller}, {Green},
  {Avison}, {Breen}, {Brooks}, {Burton}, {Chrysostomou}, {Cox}, {Diamond},
  {Ellingsen}, {Gray}, {Hoare}, {Masheder}, {McClure-Griffiths}, {Pestalozzi},
  {Phillips}, {Quinn}, {Thompson}, {Voronkov}, {Walsh}, {Ward-Thompson},
  {Wong-McSweeney}, {Yates}, \& {Cohen}}]{caswell10}
{Caswell} J.~L. {et~al.}, 2010, \mnras, 404, 1029

\bibitem[{{Caswell} {et~al}\mbox{.}(2011){Caswell}, {Fuller}, {Green},
  {Avison}, {Breen}, {Ellingsen}, {Gray}, {Pestalozzi}, {Quinn}, {Thompson}, \&
  {Voronkov}}]{caswell11}
{Caswell} J.~L. {et~al.}, 2011, \mnras, 417, 1964

\bibitem[{{Caswell} {et~al}\mbox{.}(2013){Caswell}, {Green}, \&
  {Phillips}}]{caswell13}
{Caswell} J.~L., {Green} J.~A., {Phillips} C.~J., 2013, \mnras, 431, 1180

\bibitem[{{Caswell} \& {Phillips}(2008)}]{caswell08}
{Caswell} J.~L., {Phillips} C.~J., 2008, \mnras, 386, 1521

\bibitem[{{Chambers} {et~al}\mbox{.}(2009){Chambers}, {Jackson}, {Rathborne},
  \& {Simon}}]{chambers09}
{Chambers} E.~T., {Jackson} J.~M., {Rathborne} J.~M., {Simon} R., 2009, \apjs,
  181, 360

\bibitem[{{Chen} {et~al}\mbox{.}(2012){Chen}, {Ellingsen}, {He}, {Xu}, {Gan},
  {Shen}, {An}, {Sun}, \& {Ju}}]{chen12}
{Chen} X. {et~al.}, 2012, \apjs, 200, 5

\bibitem[{{Chen} {et~al}\mbox{.}(2013{\natexlab{a}}){Chen}, {Gan}, {Ellingsen},
  {He}, {Shen}, \& {Titmarsh}}]{chen13a}
{Chen} X., {Gan} C.-G., {Ellingsen} S.~P., {He} J.-H., {Shen} Z.-Q., {Titmarsh}
  A., 2013{\natexlab{a}}, \apjs, 206, 9

\bibitem[{{Chen} {et~al}\mbox{.}(2013{\natexlab{b}}){Chen}, {Gan}, {Ellingsen},
  {He}, {Shen}, \& {Titmarsh}}]{chen13b}
{Chen} X., {Gan} C.-G., {Ellingsen} S.~P., {He} J.-H., {Shen} Z.-Q., {Titmarsh}
  A., 2013{\natexlab{b}}, \apjs, 206, 22

\bibitem[{{Comoretto} {et~al}\mbox{.}(1990){Comoretto}, {Palagi}, {Cesaroni},
  {Felli}, {Bettarini}, {Catarzi}, {Curioni}, {Curioni}, {Di Franco},
  {Giovanardi}, {Massi}, {Palla}, {Panella}, {Rossi}, {Speroni}, \&
  {Tofani}}]{comoretto90}
{Comoretto} G. {et~al.}, 1990, \aaps, 84, 179

\bibitem[{{Cragg} {et~al}\mbox{.}(1992){Cragg}, {Johns}, {Godfrey}, \&
  {Brown}}]{cragg92}
{Cragg} D.~M., {Johns} K.~P., {Godfrey} P.~D., {Brown} R.~D., 1992, \mnras,
  259, 203

\bibitem[{{Cragg} {et~al}\mbox{.}(2002){Cragg}, {Sobolev}, \&
  {Godfrey}}]{cragg02}
{Cragg} D.~M., {Sobolev} A.~M., {Godfrey} P.~D., 2002, \mnras, 331, 521

\bibitem[{{Cragg} {et~al}\mbox{.}(2005){Cragg}, {Sobolev}, \&
  {Godfrey}}]{cragg05}
{Cragg} D.~M., {Sobolev} A.~M., {Godfrey} P.~D., 2005, \mnras, 360, 533

\bibitem[{{Cyganowski} {et~al}\mbox{.}(2008){Cyganowski}, {Whitney}, {Holden},
  {Braden}, {Brogan}, {Churchwell}, {Indebetouw}, {Watson}, {Babler},
  {Benjamin}, {Gomez}, {Meade}, {Povich}, {Robitaille}, \&
  {Watson}}]{cyganowski08}
{Cyganowski} C.~J. {et~al.}, 2008, \aj, 136, 2391

\bibitem[{{Elitzur} {et~al}\mbox{.}(1989){Elitzur}, {Hollenbach}, \&
  {McKee}}]{elitzur89}
{Elitzur} M., {Hollenbach} D.~J., {McKee} C.~F., 1989, \apj, 346, 983

\bibitem[{{Ellingsen}(2006)}]{ellingsen06}
{Ellingsen} S.~P., 2006, \apj, 638, 241

\bibitem[{{Ellingsen} {et~al}\mbox{.}(2013){Ellingsen}, {Breen}, {Voronkov}, \&
  {Dawson}}]{ellingsen13}
{Ellingsen} S.~P., {Breen} S.~L., {Voronkov} M.~A., {Dawson} J.~R., 2013,
  \mnras, 429, 3501

\bibitem[{{Ellingsen} {et~al}\mbox{.}(2007){Ellingsen}, {Voronkov}, {Cragg},
  {Sobolev}, {Breen}, \& {Godfrey}}]{ellingsen07}
{Ellingsen} S.~P., {Voronkov} M.~A., {Cragg} D.~M., {Sobolev} A.~M., {Breen}
  S.~L., {Godfrey} P.~D., 2007, in IAU Symposium, Vol. 242, IAU Symposium,
  {Chapman} J.~M., {Baan} W.~A., eds., pp. 213--217

\bibitem[{{Felli} {et~al}\mbox{.}(2007){Felli}, {Brand}, {Cesaroni}, {Codella},
  {Comoretto}, {di Franco}, {Massi}, {Moscadelli}, {Nesti}, {Olmi}, {Palagi},
  {Panella}, \& {Valdettaro}}]{felli07}
{Felli} M. {et~al.}, 2007, \aap, 476, 373

\bibitem[{{Forster} \& {Caswell}(1989)}]{forster89}
{Forster} J.~R., {Caswell} J.~L., 1989, \aap, 213, 339

\bibitem[{{Gallaway} {et~al}\mbox{.}(2013){Gallaway}, {Thompson}, {Lucas},
  {Fuller}, {Caswell}, {Green}, {Voronkov}, {Breen}, {Quinn}, {Ellingsen},
  {Avison}, {Ward-Thompson}, \& {Cox}}]{gallaway13}
{Gallaway} M. {et~al.}, 2013, \mnras, 430, 808

\bibitem[{{Ginsburg} {et~al}\mbox{.}(2013){Ginsburg}, {Glenn}, {Rosolowsky},
  {Ellsworth-Bowers}, {Battersby}, {Dunham}, {Merello}, {Shirley}, {Bally},
  {Evans}, {Stringfellow}, \& {Aguirre}}]{ginsburg13}
{Ginsburg} A. {et~al.}, 2013, \apjs, 208, 14

\bibitem[{{Green} {et~al}\mbox{.}(2009){Green}, {Caswell}, {Fuller}, {Avison},
  {Breen}, {Brooks}, {Burton}, {Chrysostomou}, {Cox}, {Diamond}, {Ellingsen},
  {Gray}, {Hoare}, {Masheder}, {McClure-Griffiths}, {Pestalozzi}, {Phillips},
  {Quinn}, {Thompson}, {Voronkov}, {Walsh}, {Ward-Thompson}, {Wong-McSweeney},
  {Yates}, \& {Cohen}}]{green09}
{Green} J.~A. {et~al.}, 2009, \mnras, 392, 783

\bibitem[{{Green} {et~al}\mbox{.}(2010){Green}, {Caswell}, {Fuller}, {Avison},
  {Breen}, {Ellingsen}, {Gray}, {Pestalozzi}, {Quinn}, {Thompson}, \&
  {Voronkov}}]{green10}
{Green} J.~A. {et~al.}, 2010, \mnras, 409, 913

\bibitem[{{Green} {et~al}\mbox{.}(2012){Green}, {Caswell}, {Fuller}, {Avison},
  {Breen}, {Ellingsen}, {Gray}, {Pestalozzi}, {Quinn}, {Thompson}, \&
  {Voronkov}}]{green12}
{Green} J.~A. {et~al.}, 2012, \mnras, 420, 3108

\bibitem[{{Green} \& {McClure-Griffiths}(2011)}]{green11}
{Green} J.~A., {McClure-Griffiths} N.~M., 2011, \mnras, 417, 2500

\bibitem[{{Menten}(1991)}]{menten91}
{Menten} K., 1991, in Astronomical Society of the Pacific Conference Series,
  Vol.~16, Atoms, Ions and Molecules: New Results in Spectral Line
  Astrophysics, {Haschick} A.~D., {Ho} P.~T.~P., eds., p. 119

\bibitem[{{Minier} {et~al}\mbox{.}(2003){Minier}, {Ellingsen}, {Norris}, \&
  {Booth}}]{minier03}
{Minier} V., {Ellingsen} S.~P., {Norris} R.~P., {Booth} R.~S., 2003, \aap, 403,
  1095

\bibitem[{{Pandian} {et~al}\mbox{.}(2009){Pandian}, {Menten}, \&
  {Goldsmith}}]{pandian09}
{Pandian} J.~D., {Menten} K.~M., {Goldsmith} P.~F., 2009, \apj, 706, 1609

\bibitem[{{Peretto} \& {Fuller}(2009)}]{peretto09}
{Peretto} N., {Fuller} G.~A., 2009, \aap, 505, 405

\bibitem[{{Peretto} {et~al}\mbox{.}(2013){Peretto}, {Fuller}, {Duarte-Cabral},
  {Avison}, {Hennebelle}, {Pineda}, {Andr{\'e}}, {Bontemps}, {Motte},
  {Schneider}, \& {Molinari}}]{peretto13}
{Peretto} N. {et~al.}, 2013, \aap, 555, A112

\bibitem[{{Phillips} {et~al}\mbox{.}(1998){Phillips}, {Norris}, {Ellingsen}, \&
  {McCulloch}}]{phillips98}
{Phillips} C.~J., {Norris} R.~P., {Ellingsen} S.~P., {McCulloch} P.~M., 1998,
  \mnras, 300, 1131

\bibitem[{{Pillai} {et~al}\mbox{.}(2006{\natexlab{a}}){Pillai}, {Wyrowski},
  {Carey}, \& {Menten}}]{pillai2006}
{Pillai} T., {Wyrowski} F., {Carey} S.~J., {Menten} K.~M., 2006{\natexlab{a}},
  \aap, 450, 569

\bibitem[{{Pillai} {et~al}\mbox{.}(2006{\natexlab{b}}){Pillai}, {Wyrowski},
  {Menten}, \& {Kr{\"u}gel}}]{pillai06b}
{Pillai} T., {Wyrowski} F., {Menten} K.~M., {Kr{\"u}gel} E.,
  2006{\natexlab{b}}, \aap, 447, 929

\bibitem[{{Rathborne} {et~al}\mbox{.}(2006){Rathborne}, {Jackson}, \&
  {Simon}}]{rathborne2006}
{Rathborne} J.~M., {Jackson} J.~M., {Simon} R., 2006, \apj, 641, 389

\bibitem[{{Reid} {et~al}\mbox{.}(2009){Reid}, {Menten}, {Zheng}, {Brunthaler},
  {Moscadelli}, {Xu}, {Zhang}, {Sato}, {Honma}, {Hirota}, {Hachisuka}, {Choi},
  {Moellenbrock}, \& {Bartkiewicz}}]{reid09}
{Reid} M.~J. {et~al.}, 2009, \apj, 700, 137

\bibitem[{{Reid} {et~al}\mbox{.}(1988){Reid}, {Schneps}, {Moran}, {Gwinn},
  {Genzel}, {Downes}, \& {Roennaeng}}]{reid88}
{Reid} M.~J., {Schneps} M.~H., {Moran} J.~M., {Gwinn} C.~R., {Genzel} R.,
  {Downes} D., {Roennaeng} B., 1988, \apj, 330, 809

\bibitem[{{Rosolowsky} {et~al}\mbox{.}(2010){Rosolowsky}, {Dunham}, {Ginsburg},
  {Bradley}, {Aguirre}, {Bally}, {Battersby}, {Cyganowski}, {Dowell},
  {Drosback}, {Evans}, {Glenn}, {Harvey}, {Stringfellow}, {Walawender}, \&
  {Williams}}]{rosolowsky10}
{Rosolowsky} E. {et~al.}, 2010, \apjs, 188, 123

\bibitem[{{Sault} {et~al}\mbox{.}(1995){Sault}, {Teuben}, \&
  {Wright}}]{sault95}
{Sault} R.~J., {Teuben} P.~J., {Wright} M.~C.~H., 1995, in Astronomical Society
  of the Pacific Conference Series, Vol.~77, Astronomical Data Analysis
  Software and Systems IV, {Shaw} R.~A., {Payne} H.~E., {Hayes} J.~J.~E., eds.,
  p. 433

\bibitem[{{Sobolev} \& {Deguchi}(1994)}]{sobolev94}
{Sobolev} A.~M., {Deguchi} S., 1994, \aap, 291, 569

\bibitem[{{Szymczak} {et~al}\mbox{.}(2007){Szymczak}, {Bartkiewicz}, \&
  {Richards}}]{szymczak07}
{Szymczak} M., {Bartkiewicz} A., {Richards} A.~M.~S., 2007, \aap, 468, 617

\bibitem[{{Szymczak} {et~al}\mbox{.}(2005){Szymczak}, {Pillai}, \&
  {Menten}}]{szymczak05}
{Szymczak} M., {Pillai} T., {Menten} K.~M., 2005, \aap, 434, 613

\bibitem[{{Taquet} {et~al}\mbox{.}(2013){Taquet}, {Peters}, {Kahane},
  {Ceccarelli}, {L{\'o}pez-Sepulcre}, {Toubin}, {Duflot}, \&
  {Wiesenfeld}}]{taquet13}
{Taquet} V., {Peters} P.~S., {Kahane} C., {Ceccarelli} C., {L{\'o}pez-Sepulcre}
  A., {Toubin} C., {Duflot} D., {Wiesenfeld} L., 2013, \aap, 550, A127

\bibitem[{{Titmarsh} {et~al}\mbox{.}(2013){Titmarsh}, {Ellingsen}, {Breen},
  {Caswell}, \& {Voronkov}}]{titmarsh13}
{Titmarsh} A.~M., {Ellingsen} S.~P., {Breen} S.~L., {Caswell} J.~L., {Voronkov}
  M.~A., 2013, \apjl, 775, L12

\bibitem[{{Voronkov} {et~al}\mbox{.}(2006){Voronkov}, {Brooks}, {Sobolev},
  {Ellingsen}, {Ostrovskii}, \& {Caswell}}]{voronkov06}
{Voronkov} M.~A., {Brooks} K.~J., {Sobolev} A.~M., {Ellingsen} S.~P.,
  {Ostrovskii} A.~B., {Caswell} J.~L., 2006, \mnras, 373, 411

\bibitem[{{Voronkov} {et~al}\mbox{.}(2014){Voronkov}, {Caswell}, {Ellingsen},
  {Green}, \& {Breen}}]{voronkov14}
{Voronkov} M.~A., {Caswell} J.~L., {Ellingsen} S.~P., {Green} J.~A., {Breen}
  S.~L., 2014, ArXiv e-prints

\bibitem[{{Walsh} {et~al}\mbox{.}(2011){Walsh}, {Breen}, {Britton}, {Brooks},
  {Burton}, {Cunningham}, {Green}, {Harvey-Smith}, {Hindson}, {Hoare},
  {Indermuehle}, {Jones}, {Lo}, {Longmore}, {Lowe}, {Phillips}, {Purcell},
  {Thompson}, {Urquhart}, {Voronkov}, {White}, \& {Whiting}}]{walsh11}
{Walsh} A.~J. {et~al.}, 2011, \mnras, 416, 1764

\bibitem[{{Walsh} {et~al}\mbox{.}(1998){Walsh}, {Burton}, {Hyland}, \&
  {Robinson}}]{walsh98}
{Walsh} A.~J., {Burton} M.~G., {Hyland} A.~R., {Robinson} G., 1998, \mnras,
  301, 640

\bibitem[{{Wilson} {et~al}\mbox{.}(2011){Wilson}, {Ferris}, {Axtens}, {Brown},
  {Davis}, {Hampson}, {Leach}, {Roberts}, {Saunders}, {Koribalski}, {Caswell},
  {Lenc}, {Stevens}, {Voronkov}, {Wieringa}, {Brooks}, {Edwards}, {Ekers},
  {Emonts}, {Hindson}, {Johnston}, {Maddison}, {Mahony}, {Malu}, {Massardi},
  {Mao}, {McConnell}, {Norris}, {Schnitzeler}, {Subrahmanyan}, {Urquhart},
  {Thompson}, \& {Wark}}]{wilson11}
{Wilson} W.~E. {et~al.}, 2011, \mnras, 416, 832

\bibitem[{{Xu} {et~al}\mbox{.}(2008){Xu}, {Li}, {Hachisuka}, {Pandian},
  {Menten}, \& {Henkel}}]{xu08}
{Xu} Y., {Li} J.~J., {Hachisuka} K., {Pandian} J.~D., {Menten} K.~M., {Henkel}
  C., 2008, \aap, 485, 729

\end{thebibliography}

\end{document}